\documentclass[twocolumn]{aastex631}
\usepackage{float}
\usepackage{amsmath}
\newcommand\norm[1]{\left\lVert#1\right\rVert}
\usepackage[caption=false]{subfig}
\usepackage{hyperref} 
\usepackage{makecell}
\usepackage{array}
\usepackage{dsfont}
\usepackage{gensymb}
\usepackage{longtable}
\hypersetup{
    colorlinks=true,
    linkcolor=blue,
    filecolor=magenta,      
    urlcolor=cyan,
    pdftitle={Overleaf Example},
    pdfpagemode=FullScreen,
    }  
\DeclareRobustCommand{\VAN}[3]{#2}
\let\VANthebibliography\thebibliography
\def\thebibliography{\DeclareRobustCommand{\VAN}[3]{##3}\VANthebibliography}

\shorttitle{Measurement of stellar rotations}
\shortauthors{Kamai \& Perets}

\begin{document}

\title{Accurate and Robust Stellar Rotation Periods catalog for 82771 Kepler stars using  deep learning}

\author[0009-0008-5080-496X]{Ilay Kamai}
\affiliation{Physics Department, Technion: Israel Institute of Technology,
Haifa 32000,
Israel}

\author[0000-0002-5004-199X]{Hagai B. Perets}
\affiliation{Physics Department, Technion: Israel Institute of Technology,
Haifa 32000,
Israel}
\affiliation{ACRO, Open University of Israel, R'anana, 
Israel}

\begin{abstract}
We propose a new framework to predict stellar properties from light curves. We analyze the light-curve data from the Kepler space mission and develop a novel tool for deriving the stellar rotation periods for main-sequence stars. Using this tool, we provide rotation periods for more than 80K stars. Our model, LightPred, is a novel deep-learning model designed to extract stellar rotation periods from light curves. The model utilizes a dual-branch architecture combining Long Short-Term Memory (LSTM) and Transformer components to capture temporal and global data features. We train LightPred on self-supervised contrastive pre-training and simulated light curves generated using a realistic spot model. Our evaluation demonstrates that LightPred outperforms classical methods like the Autocorrelation Function (ACF) in terms of accuracy and average error. We apply LightPred to the Kepler dataset, generating the largest catalog to date. Using error analysis based on learned confidence and consistency metric, we were able to filter the predictions and remove stellar types with variability which is different than spot-induced variability. Our analysis shows strong correlations between error levels and stellar parameters. Additionally, we confirm tidal synchronization in eclipsing binaries with orbital periods shorter than 10 days. Our findings highlight the potential of deep learning in extracting fundamental stellar properties from light curves, opening new avenues for understanding stellar evolution and population demographics.

\end{abstract}

\keywords{astronomy --- Machine learning --- light curve ---star spots}

\section{Introduction} \label{sec:intro}
The study of stellar rotation periods has become increasingly important in our understanding of stellar evolution, activity cycles, and exoplanet detection. With the advent of space-based photometry missions like Kepler, CoRoT, and TESS, we now have access to high-precision, long-duration light curves for hundreds of thousands of stars. These light curves often exhibit periodic variations due to starspots rotating in and out of view, allowing us to measure stellar rotation periods with unprecedented accuracy.

Classical methods for measuring rotation periods, such as Lomb-Scargle periodograms and autocorrelation functions, have been successful (\cite{McQuillan2013}, \cite{Reinhold2013}, \cite{Walkowicz2013} \cite{Nielsen2013}, \cite{McQuillan2014},
\cite{Garcia_2014}, \cite{Santos_2019},
\cite{Santos2021}, \cite{Reinhold2023}) but can be time-consuming and challenging to apply to large datasets. Furthermore, these methods often struggle with complex light curves that exhibit multiple periodicities or evolving spot patterns. In recent years, machine learning techniques have shown promise in automating and improving the accuracy of rotation period measurements. Deep learning models, in particular, have demonstrated the ability to extract complex features from time series data and make accurate predictions across a wide range of stellar types and rotation periods.

In this paper, we present a novel deep-learning approach for measuring stellar rotation periods using Kepler light curves. Our method, which we call LightPred, combines different neural network architectures to capture both local and global temporal features in the light curves. We train our model on a combination of simulated and real Kepler data, allowing it to generalize well to a variety of stellar types and activity patterns.

In the following, we first introduce the necessary background and context for our analysis, discuss previously used methodologies, and then present our analysis and results. Our code is open-sourced at \url{https://github.com/IlayMalinyak/lightPred}.\\

In the following subsections, we present existing works that use classical and data-driven approaches for period determination. We note that this is not intended to be a complete survey of the field, but rather a few pertinent examples.
\subsection{Classic rotation measurements analysis methods}
Various works have been done to directly infer the rotation period from light curves. We refer to those methods, which do not use learning or optimization processes, as "classical methods".
\cite{Basri2011}, \cite{Reinhold2013} and \cite{Nielsen2013} used Lomb–Scargle periodogram \citep{Lomb1976,Scargle1982} to detect the period. A different method was used in \cite{McQuillan2013} and \cite{McQuillan2014}, and called Autocorrelation function (ACF) - 
Given a time series data $x={x_1...x_N}$, ACF measures the correlation between the data and a lagged version of it and is calculated as follows: 
\[r_k = \frac{\sum_{i=1}^{N-k}(x_i-\bar{x})(x_{i+k}-\bar{x})}{\sum_{i=1}^N(x_i-\bar{x})^2}.\]
where $r_k$ is the autocorrelation coefficient at lag k. It can be seen that $r_k$ is high if the input has a periodicity of length $k$. After applying ACF for a range of lags, the result is another time series data with peaks at the period and its multiplications. \\
another common method is the wavelet transform \citep{Morlet1984,Torrence1998}. Wavelet transform is the result of multiplying a function with some scaled and translated version of a specific response function (usually called the "wavelet mother"). The Continuous Wavelet Transform (CWT) of a function $x(t)$ at scale $a$ and translation $b$ is the following 2 parameters function: \[W(a,b) = \frac{1}{|a|^{\frac{1}{2}}}\int_{-\infty}^\infty x(t)\psi^\ast(\frac{t-b}{a})dt\]
where $\psi^\ast(t)$ is the complex conjugate of the mother wavelet. Wavelet transform can be seen as Short Time Fourier Transform (STFT) with varying window sizes. \\
To compare different methods in a controlled process, \cite{Aigrain2015} conducted a blind test to compare the different methods for period calculation. It was done using a dataset of 1000 simulated light curves where the underlying parameters are known. Among the suggested methods, the ACF method was the only method that was applied to the entire dataset and achieved superior results relative to other methods. Attempts were also made to combine multiple methods; \cite{Garcia2014},\cite{Santos_2019}, and  \cite{Santos2021} have used both Wavelet transform and ACF. \cite{Santos2021} also used a machine learning algorithm (suggested by \cite{Breton2021}) to select the most reliable period between the different results. The probabilistic approach was also tested; \cite{Angus_2017} used Gaussian processes to predict periods of 1132 Kepler object of interest {\cite{Shapiro2020} which suggested a variant of the wavelet method is also worth mentioning}. This method, called Gradient Power Spectrum (GPS), takes the highest value of the gradient of the wavelet transform (inflection point) as an indicator for the period. \cite{Reinhold2023} used both ACF and GPS to create the largest current sample of stellar periods. They used a scoring function to decide whether to use the ACF prediction or the GPS. This suggests that ACF and GPS are the preferred methods to analyze periods of large samples in a non-learning fashion.
One shortcoming of classical methods is their inability to provide uncertainty estimates for predictions, which are essential for evaluating the reliability of the results. To address this, statistical measures are often applied. \cite{Nielsen2013} used the median absolute deviation (MAD) of the quarterly predicted period from the median period to estimate uncertainty. \cite{McQuillan2013} and \cite{McQuillan2014} also used a variant of the MAD estimation. \cite{Santos_2019}, \cite{Santos2021} used the half width at half maximum (HWHM) of a Gaussian profile fitted to the highest peak of the wavelet power spectrum, and \cite{Reinhold2023} used the difference between the power spectrum at the inflection point and the minimum power as a signal to noise estimation. A different and important example is \cite{Angus_2017}, which used a probabilistic model that outputs uncertainties inherently.

\subsection{Machine-learning analysis methods}
 A different approach is a data-driven approach. In this approach, one uses a dataset of light curves with known parameters to train a model and predict the desired properties. Then, using the trained model, one can predict properties on real data samples. Several works have been done in the field of period predictions: \cite{Lu2020} used a machine learning model that was trained to predict long rotation periods from short-duration measurements. They used MAD as an uncertainty estimation. \cite{Breton2021} used Random Forest algorithms to perform classification and vetting - first predicting if the sample is periodic enough, then predicting if it is not polluted by Pulsators and binaries, and lastly, choosing the best period from different classical methods. For uncertainty estimation, they used HWHM. For the task of period prediction, few attempts were made to use neural networks- \cite{Blancato2022} used a simple CNN network to predict periods among other properties of Kepler stars. For Labels, they used values that were calculated using the ACF method from \cite{McQuillan2014}. As such, the model is restricted to the ACF results. \cite{Claytor2022} used simulations to train a 2d CNN model, with wavelet transform as input, and predicted periods of TESS-like simulated samples. based on that work, \cite{Claytor_2024} used the same model to predict the period of 7245 TESS samples. Their model is optimized to predict, in addition to the period,  uncertainty based on MAD.\\
Each of the approaches, data-driven and classical, has pros and cons. While the data-driven approach relies on a model that can learn complex high-dimensional functions, it is much less interpretable. It assumes that the simulated data represent the real data sufficiently. Classical methods are more sensitive to parameter tuning and usually less robust as they are applied only to a subset of the samples. \\
In this work, we suggest a data-driven method to derive the inherent parameters of stars using a deep learning model. Our work brings the following contributions
\begin{itemize}
    \item We use a simulation-based deep learning model combined with self-supervised learning. Simulation-based learning for period determination was done only by \cite{Claytor2022} and \cite{Claytor_2024} for predictions of TESS samples and not for \textit{Kepler} samples. To the best of our knowledge, the use of self-supervised learning for period determination was not done at all.
    \item We use new error metric which combines a learned confidence level, optimized together with the predictions, and a consistency measure based on the ability to reproduce the same predictions on different segments of the same sample.
    \item We predicted periods for a large sample of main sequence stars for the first time with better accuracy compared to established methods. 
\end{itemize}

The paper is organized as follows:
In section \ref{sec:Data}, we discuss the creation of the mock dataset for the model's supervised training, as well as the pre-processing stages for both simulated and \textit{Kepler} data. Section \ref{sec:Model} discusses the model's architecture. In section \ref{sec:Training}, we discuss the training pipeline on both simulations and \textit{Kepler}, and in Section \ref{subsec:Results}, we present the results on both simulated and \textit{Kepler} data, analyze the model's errors, discuss possible pollution in the dataset, and different ways to constrain the results. Section \ref{sec:conclusions} contains the conclusion and discussion. The appendix contains supplementary material with further examples, implementation details, additional figures, and tables.

\section{Data} \label{sec:Data}
\subsection{Mock Dataset} \label{subsec:mock_data}
To create a dataset for supervised training we used simulated light curves using \textbf{Butterpy}, a python package described in \cite{Claytor2022}. The simulation is based on a simplistic spots model that alters the flux observed from a star, based on the spot's activity and configuration, and creates a unique light curve. The spots emerge and decay as Gaussian with characteristic emergence and decay time scales. In addition, each spot is located at a random latitude (within a range that is chosen from prior distribution) and longitude and can either decay towards the equator (to create the known butterfly shape) or not. The suggested model implies an inherent degeneracy between inclination and spot location. This was already observed by \cite{Walkowicz2013}, \cite{Basri2020}, and \cite{Hogg2021}, who suggest that without knowing the full spot distribution of each star, learning the stellar inclination is limited. As stated in \cite{Hogg2021}, This implies that the model is prior-depended: the distribution of spots in the model has a large effect on the results. Our sun has unique spot distribution - the spots emerge on the latitude of maximum $\sim 30$ degrees and decay towards the equator \citep{Hathaway2011}. 
This decay, with a time scale of the solar cycle and some overlap between cycles, creates the known butterfly effect. However, it is not well known whether this pattern (latitude distribution, decay, and cycle overlap) is consistent with other stars. While butterfly pattern was observed in some stars \citep{Netto2020,Bazot2018}, other stars showed no decay pattern and wide latitude distributions. This was found in both solar-like stars \citep{Thomas2019} and in younger and more active stars \citep{Mackay2004}. Observations of wider latitudes on more rapidly rotating stars \cite[as seen in][]{Mackay2004} suggest that there's a relation between stars' period and spot distribution. Indeed, \cite{Morris2020} suggested a relation between stellar age and spot total coverage. Combining this with the known period-age relationships (\cite{Skumanich1972}, \cite{barnes2003}, \cite{barnes2007}, \cite{Barnes2010}, \cite{Mamajek2008}, \cite{Angus2019},\cite{Bouma2023}), can lead to a relation between stellar period and spot total coverage. However, as stated in \cite{Morris2020}, this relation is of a statistical nature, and applying this relation to the spot distribution of a specific star is not trivial. As such, it was not implemented in our simulation. We will elaborate on this in section \ref{subsec:Results}. 

From the above discussion, it is clear that identifying the line-of-sight inclination of a star is potentially much more challenging compared with identifying a stellar rotation. Although both depend on the spot distribution and behavior, the inclination determination is more sensitive to the spot distribution of a given star, of which we still have a limited understanding. We therefore treat inclination and period predictions separately; in this paper, we investigate the period predictions and treat inclinations as an auxiliary variable, while in a future paper, we will focus on the more complex issue of inclination determination. Nevertheless, the information from the inclination of samples is still valuable for this work. By training a model to predict periods and inclinations at the same time we rely on the fact that mutual learning can benefit both aspects independently. In section \ref{subsec:ablation} we show this explicitly for the period determination, but this is also backed with simple intuition: On the one hand, the spots distribution and the stellar inclination will determine which and how many spots will rotate in and out of view and hence determine the amplitude of the periodic-changes in the light curve; these would also affect aspects of differential rotation, as the period might change at different latitudes. When predicting the equatorial rotation, the inclination information might help with "locating" the equator and determining the noise level from spots, which do not rotate out of view. On the other hand, the information about inclination is affected by the spots that rotate with the stellar period, and the amplitude of the periodic signal related to the ratio between regions in the stars that go in and out of view and those that do not, which depend on the inclination. Indeed, such inclination measurement aspects using the rotation measurement amplitude were demonstrated by \cite{Mazeh2015}.

As mentioned above, prior distributions have a large effect on the results and we chose slightly different parameters than \cite{Claytor2022} and \cite{Aigrain2015}. specifically, for the inclination of the star, $i$, we used an isotropic distribution which translates to a uniform distribution in $cos(i)$ instead of an isotropic uniform in $sin^2(i)$, chosen in the original papers. This is a result of observational bias - for isotropic distribution of inclinations, we will see many more equator-on stars than edge-on. For the period, we used the same distribution as in \cite{Aigrain2015}. The full parameters of the simulation distributions are specified in Table 1. 

\subsection{pre-processing} \label{subsec:pre_process}
We pre-processed each sample in the following way: First, we cut the first 200 days to reach a steady state for spot evolution. Then we randomly chose a 720-day subset of the light curve (corresponding to 8 quarters of Kepler data). Next, we injected noise from real Kepler data. To do this we followed the same procedure as \cite{Aigrain2015} and used the results from \cite{McQuillan2014} which assigned the 'periodicity' parameter $w$. $w$ is a parameter between $0$ to $1$ which is a function of the relative height of the largest peak of the ACF (Local Peak Height - LPH), the effective temperature, and the period predicted by the ACF. In their work, \cite{McQuillan2014} considered only samples with $w > 0.25$ for period calculation. We chose samples with $w < 0.04$ which results in 6981 stars which we refer to as "noise". For each simulated sample, we randomly chose a star from the "noise" samples and median normalized it. The addition of the noise was performed in two ways. The first is simply by adding the noise to the sample. The second involves scaling the noise sample such that $std(noise)=\alpha*std(sample)$, where $\alpha$ is a random number between $0.02-0.05$, before adding it. In this way, we inject \textit{Kepler} data characteristics into the light curve while keeping the signal-to-noise ratio small. The reason we implemented two approaches is to be able to directly compare our results with \cite{Aigrain2015}. We refer to samples that were processed through the first method as "noisy samples" and samples that were processed using the second method as "noiseless samples". We then applied a Savitzky-Golay filter with a window of 13 time-points ($6h$) and created a time series of differences between consecutive time stamps to reduce long-term variations. We decided to use the differences of a light curve instead of the original light curve; While computing the time series differences might be a non-trivial choice, it is a known technique to remove trends \citep{Hyndman2018} and can be thought of as an approximation of the derivative. As such it removes linear trends while keeping the periodicity. We also justify this choice experimentally in \ref{subsec:ablation}. We refer to the result of this stage as Diff-Lc. Next, we calculated the ACF of the Diff-Lc with all possible lags and filled missing values by linear interpolation. Note that the calculation of ACF on the different light curves is different from that of the common use of ACF (where it is calculated on the light curve itself). We therefore use the term Diff-ACF to emphasize this difference. Lastly, we normalized both the Diff-ACF and Diff-Lc to zero mean and unity standard deviation. The result is a two-channel time series which is the input for the model. An example of the simulated light curve and the pre-processing stages is presented in Figure 1. More examples can be found in Appendix \ref{Appendix_A}
\begin{table*}
\centering
\begin{tabular}{||c c c||} 
 \hline
 parameter & range & distribution  \\ [0.5ex] 
 \hline \hline
 Inclination $i$ & 0\textdegree-90\textdegree & uniform in cos(i) \\ 
 \hline
 Period $P_{eq}$ & 10-50 (90\%) 0-10 (10\%) days & log-uniform \\
 \hline
 Spot lifetime $\tau_{spot}$ & 1-10 Period & log-uniform \\
 \hline
 Activity Cycle length $T_{cycle}$ & 1-10 Years & log-uniform  \\
 \hline
 Activity Cycle overlap $T_{overlap}$ & 0.1-$T_{cycle}$ & log-uniform \\ [1ex] 
 \hline
\end{tabular}

\caption{\label{demo-table}Parameters for simulated light curves. parameters that are not specified here are taken from \cite{Aigrain2015}.}
\end{table*}

\begin{figure*}
    \centering
    \includegraphics[scale=0.3]{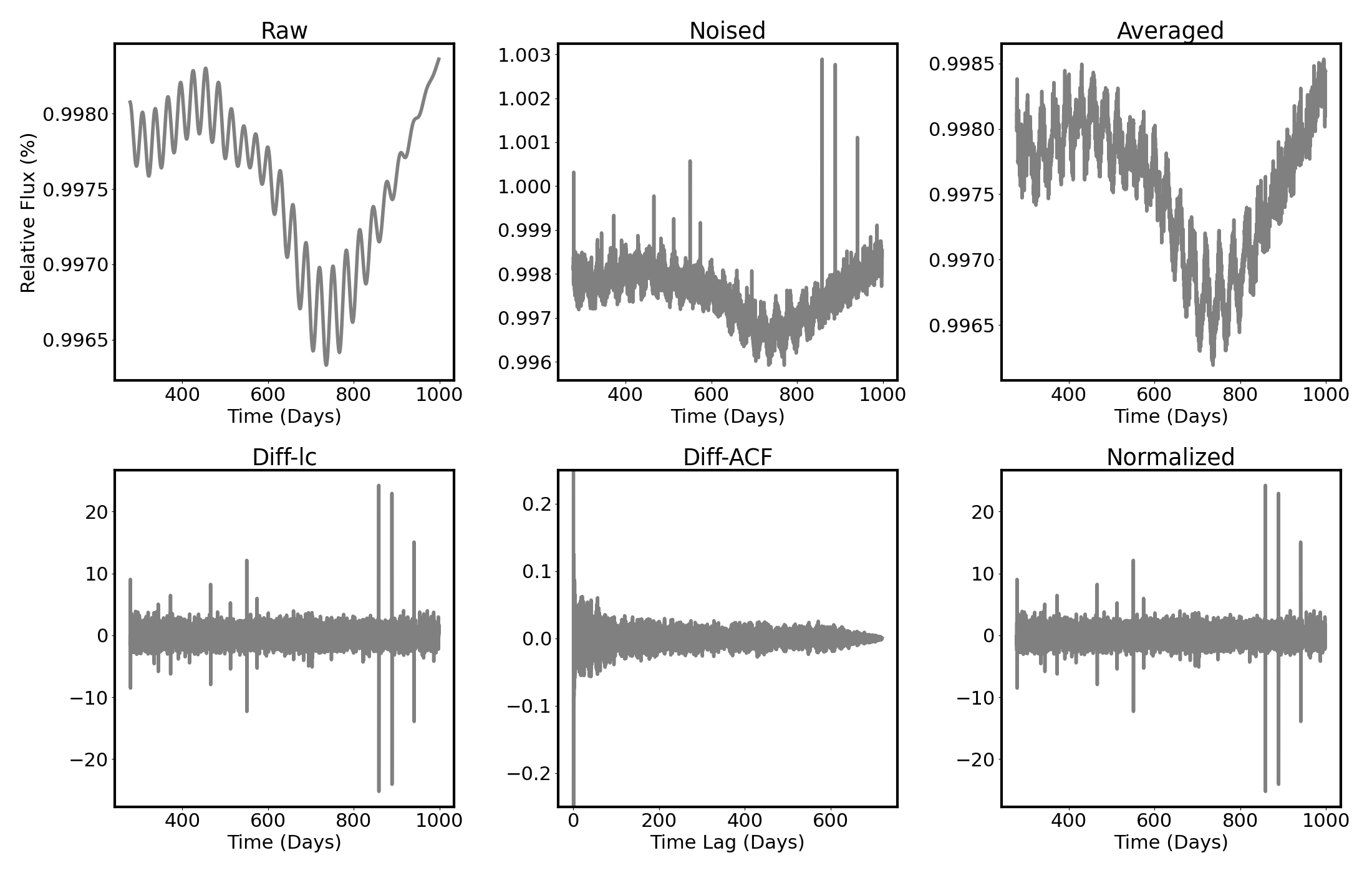}
    \caption{Example of a simulated light curve and the different pre-processing stages. The stages are shown sequentially from the upper left to the lower right. On the upper left panel is the raw light curve, randomly cropped to the 720-day segment. The upper middle panel shows the light curve after the addition of \textit{Kepler} noise, and the upper right panel shows the resulting light curve after averaging with a 6h window. The lower left panel shows the light curve differences (Diff-Lc), and the lower middle panel shows the resulting normalized Diff-ACF. The lower right panel shows the normalized differences light curve. The input for the model is a 2-channel time series corresponding to the last two panels.}
    
\end{figure*}

\subsection{Kepler dataset}\label{subsec:kepler data}
In addition to the simulated light curves, we also created a dataset from Kepler data. First, we downloaded the long cadence \textit{Kepler} dataset Data Release 25 \citep{kepler_dr25}. The data is available at \url{https://doi.org/10.17909/T9488N} \\
Similarly to \cite{McQuillan2014} and \cite{Reinhold2023}, we omitted  Q0-Q2 and Q17 and used the data corrected for instrumental systematic using PDC-MAP \citep{Stumpe2012}, \citep{Smith2012}. We then divided each quarter by the maximum value, normalized it so that all quarters would have the same average, and stitched together all the quarters. We then pre-processed the sample in the same way as the mock data, except for noise addition. For the contrastive learning setting, which we will elaborate on later, we added the following data augmentations: \\
\textbf{Masking} similarly to \cite{Morvan2022}, we masked $20\%$ of the time points and replaced $20\%$ of the masked points with a random value from a uniform distribution between $-2$ and $2$. \\
\textbf{Gaussian Noise} we added Gaussian noise with standard deviation of $10^{-4}$. \\
\textbf{Shuffle} this augmentation shuffles different quarters of the same \textit{kepler} sample. We acknowledge that shuffling quarters (90 Days) might remove phase information and potentially hinder the detection of long periods. However, we believe that the benefits of shuffling in terms of improving the model's ability to generalize outweigh the potential drawbacks. 

For each sample in a self-supervised contrastive setting, in addition to the pre-processing stage, we chose one of the following augmentations: masking, Gaussian noise, shuffle, or identity (no augmentation).  
The result is a 2-channel time series data, similar to the simulated data. 

\section{Machine Learning Model} \label{sec:Model}
In recent years, deep learning models have shown promising results on a wide variety of time series tasks (for reviews on forecasting and classification you can look at \cite{Lara-Ben2021}, \cite{fawaz2018}). To find an optimal model, we started by using classical architectures for time series analysis, namely, CNN and RNN. CNNs are well known to work well on time series data. Indeed, both \cite{Blancato2022} and \cite{Claytor_2024} used them as an architecture to analyze light curves. RNNs (Recurrent Neural Networks), with their sequential nature, are also effective in analyzing time series. however, the vanilla architecture of RNNs suffers from vanishing/exploding gradient as noted by \cite{bengio94}. An Alternative that aims to improve this problem is the Long-Short-Term Memory (LSTM) network \citep{lstm} which adds gates that selectively retain or discard information. Using this intuition, our basic approach used a combination of CNN and bi-directional LSTM. This showed good performance on noiseless samples but was less ideal when adding noise (as explained in \ref{subsec:pre_process}). To better handle noisy samples we added another branch which is transformer-based. Transformers, which revolutionized the field of natural language processing, were also shown to be very effective for time series analysis. This gave rise to a family of transformer-based architectures including Informer \citep{Zhou2021}, Autoformer \citep{Wu2021}, and Conformer \citep{gulati2020}. Indeed, \cite{Morvan2022} used a transformer-based model to denoise light curves, and \cite{Pan2024} used them to predict \textit{logg} from light curves. We therefore added a transformer-based branch which is motivated by the work of \cite{Pan2024}. The addition of the extra branch performed better than each branch separately and improved the results on noisy samples. In section \ref{subsec:ablation} we present the metrics used to evaluate different models and show an ablation study for different parts of the model.

\subsection{lightPred model} \label{subsec:lightpred}
In the following we provide an overview of the model; detailed model schematics can be found in Appendix A.

\textbf{Temporal branch} - consists of an "embedding" convolution block with max pool, dropout, batch normalization, skip connection, a bi-directional LSTM block, and a non-learnable self-attention block. The convolution block gets the Diff-ACF, transforms it into a multi-channel 2d input, and serves as an effective embedding layer. The LSTM block processes the short and medium-range dependencies and the attention block (with the LSTM's final cell state as queries and the features of the last layer as keys and values) enhances the long-term memory of the network.

\textbf{Global branch} - here we follow the recent work of \cite{Pan2024} called \textbf{Astroconformer}. They used a modified version of the Conformer \citep{gulati2020} architecture which combines multi-head self-attention with convolution. The main change in \cite{Pan2024} is the use of rotary positional embedding \citep{Su2023}. We used a slight modification of Astroconformer; in the conformer block, instead of 2 convolution layers after the multi-head self-attention (MHSA), we used one convolution and another MHSA. This showed to have better performance. The input of this branch is Diff-Lc.

The features from the two branches are concatenated and sent into the prediction head; two linear layers with Gaussian Error Linear Unit (GELU) activation function and a dropout between them. GELU \citep{Hendrycks2016} is an activation function similar to the well-known Rectified Linear Unit (RELU). However, GELU is smoother, has non-zero values on the negative regime, and is differentiable everywhere which makes it a better alternative. The output of the model is four numbers: rotation period, inclination, period confidence score, and inclination confidence score. To find optimal hyper-parameters for the model we followed a two-step search. First, we tuned the temporal branch using \textbf{optuna} \citep{optuna_2019}, a Python package for hyperparameters optimization in deep learning frameworks. 
Then, we tuned the parameters of the entire model manually. The resulting hyperparameters are listed in Appendix \ref{Appendix_B}. Figure 2 shows a high-level architecture of the model. A more detailed diagram of the different parts of the model can be seen in Appendix \ref{Appendix_B}.

\begin{figure*}
    \centering
    \includegraphics[scale=0.6]{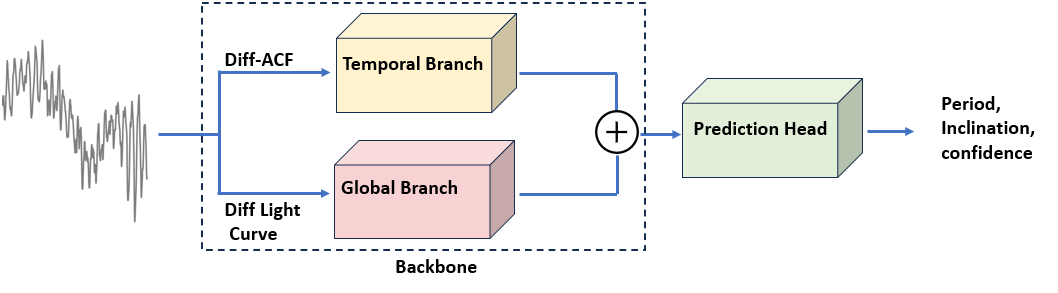}
    \caption{High-level diagram of the lightPred model.}
\end{figure*}

\section{Training} \label{sec:Training}
\subsection{Contrastive Learning} \label{subsec:contrastive}
Before presenting the training scheme we used, we give a short background on Self Supervised Contrastive Learning. Contrastive learning is a branch of representation learning. The goal of representation learning is to map high-dimensional data into meaningful low-dimensional representation. Maybe one of the most well-known examples of a simple representation algorithm is principal component analysis (PCA). Since the advances of deep learning models, representation learning has become a very common task for deep neural networks. 
Generally speaking, deep learning models can be divided into two families - \textit{generative} and \textit{discriminative} models. Given a dataset, $X$, and labels $Y$, generative models try to learn the probability distribution of the data, $\{x \in X\}$, namely $p(x)$ or $p(x|y)$ (with or without labels) using latent representation, $z$, of the input, $x \in X $. An example of generative models is Generative Adversarial Network (GAN) \cite{goodfellow2014} and Variational AutoEncoders (VAE) \cite{kingma2022}]. 
Discriminative models, on the other hand, learn the conditional probability $p(y|x)$. Contrary to generative models, which can be trained with or without labels, discriminative models require labels. Self-supervised learning tries to apply a discriminative approach to the case where no labels are available. It is done using a \textit{pretext task}, i.e. a task that the model can solve without the true labels. A good pretext task is a task that will enforce the model to learn meaningful representation during training that will be easily adapted to other downstream tasks. An example of a pretext task is to predict an angle in which an image is rotated \citep{gidaris2018}, or prediction of masked elements \citep{devlin2019}. 
Contrastive learning is a type of self-supervised learning where a specific pretext task is used. The task in contrastive learning is to find representations that are invariant to different views of the input. Usually, input is divided into \textit{positive} and \textit{negative} pairs and different views can be seen as different augmentations of the elements in the pair. The network would then be optimized such that positive pairs are close in the latent space and negative pairs are far away. This is usually done using a loss function that maximizes the similarity for positive pairs and minimizes the similarity for negative pairs. A well-known example is the SimCLR network \citep{chen2020}. A modification for the SimCLR network was suggested by \cite{grill2020} (BYOL) and \cite{chen2020simsiam} (SimSiam) which can be seen as "SimCLR with positive pairs only". Those networks only try to maximize the similarity between augmented views of positive pairs. To avoid collapsing to a trivial solution, \cite{chen2020simsiam} suggested the following architecture (in the following we will use the notations of \cite{chen2020simsiam}): given two randomly augmented views $x_1$ and $x_2$ of the same input $x$, the two views are processed through an encoder network, $f$, consisting of a backbone and Multi-Layer Perceptron (MLP) head. A prediction head, $h$, which is another MLP, transforms the output of one view and matches it to the other view. Given $p_1 = h(f(x_1)$ and $z_2 = f(x_2)$ we define a loss function that minimizes the negative cosine similarity between them: \[\mathcal{D}(p_1, z_2) = -\frac{p_1}{\norm{ p_1}_2}\cdot \frac{z_2}{\norm{ z_2}_2}\] an important character of this loss function is called the "stop gradient". It means that gradient would propagate only through $p_1$, and not through $z_2$. The total loss function of the model is the symmetric loss: \[\mathcal{L} = \frac{1}{2}\mathcal{D}(p_1, z_2) + \frac{1}{2}\mathcal{D}(p_2, z_1)\]. In their paper, \cite{chen2020simsiam} show that using this loss function together with the stop gradient property the model can converge to a non-trivial solution.
Many works explored self-supervised learning for time series data. Recently, \cite{zhang2024} published a review on the different methods in time series self-supervised learning. In the field of self-supervised learning in astrophysics, most of the work has been done on images and spectra as reviewed in \cite{Huertas2023}. It is worthwhile to mention the study by \cite{Morvan2022} which used a self-supervised framework to reduce noise from TESS light curve samples using predictions of masked time stamps.

\subsection{2-steps Training} \label{subsec:pipeline}
To train the model, we used a 2-step pipeline; First, we trained the model, without the prediction head, in a self-supervised fashion, with \textit{Kepler} samples. We call this step a Self Supervised Learning (SSL) step. Next, we used the weights from the SSL Step as an initialization for supervised training with the mock dataset. We call this step Simulation Based Supervised Learning (SBSL) step. The SSL step was done using the Contrastive framework that was suggested by \cite{chen2020simsiam} as explained above. As mentioned in section \ref{subsec:kepler data}, for the SSL step we used special augmentations in addition to the regular ones: Shuffle, Gaussian Noise, and Masking. For each sample, we randomly chose 2 special augmentations (including Identity - no augmentation) and created 2 different views of the same sample. It is important to emphasize that we use those augmentations only at the SSL step and not in later steps (SBSL Step and inference on \textit{Kepler} data). The reason is that those augmentations improve the latent representation founded by the model but it is not intuitive that they would benefit the actual predictions in a supervised setup.\\
As explained in \ref{subsec:contrastive}, In the SSL step the model consists of the backbone (LightPred without prediction head), projection Multi-Layer Perceptron (MLP), and prediction MLP. For the projection MLP we used 3 Layers with dimension of 64. For the prediction MLP, we used 2 layers with dimensions 32 and 64. Figure 3 shows a diagram of the training pipeline. \\ After the SSL step we used the trained LightPred backbone as a starting point for the SBSL. This was done on a dataset of 50000 simulated samples of 1000 days that were generated as explained in \ref{subsec:mock_data} and \ref{subsec:pre_process}. We then split the dataset into training ($81 \%$), validation ($9 \%$), and test sets ($10 \%$). This gave us 5000 samples for testing and 4500 samples for validating the model which is representative. While this is more than the 1000 samples used by \cite{Aigrain2015} to test their models, this is much lower than \cite{Claytor2022} and \cite{Claytor_2024} which used 1 million samples for training. We experimented with doubling the dataset size to 100000 samples and found no significant change in the results so we decided to keep the dataset size with 50000 samples. \\ During training, we scaled the ground truth labels to the range $0-1$ and used the following loss function
\[\mathcal{L} = \mathcal{L}_1(\hat{c}, c) + \mathcal{L}_1(\hat{p}, p),\]
where $\mathcal{L}_1$ represents the L1 loss, $c$ and $p$ are the absolute deviation and ground truth values respectively, and $\hat{c}$, $\hat{p}$ are the models' predictions for those values. We then defined the confidence of the model as $1-\hat{c}$. \\ During training, We used AdamW optimizer \citep{adamw}. When optimizing a neural network using gradient descent, one of the most critical hyperparameters is the learning rate, which controls the step size in the direction opposite to the gradient. A low learning rate results in slow convergence, while a high learning rate can cause the optimization to oscillate without settling into a minimum. Although various techniques exist to find an optimal learning rate and adjust it dynamically during training, in this work, we used a fixed learning rate of $5 \times 10^{-4}$, determined experimentally.

\subsection{Ablation Study} \label{subsec:ablation}
Before comparing our model to known results, We performed an ablation study to test the different parts of the model. An ablation study is a systematic approach used to understand the contribution of individual components within the model. This is done by selectively removing or modifying specific parts of the model or its training process and evaluating the resulting performance. In this study, we would like to test not only the high-level architecture but also the use of Diff-Lc, Diff-ACF, and the use of inclination as an auxiliary prediction, since they are all nontrivial choices that are worth justification. To perform the study we trained multiple models using the pipeline described in \ref{subsec:pipeline} on the same dataset, and compared the results. The metric we used to evaluate the models is the 10 percent accuracy - the percentage of points with predictions within 10\% absolute error from the ground truth. the different models we tested are:
\begin{enumerate}
    \item LightPred: The full model as shown in \ref{subsec:lightpred}.
    \item LightPred no-diff: The same as LightPred with a change in the pre-processing; instead of using Diff-Lc and Diff-ACF, we used the regular light curve and ACF.
    \item Temporal Branch: Using only the Temporal branch; LSTM-based network with Diff-ACF as input.
    \item Conformer Branch: Using only the Conformer branch; a convolution-transformer network with Diff-Lc as input.
    \item LightPred period only: the same as LightPred but without inclination predictions. Namely, predictions of only the period. 
\end{enumerate}
Table 2 shows the results of the study. We can see that the best performance is achieved by the full model as described in \ref{subsec:lightpred}. In addition, we see that both the use of Diff-Lc and Diff-ACF and the use of inclination predictions improved the performance of the model. In the next sections, by using the term \textit{LightPred} we mean the full model, with Diff-Lc, Diff-ACF, and inclination predictions as explained in \ref{subsec:lightpred}. 

\begin{figure*}
    \begin{center} 
    \includegraphics[scale=0.7]{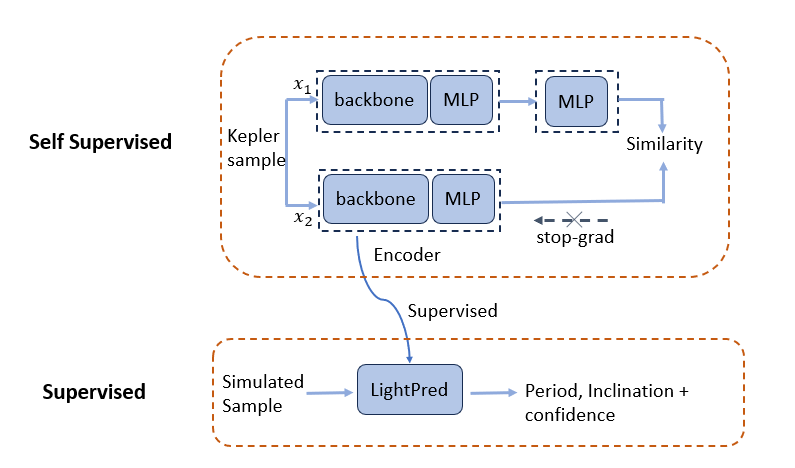}
    \caption{Training Pipeline. In the Self Supervised Step, the LightPred model is trained without the prediction head.}
    \end{center}
\end{figure*}

\begin{table*}
\centering
\begin{tabular}{||c c||} 
 \hline
 model & acc10p (\%) \\ [0.5ex] 
 \hline\hline
 LightPred no-diff & 76   \\
 \hline
 Temporal Branch  & 75 \\
 \hline
 Conformer Branch & 76 \\
 \hline
 \textbf{LightPred}  & \textbf{78}  \\
 \hline
 LightPred period only  & 73  \\
 \hline
 \hline
\end{tabular}
\caption{Ablation study results.}
\end{table*}

\section{Results} \label{subsec:Results}

\subsection{Results on simulated data} \label{subsec:Results on simulated data}
To compare our results with \cite{Aigrain2015} we created another 50000 samples dataset, this time with the same prior distributions as in \cite{Aigrain2015}. Note that the main difference in the distributions is in the inclinations distribution, where \cite{Aigrain2015} used uniform in $\sin^2(i)$, while we used uniform in $cos(i)$. We also adopted the same metrics - the percentage of samples that lie within 10\% error (\textbf{acc10p}) and the percentage of samples that lie within 20\% error (\textbf{acc20p}). In addition, we used the average absolute error in days (\textbf{mean\_err}).

We compare our results to the results reported by the Tel-Aviv team in \cite{Aigrain2015}. We chose the Tel-Aviv team, as it obtained the best overall results in the comparison study; in addition, they were the only ones who used the ACF method and predicted periods for the entire dataset. The results are shown in Table 3. It can be seen that our results are better for both noisy and noiseless samples.\\
To take into account changes between the dataset provided in \cite{Aigrain2015} and the dataset we used, and to enable a more detailed comparison, we also implemented ACF and GPS methods and tested them on our test set. 
Figure 4 shows prediction plots of our model, our ACF, and our GPS implementations. Comparing the plots visually we can see that our model is more robust; both the ACF and the GPS models have more outliers and a non-negligible number of points that were effectively not predicted. In the case of ACF, when there are no high enough peaks we set the period to zero (essentially the period can not be determined). In the case of GPS, many samples were predicted with the maximum possible period (probably due to long-term trends in the data). None of this is apparent in our model's result.
To compare our model and ACF/GPS qualitatively we used two sets - the entire test dataset and a subset of only valid points. In the case of ACF, valid points are points with period $>0$ and in the case of GPS we took periods that are lower than the maximal period. Table 3 shows this comparison with ACF and Table 4 shows the same for GPS, both for noisy samples. It can be seen that both ACF and GPS predicted effectively $\sim 85\%$ of the dataset while our model predicted the entire dataset. For a subset of valid predictions, we see that the performance of ACF is comparable to our model in terms of acc10p and acc20p but our model shows significantly lower average error. In the case of GPS, our model is better in all aspects for both the entire dataset and the subset of valid predictions. We conclude that our model is more robust than current established methods. On a subset of points that are detectable by ACF, the performance of our model and that of ACF are comparable, and both our model and ACF have shown to be better than GPS.

An important difference between our model and other methods is the confidence level. In our model, a confidence estimate is a natural product. As explained in \ref{subsec:pipeline}, we train the model to predict, in addition to the period and inclination, the $L1$ loss, normalized to the range $0-1$, for each label (period/inclination) and for each sample which we define as $\hat{c}$.  The confidence is then defined as $1-\hat{c}$. To see the relation between the confidence value and the model's performance we show in Figure 5 the period confidence value vs the absolute error (absolute value of the difference between true period and predicted period). The trend of high confidence and low error, which can be easily seen in the figure, implies that the model was able to infer which samples are "harder" for it to determine the stellar properties. It is possible to get confidence functions also from ACF and GPS, but usually, it requires extra processing of the results. \\

\begin{figure*}
    \begin{center}
  \begin{minipage}[b]{0.4\textwidth}
    \includegraphics[width=\textwidth]{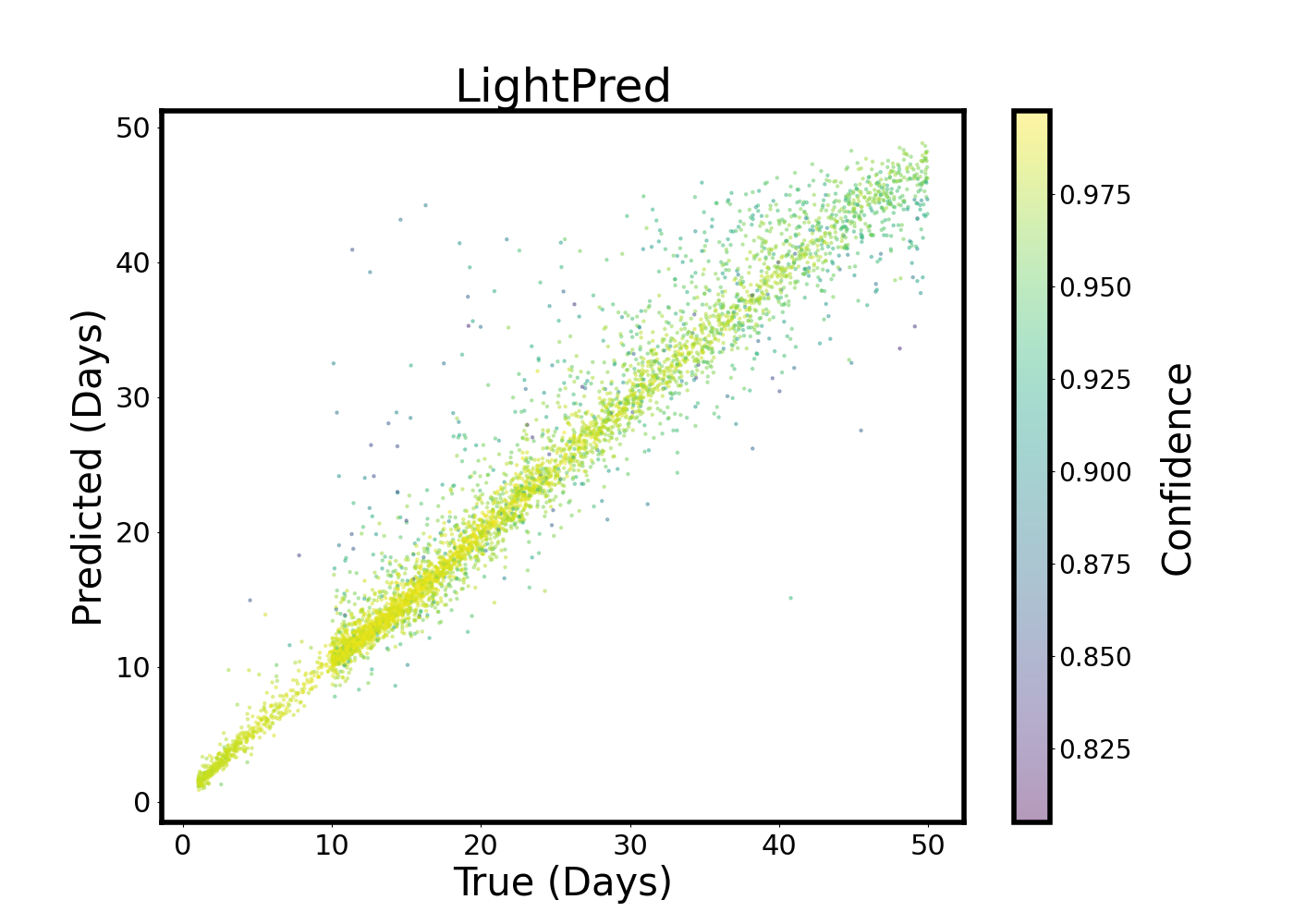}
  \end{minipage}
  \begin{minipage}[b]{0.4\textwidth}
    \includegraphics[width=\textwidth]{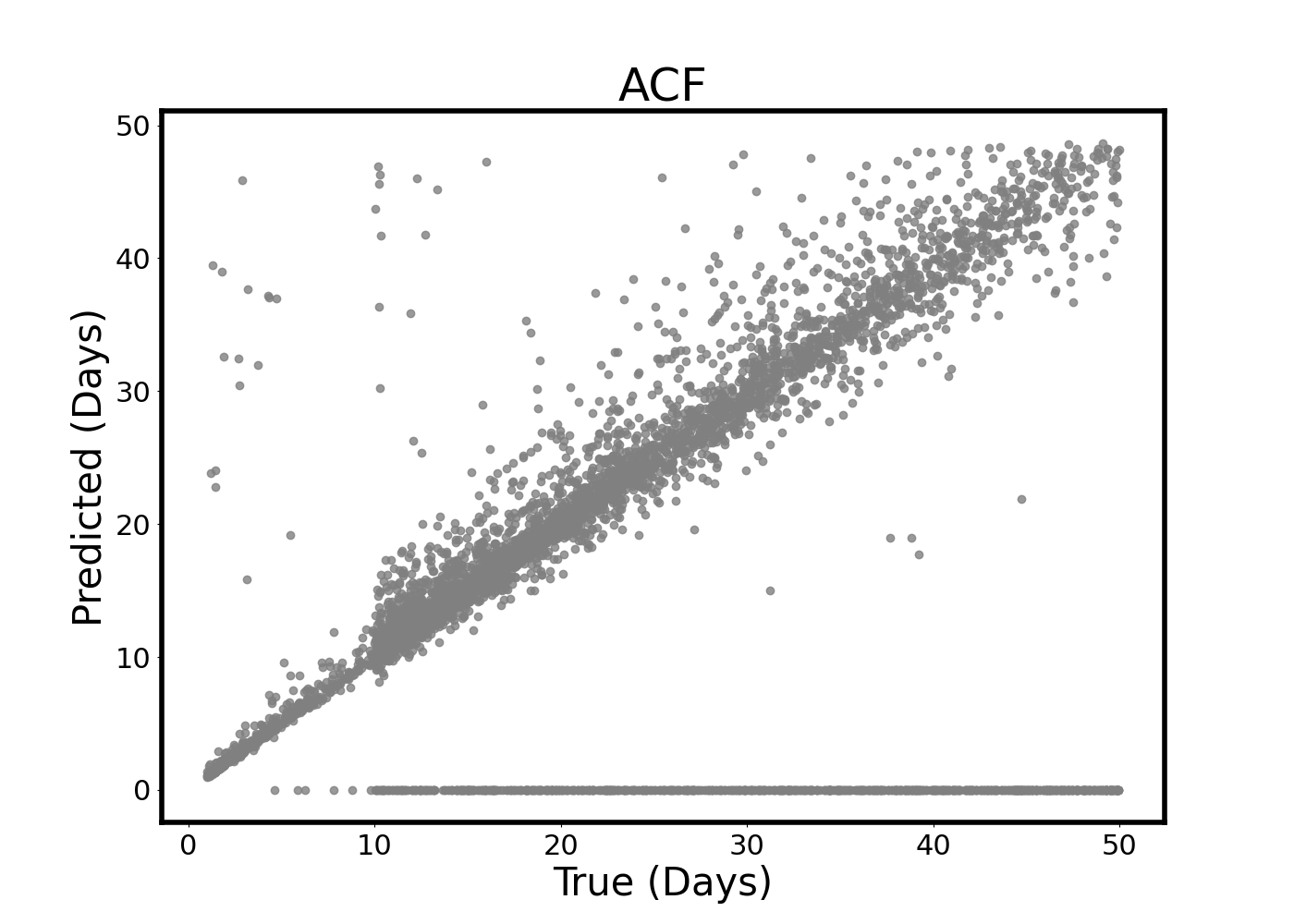}
  \end{minipage}
  \begin{minipage}[b]{0.4\textwidth}
    \includegraphics[width=\textwidth]{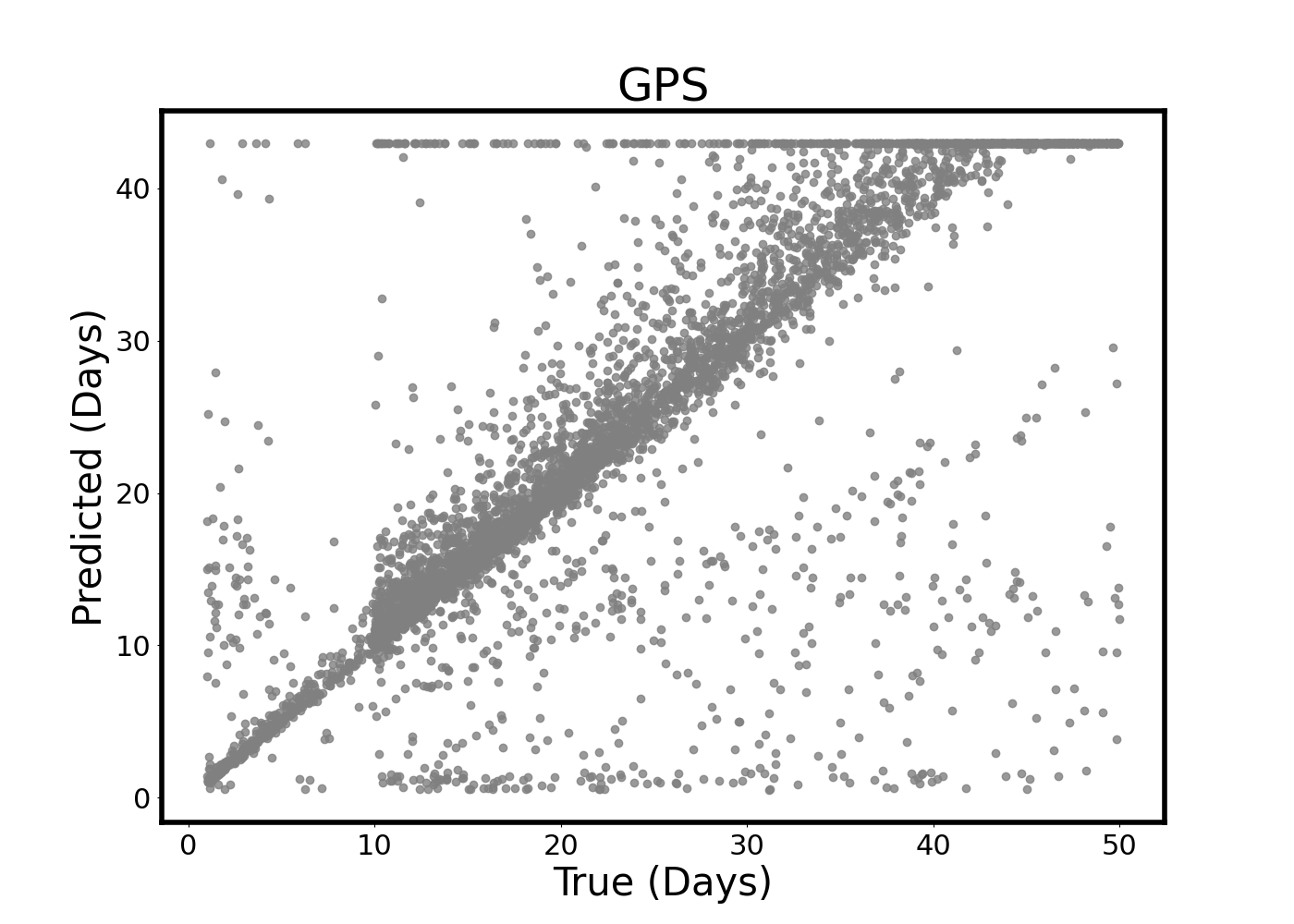}
    \end{minipage}
\caption{Results of our model (upper left) vs implementations of ACF (upper right) and GPS (bottom) on our test dataset. on the left panel, the color represents the model's confidence levels. The sharp drop of samples, seen at 10 days, is a result of the prior distribution of periods (see Table 1). This distribution was also implement in \cite{Aigrain2015} and motivated by the bimodality found by \cite{McQuillan2013}, \cite{McQuillan2014}}
    \end{center}
\end{figure*}

\begin{figure*}
    \begin{center} 
    \includegraphics[scale=0.3]{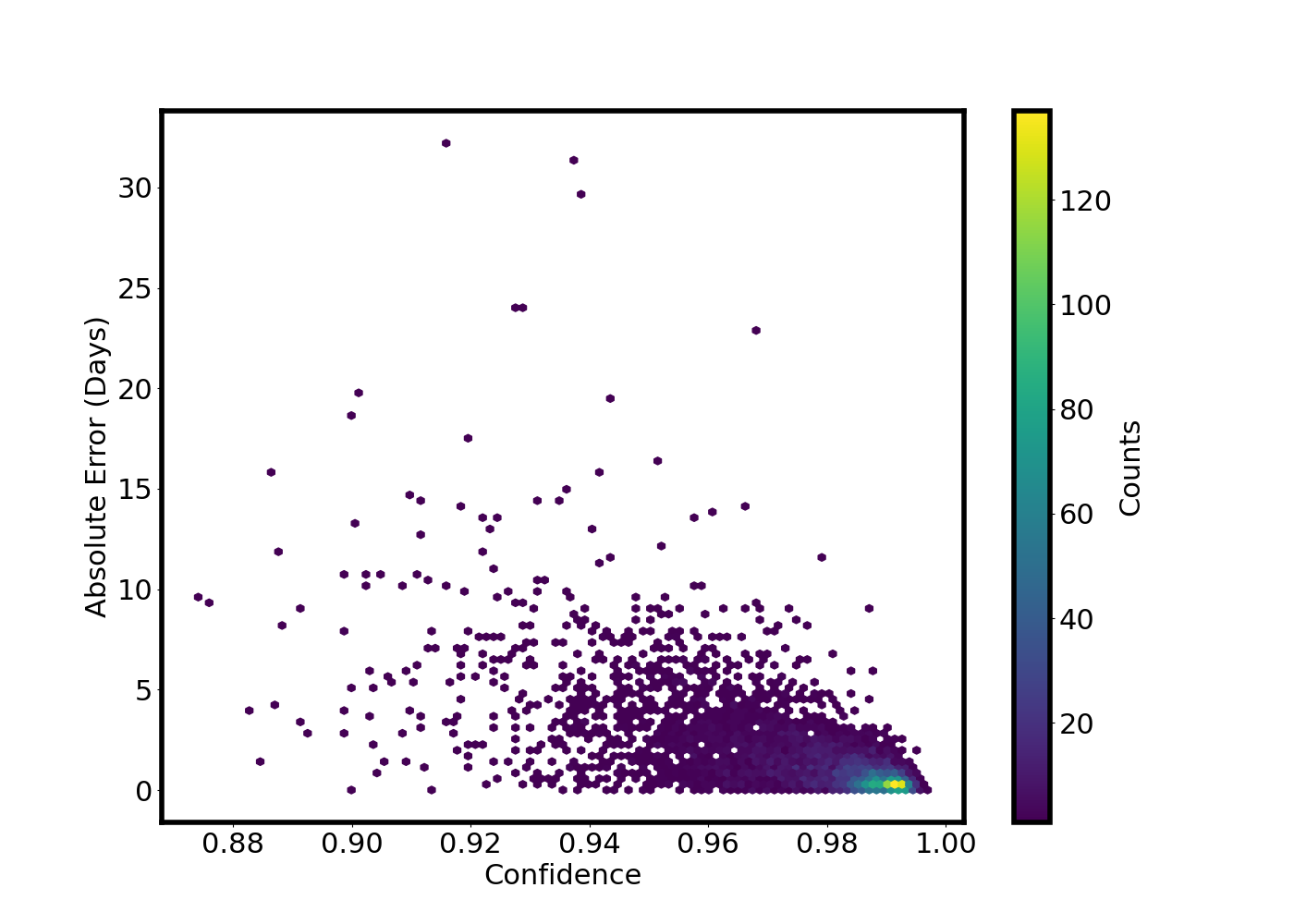}
    \caption{Period confidence value vs absolute period error of the model.}
    \end{center}
\end{figure*}

\begin{table*}
\centering
\begin{tabular}{||c c c||} 
 \hline
 model & acc10p (\%) & acc20p (\%)  \\ [0.5ex] 
 \hline\hline
 \textbf{LightPred Noiseless 8q} & \textbf{80} & \textbf{93}  \\
 \hline
 ACF Noiseless Tel-Aviv  & 75 & 90  \\
 \hline
 \textbf{LightPred, Noisy 8q}  & \textbf{76} & \textbf{90}
  \\
 \hline
 ACF Noisy Tel Aviv  & 68 & 80  \\
 \hline
 \hline
\end{tabular}
\caption{Comparison between the LightPred model and the ACF model as reported by the Tel-Aviv team at \cite{Aigrain2015}}
\end{table*}

\begin{table*}
\centering
\begin{tabular}{||c c c c c||} 
 \hline
 model & percent predicted (\%) & \makecell{acc10p (\%) \\ all (subset)} & \makecell{acc20p  (\%) \\ all (subset)} & \makecell{average error (Days) \\  all (subset)}\\ [0.5ex] 
 \hline\hline
 \textbf{LightPred Noisy 8q}  & \textbf{100} & \textbf{76} \textbf{(78)} & \textbf{90} (91)
 & \textbf{1.60} \textbf{(1.28)} \\
 \hline
 ACF Noisy Ours & 83  & 65 (78) & 76 \textbf{(92)} & 6.86 (1.62) \\
 \hline\hline
\end{tabular}
\caption{Comparison between the LightPred model; and our implementation of the ACF model. Each column presents the result on the entire dataset and the result on a subset of valid ACF predictions (in parenthesis), Best results are bolded.}
\end{table*}

\begin{table*}
\centering
\begin{tabular}{||c c c c c||} 
 \hline
 model & percent predicted (\%) & \makecell{acc10p (\%) \\ all (subset)} & \makecell{acc20p  (\%) \\ all (subset)} & \makecell{average error (Days) \\  all (subset)}\\ [0.5ex] 
 \hline\hline
 \textbf{LightPred Noisy 8q}  & \textbf{100} & \textbf{76} \textbf{(76)} & \textbf{90} \textbf{(90)}
 & \textbf{1.60} \textbf{(1.36)} \\
 \hline
 GPS Noisy Ours  & 86 & 61 (62) & 77 (77) & 4.06 (3.65) \\
 \hline
\end{tabular}
\caption{Comparison between the LightPred model and our implementation of the GPS model. Each column presents the result on the entire dataset and the result on a subset of valid GPS predictions (in parenthesis). The best results are emphasized in bold.}
\end{table*}

\subsection{Inference of stellar rotations in Kepler data\label{subsec:Inference on Kepler}}
Next, we predicted periods using our trained model on the \textit{Kepler} dataset. The model was trained as described in \ref{subsec:pipeline}. Graphs of the loss of the SSL and SBSL training steps can be found in Appendix \ref{Appendix_C}. We then used the trained model to predict periods from the \textit{Kepler} dataset. Since the model gets fixed-sized samples and we worked with 8-quarters light curves, we filtered out any sample that did not appear in 8 consecutive quarters. This left us with 130868 which serves both for the SSL step and the inference. As mentioned in \ref{subsec:pre_process} It is important to note that while we used the same samples for SSL training and inference, the pre-processing steps are different - specifically, we did not use in this step the special augmentations (masking, Shuffle, Gaussian Noise) that we used in the SSL part.\\  
\subsubsection{Period predictions}
\label{subsec:period_pred}
For inference, we performed an additional selection process. Both \cite{McQuillan2014} and \cite{Reinhold2023} took stars with $Teff < 6500K$. We used a slightly larger threshold of $7000K$ to include stars with a radiative exterior that tends to rotate faster (\cite{Albrecht2022}). Therefore, Out of the 130868 predictions we filtered out stars with $Teff < 7000K$  
which gave us 126029 samples.
Next we filter out known contaminants in the dataset - 
$\delta$ Scuti, $\gamma$ Durados and Hybrids (\cite{Uytterhoeven2011}, \cite{Bradley2015}, \cite{Murphy2019}, \cite{Van_Reeth2018}, \cite{Li2019a}, \cite{Li2019b}); RR-Lyrae stars (\cite{Benkco2010}, \cite{Nemec2011}, \cite{Nemec2013}, \cite{Forro2022});
Giants based on the criterion given by \cite{Ciardi2011}; this leaves us with 108096 samples.\\
To estimate the error we used two methods:
\begin{itemize}
    \item \textbf{Simulation Error}: To relate the error from the simulation to \textit{Kepler} predictions, we binned the simulation test-set results to 1-day bins and defined the \textit{bin error} to be the average absolute value difference from the true values per bin. We then rounded the \textit{Kepler} predictions to integers and found the corresponding bin for each prediction. this way we assigned a period-dependent error for each sample. This approach gives an average error per period bin but does not differentiate between samples with similar periods and is therefore limited.
    \item \textbf{Observational Error}: Since each \textit{Kepler} light curve sample is potentially longer than eight quarters, we can divide the full light curve into 8-quarter segments and check the consistency of the model with respect to different segments (for example, the consistency between predictions for Q3-Q10, Q4-Q11, etc.). \cite{McQuillan2014} used a similar approach, with different segment sizes to choose valid ACF predictions. In our case, different segment sizes are not possible since our model uses fixed-length input so we used a fixed size of eight quarters per segment. Since we have seven segments, there are in total, 21 quarter pairs for each sample. Figure 6 shows the distribution of differences in predictions between all the pairs in the dataset. It can be seen that the differences behave similarly to a Normal distribution. We therefore construct the following observational error; for each sample, we fit a normal distribution using all the pairs' differences and take the $1\sigma$ value as the observational error. 
\end{itemize}

\begin{figure*}
    \begin{center}
    \includegraphics[width=0.4\textwidth]{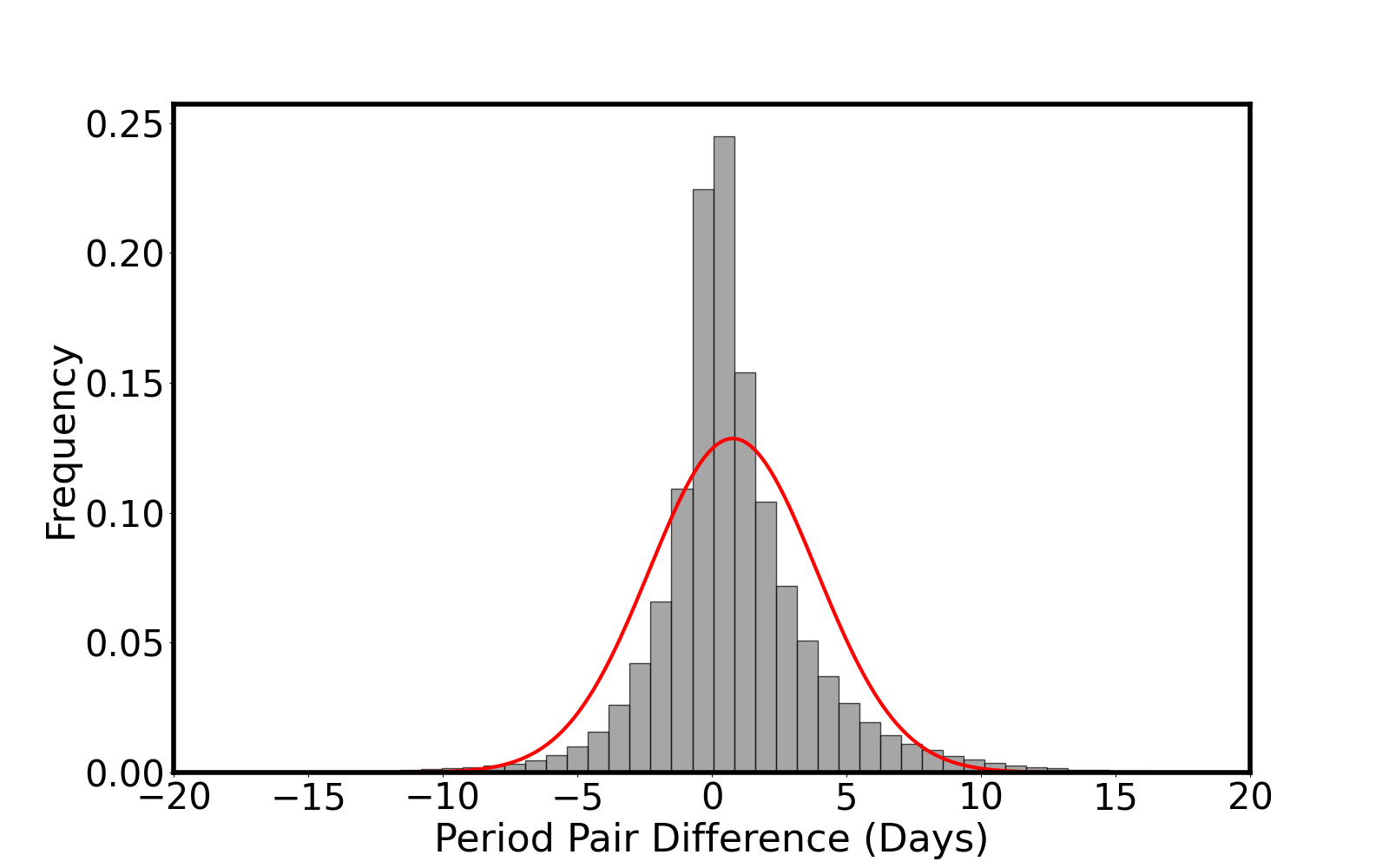}
    \caption{Distributions of predictions-difference between pairs of different quarters of the same sample for \textit{Kepler} samples. The orange curve represents the fit to a normal distribution.} 
    \end{center}
\end{figure*}

To see if the two errors are related, we binned the \textit{Kepler} predictions to 1-day bin, averaged the observational errors of each bin, and compared them to the simulation error of that bin. Figure 7 shows this comparison. It can be seen that for prediction $< 40$ days, there is a clear monotonic relationship between simulation and observational errors. For predictions $<25$ days, the relation appears linear. This is an important evidence for the relation between simulation and real data predictions. The fact that we see such a clear relation between the two suggests that, at least for periods $<25$ days, the \textit{Kepler} data is sufficiently represented by the simulations for period predictions. For longer periods the correlation is different which might be related to a drop in the activity for older stars.

\begin{figure*}
    \begin{center}
    \begin{minipage}[b]{0.4\textwidth}
        \includegraphics[width=\textwidth]{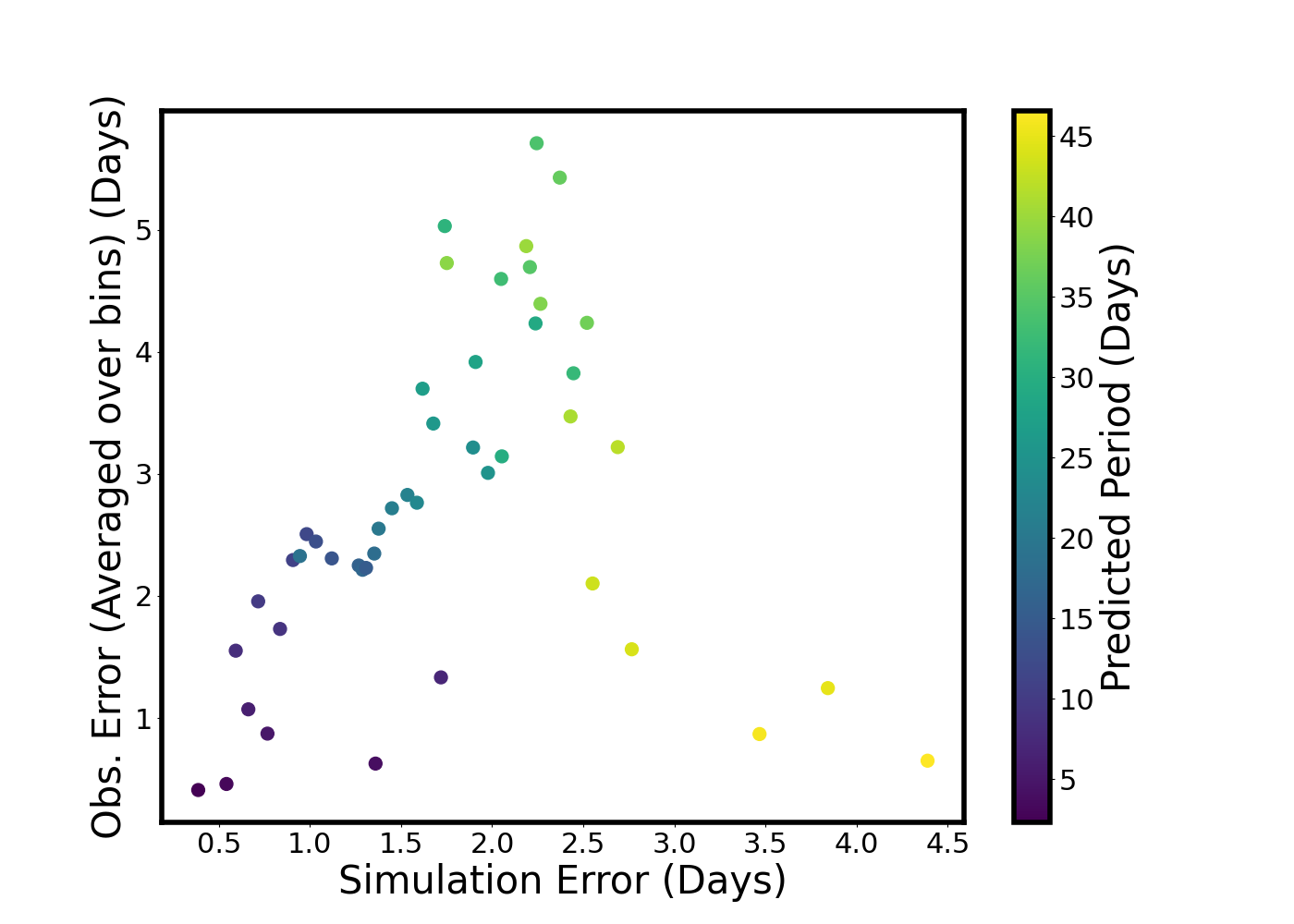}
    \end{minipage}
    \begin{minipage}[b]{0.4\textwidth}
       \includegraphics[width=\textwidth]{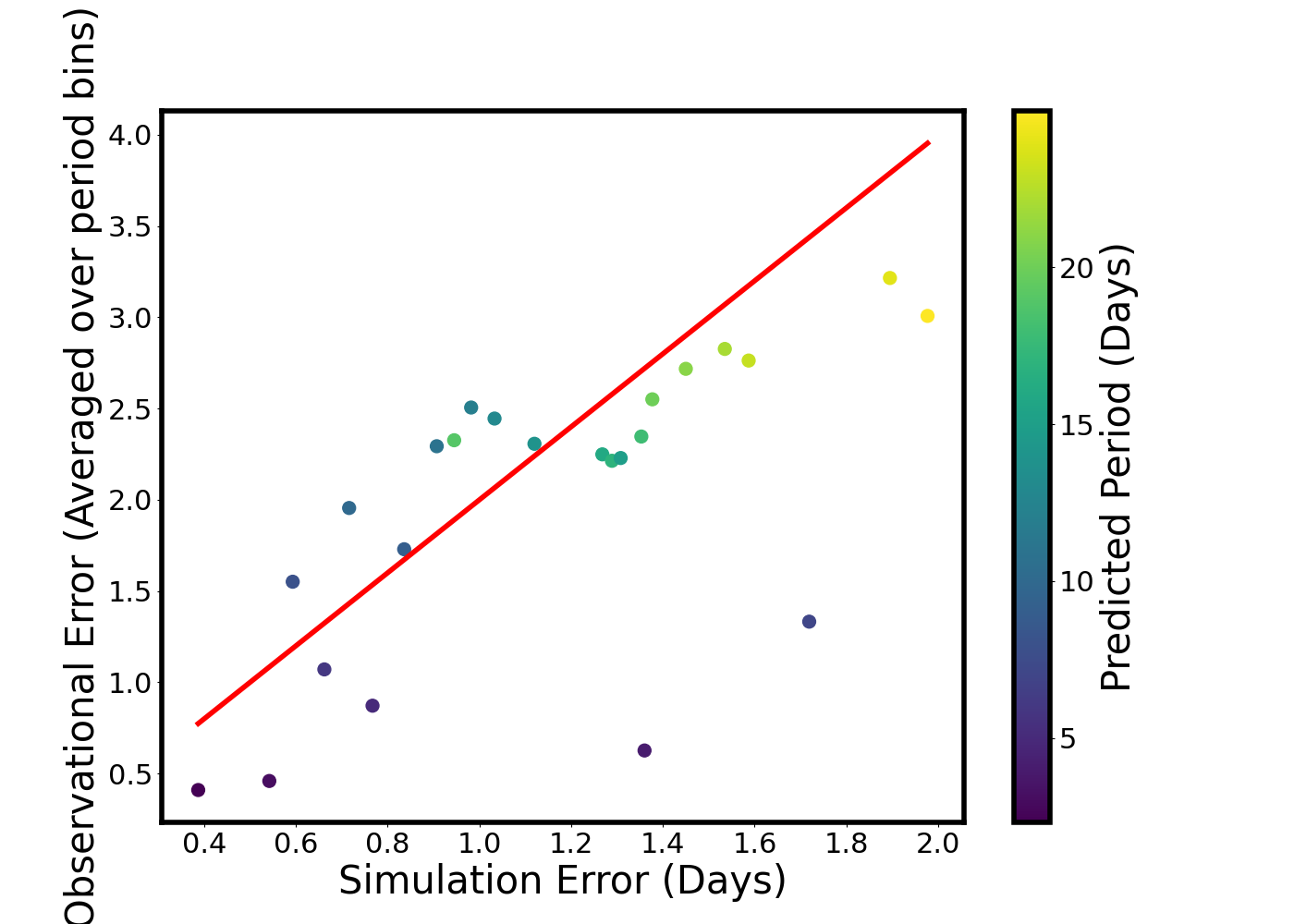}
    \end{minipage}
\caption{Simulation error vs Observational error (averaged over 1-day bin). The left panel shows all the samples and the right panel shows only samples with predicted periods smaller than 25 days. The red line represents a line with a slope of 2.}
\end{center}
\end{figure*}

\begin{figure*}
    \centering
    \begin{minipage}[b]{0.4\textwidth}
        \includegraphics[width=\textwidth]{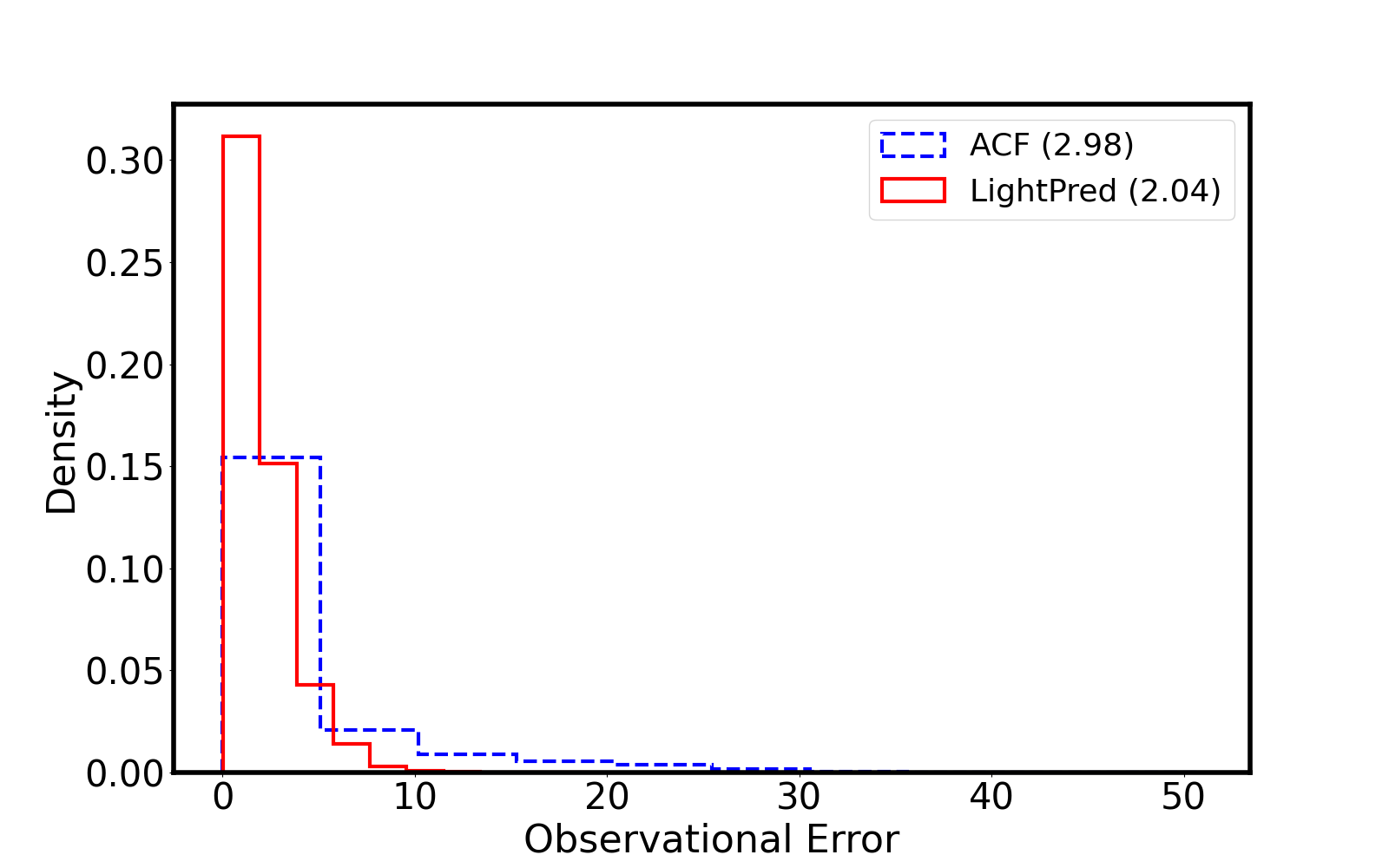}
    \end{minipage}
    \begin{minipage}[b]{0.4\textwidth}
       \includegraphics[width=\textwidth]{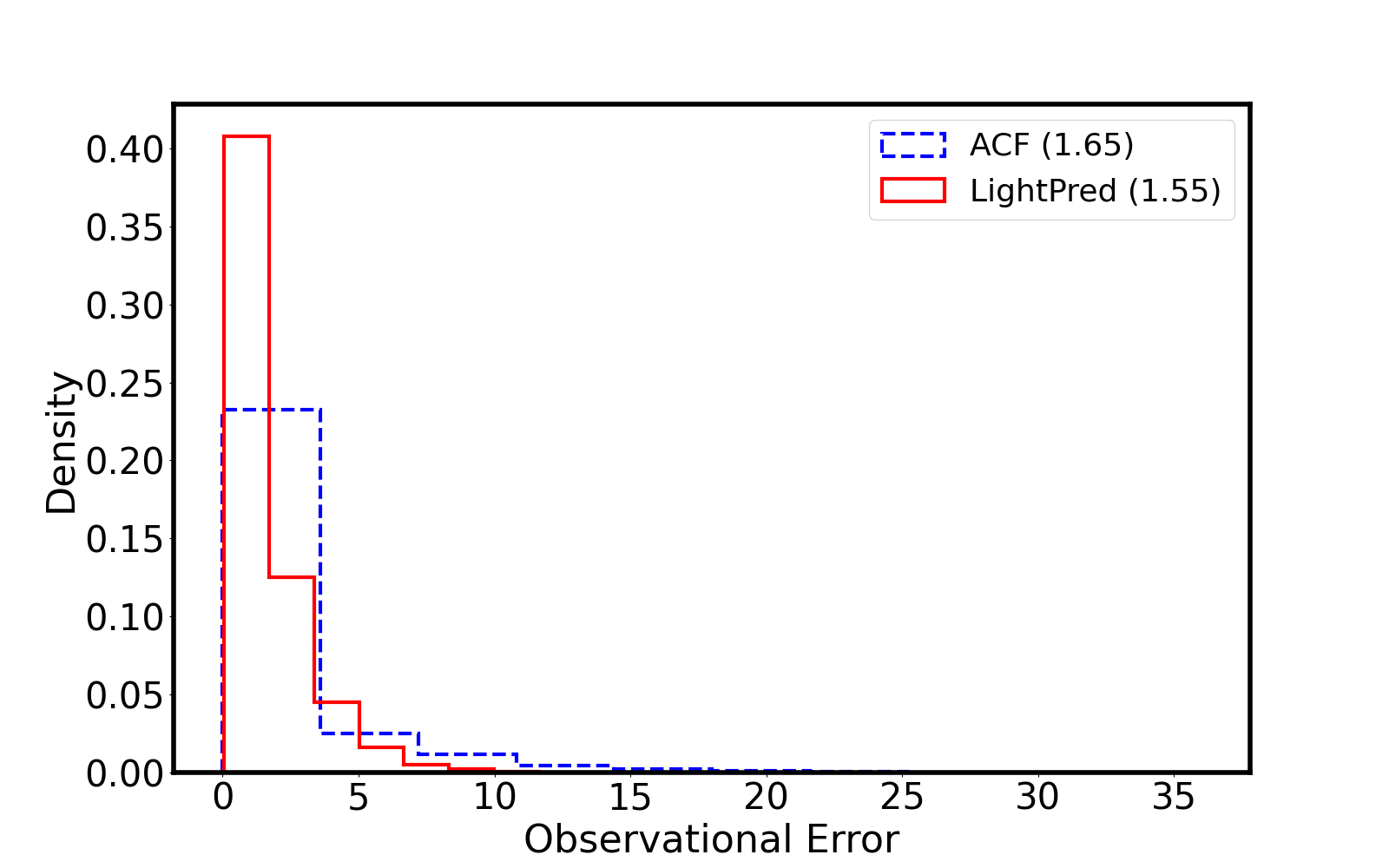}
    \end{minipage}
\caption{Observational error distributions of the LightPred model vs the ACF model. The left panel shows all the samples (108785 samples) and the right panel shows a sub-sample of 36154 points with no planet-hosting stars, eclipsing binaries, and where ACF was able to predict all the segments, i.e. the ACF effectively could not predict a rotation period for the majority of the data.}
\end{figure*}

\subsubsection{Comparison with ACF}
Next, we want to compare our method with classical methods on the \textit{Kepler} data. Our experiments on mock data with ACF and GPS showed that while GPS is very sensitive to parameter tuning, ACF was more robust and performed better. We therefore choose ACF for that comparison. To compare our method with ACF, we created an ACF prediction on different segments for the entire sample. We chose the parameters of the ACF such that the results are as similar as possible to the samples published by \cite{McQuillan2014} and calculated the observational error for the ACF, as explained \ref{subsec:period_pred}. When comparing observational error results, it is worth noting an inherent bias in the ACF method - while our model predicts a value for each sample, ACF is different. As discussed in section \ref{subsec:Results on simulated data}, there are samples for which the ACF is not able to predict any rotation period at all (when the ACF signal has too shallow peaks) in which case we assign a predefined value. The fact that we assign the same value for all the non-predicted segments creates a falsely high consistency for those samples. This is why we conduct two comparisons with ACF - a comparison where we used all the samples and a comparison where we used a subset of samples where we follow the general selection rules of \cite{McQuillan2014}, i.e. we remove planet-hosting stars and eclipsing binaries and take only samples where ACF was able to predict all the segments. The size of the second sub-sample is 36154, i.e. only $\sim \frac{1}{3}$ of the data. 

Figure 8 shows a comparison of Observational error distribution between the LightPred model and the ACF model for the two sets of samples. In both cases, the ACF shows a tail of very large errors. Our model shows more concentrated distributions and when considering the entire samples (left panel), our model has much lower error. When considering only ACF-valid samples (right panel), The errors are comparable, with LightPred error slightly lower than ACF error. When considering the relative observational error (dividing by the predicted period) we get the same error on average (11\%).  This suggests for \textit{Kepler} data what we already saw for simulated data, that is, our model is not only able to determine the rotational period for more samples but is also at least as robust as ACF. Table 6 summarizes this comparison.

We also want to test the efficiency of the self-supervised phase. Table 7 compares the results both on simulation data and \textit{Kepler} data for training with and without the self-supervised pre-training phase. We see that self-supervised training improves all the metrics both on simulation and \textit{Kepler} data. While the improvement on simulation metrics is modest, the improvement on \textit{Kepler} data is significant ($\sim 3$ times lower observational error). The significant change in observational error can be understood by the fact that there is a relation between the contrastive task and the observational error - the latter can be seen as a special case of similarity between different views of the same sample. Note that the results of the simulation data in Table 7 are slightly different than the results shown in Section \ref{subsec:Results on simulated data}. This is because here we used different inclination distributions as explained in section \ref{subsec:mock_data}.

\begin{table*}
    \begin{center}
    \begin{tabular}{||cccc||}
    \hline
        \textbf{Model} & \textbf{percent predicted (\%)} & \makecell{\textbf{Average Observational Error}\\ \textbf{(All Data) (Days)}} & \makecell{\textbf{Average Observational Error}\\ \textbf{(ACF Data) (Days)}}  \\  [0.5ex]  \hline\hline
        \textbf{LIghtPred} & \textbf{100} & \textbf{2.01} & \textbf{1.55}  \\ \hline
        ACF & 33 & 2.98 & 1.65 \\ \hline \hline
    \end{tabular}
    \caption{Prediction results of LightPred model and ACF model. Average errors refer to observational errors. Best results are bolded}
\end{center}
\end{table*}

\begin{table*}
    \begin{center}
    \begin{tabular}{||ccccc||}
    \hline
    \textbf{Method} & \textbf{acc10p} & \textbf{acc20p} & \makecell{\textbf{Average Simulation Error}\\ (Days)} & \makecell{\textbf{Average Observational Error}\\ \textbf{(All Data) (Days)}}\\
    [0.5ex]  \hline\hline
    w/ Self Supervised  & 78 & 92 & 1.58 & 2.01 \\
    \hline
    w/o Self Supervised  & 71 & 88 & 1.80 & 6.77 \\
    \hline \hline
    \end{tabular}
    \caption{Comparison between training with and without self-supervised pre-training}
    \end{center}
    
\end{table*}
Figure 9 shows a comparison between our model and the three largest available catalogs of \textit{Kepler} stellar periods: \cite{McQuillan2014}, \cite{Reinhold2023} and \cite{Santos2021}. The colors of the points refer to the learned confidence of the model (the output of the model). It can be seen that in general, more confident samples have better agreement. This implies that the learned confidence represents physical properties and is somehow similar to the scoring functions used in \cite{McQuillan2014} and \cite{Reinhold2023}. Another interesting observation is the fact that almost all the points that disagree are located in two distinct regions - the first region is around the half period line Which suggests a 'double period mistake'. The second region is where the reference period is close to zero but our model's predictions aren't. This can be seen as a vertical line of points close to $x = 0$. Both regions are apparent in all three comparisons. We investigate these non-agreement samples more closely in \ref{subsec:false_positive}. 
\begin{figure*}
    \begin{center}
  \begin{minipage}[b]{0.4\textwidth}
    \includegraphics[width=\textwidth]{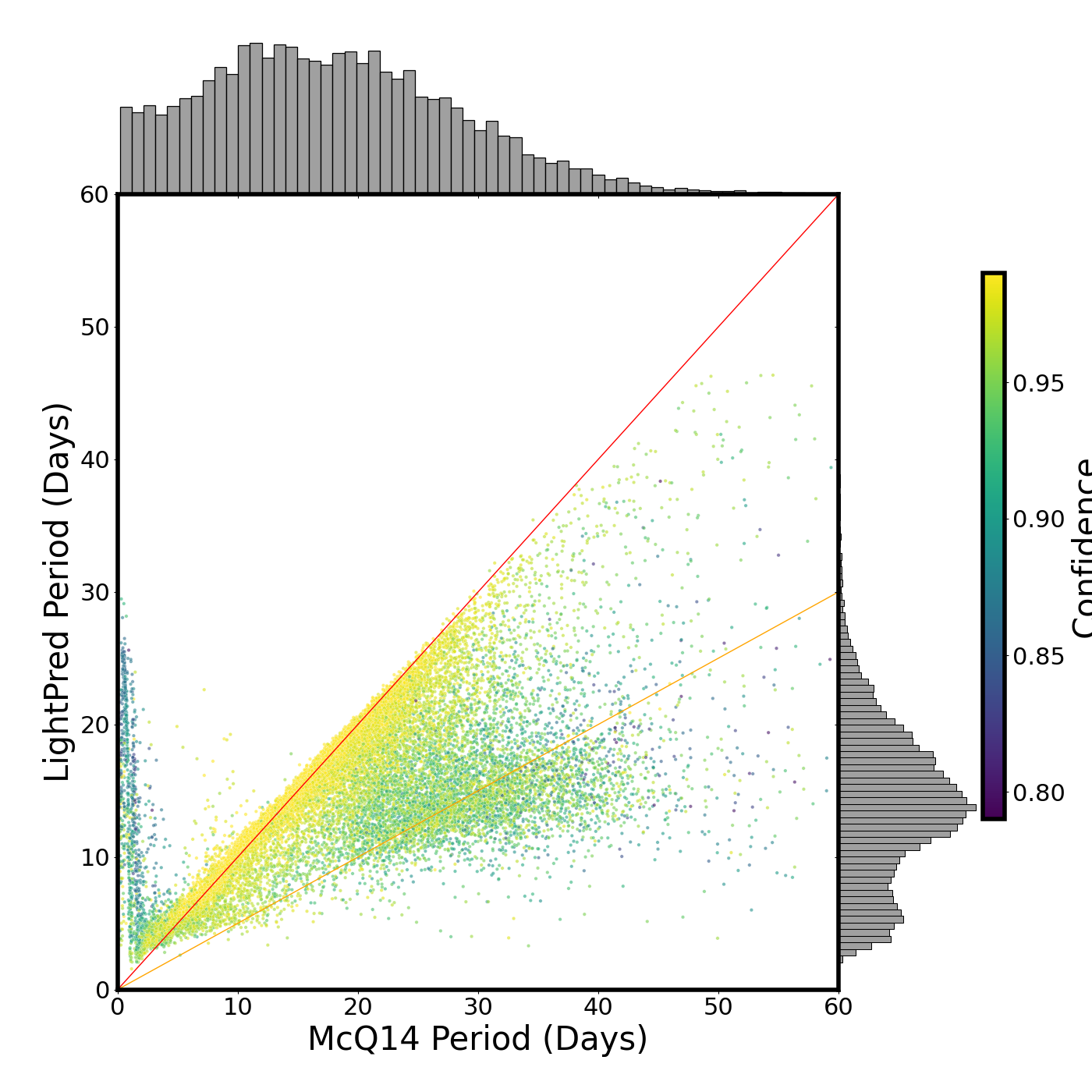}
  \end{minipage}
  \begin{minipage}[b]{0.4\textwidth}
    \includegraphics[width=\textwidth]{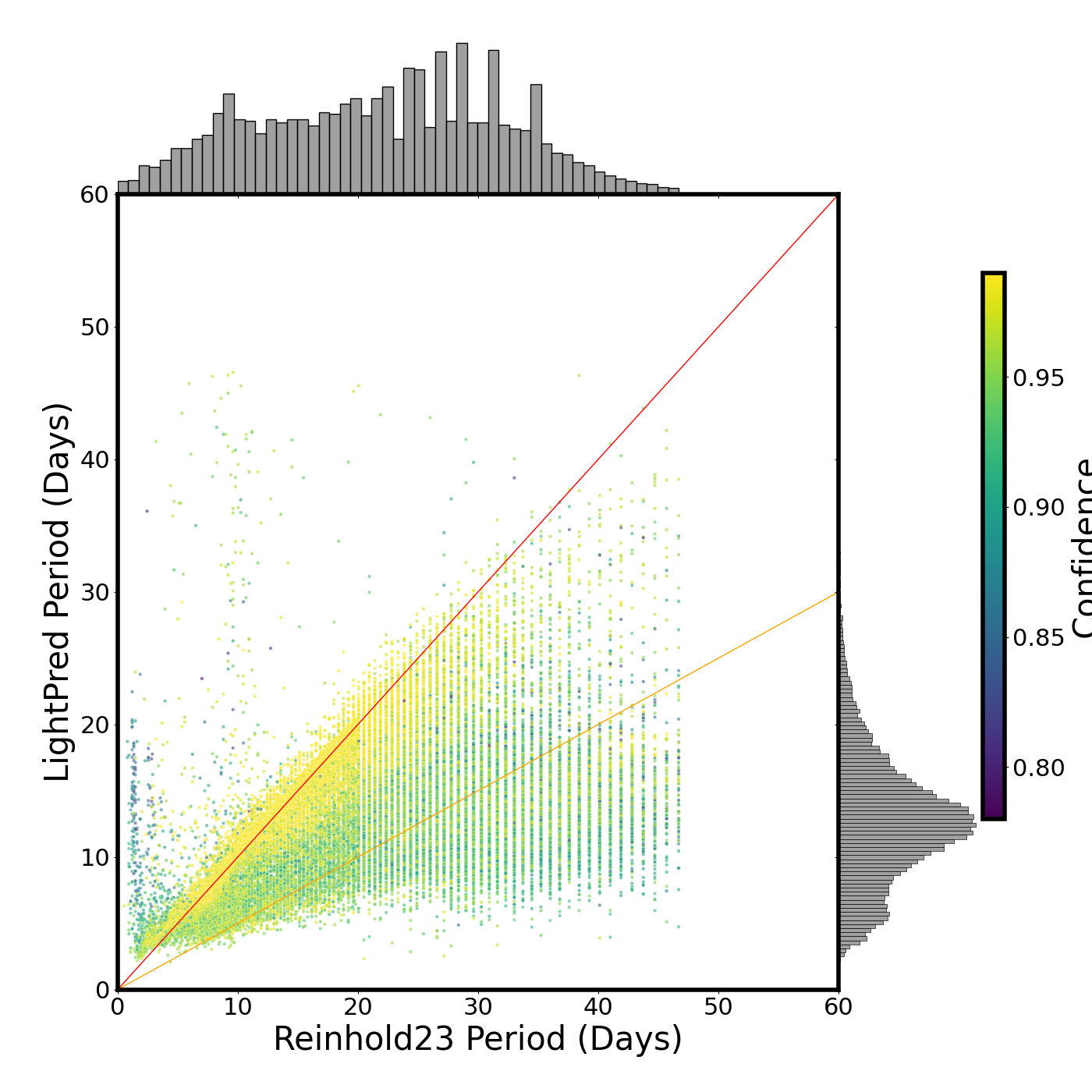}
  \end{minipage}
  \begin{minipage}[b]{0.4\textwidth}
    \includegraphics[width=\textwidth]
    {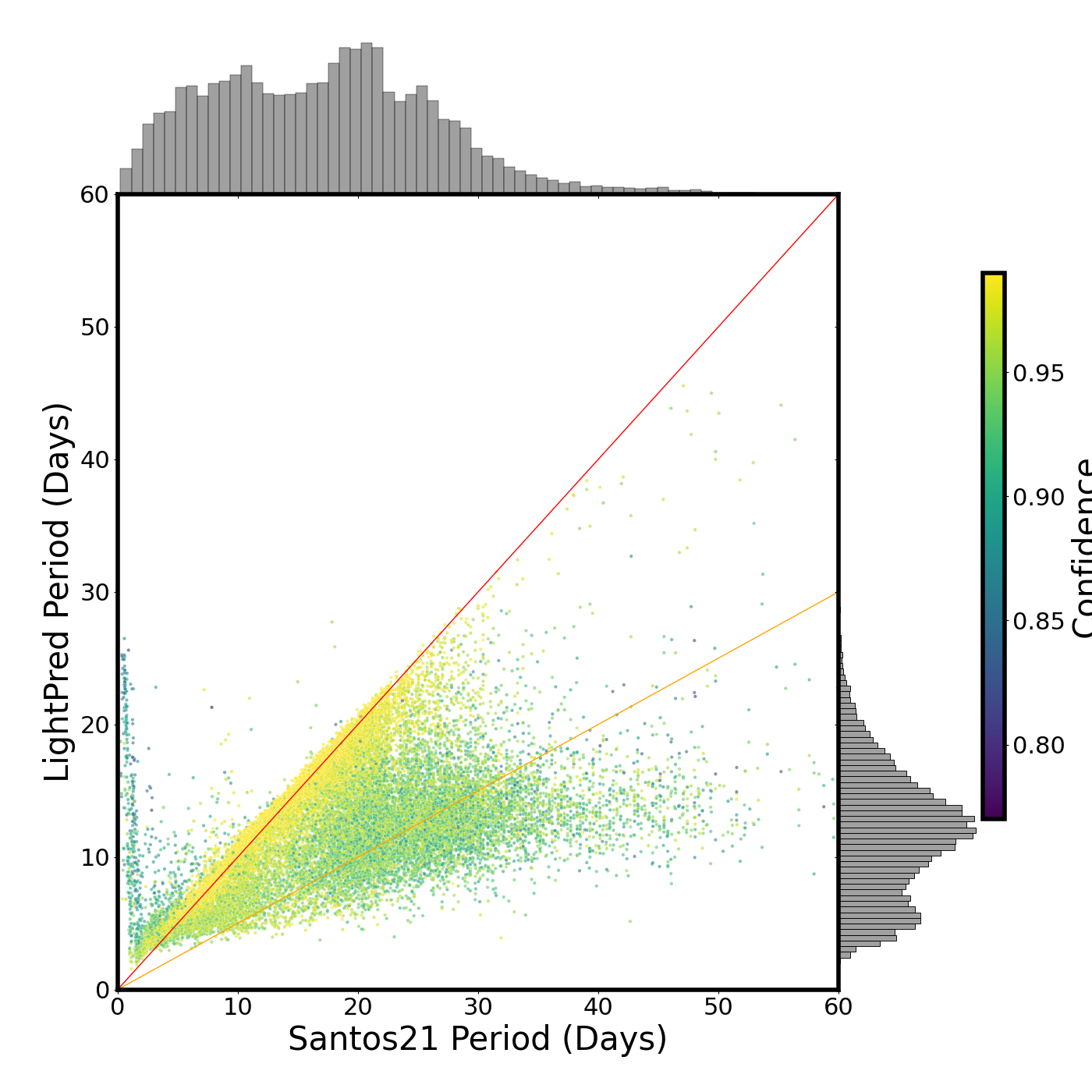}
  \end{minipage}
\caption{Comparison of the period determinations between LightPred model, \cite{McQuillan2014} (left panel),  \cite{Reinhold2023} and  \cite{Santos2021}. Colors represent model confidence scores. The red and orange lines represent slopes of $1$ and $0.5$ respectively. The inconsistencies, seen on short periods and half period regimes, are discussed in \ref{subsec:false_positive}}
\end{center}
\end{figure*}

\subsubsection{Error analysis and constraints}
\label{subsec:error_analysis}
\textit{Kepler} observations are diverse in terms of parameters such as magnitude, distance, stellar activity, etc. Since those parameters affect the observed light curve they affect the predictions. Therefore, we would like to investigate correlations between our errors and different parameters. This is important both as a sanity check for our metrics and to be able to better constrain our results and remove potential biases. In \ref{subsec:period_pred} we saw that the observational error and the simulation error are related and both depend on the period. It is beneficial to normalize the error by the predicted period to remove biases related to the length of the period. In addition, our model outputs confidence for each sample which serves as another 'error' estimation. The observational error and the model's confidence represent different types of uncertainties. The observational error represents uncertainty related to the data (sometimes referred to as Aleatoric uncertainty) and the confidence represents uncertainty of the model (sometimes referred to as Epistemic uncertainty). In order to take into account all of the above, we use the following \textit{total error}:
\[e_{tot} = \frac{e_{obs}}{\hat{p}(1-\hat{c})}\],
where $e_{obs}$ is the observational error, $\hat{p}$ is the predicted period and $1-\hat{c}$ is the model's confidence. \\
We now want to see if the total error correlates with stellar parameters. If such a correlation exists, the total error can help us constrain the predictions to a meaningful subset of the samples. For such comparison we used effective temperature, distance, metallicity, and luminosity from  \cite{berger2020}. In addition, we used the Kepler magnitude from the Kepler input catalog (KIC) and two different photometric proxies for activity.  The first is called $R_{var}$ and was suggested by \cite{Basri2011}; this is the difference between the $5^{th}$ and the $95^{th}$ percentiles of the flux at a given segment. Indeed, $R_{var}$ was used by \cite{Reinhold_2017} to calculate activity cycles for 3203 \textit{Kepler} stars.  To calculate $R_{var}$, we calculated the $5^{th}$ and $95^{th}$ percentiles on windows of length 90 days with an overlap of 45 days. We then defined $R_{var}$ as the difference between the maximum and the minimum elements in this time series. Another activity proxy is $S_{ph}$ that was suggested by \cite{Mathur2014} and showed to be correlated with chromospheric activity \citep{Salabert2016}. $S_{ph}$ is defined as the mean standard deviation of the light curve over a window of five periods.

To see the correlations between different variables and the total error we did the following experiment; for each stellar property, we divided the samples into 200 bins within the range of the property, and calculated the average total error for each bin, neglecting bins with less than 30 points. Figure 10 shows these calculations for the effective temperature, distance, luminosity, and magnitude, $S_{ph}$ and $R_{var}$. We see that while there is a clear correlation between the error and some parameters, for others there is weak/no correlation. The most striking example is $R_{var}$ which looks almost periodic with the error. The problem with this test is that there are internal correlations between the properties. To take that into account we need to look at higher dimensional correlations such as 2d correlations. Such correlations were calculated in the following way: for a pair of properties, we first normalized each of the properties by subtracting the minimum value and dividing by the maximum value to get a dimensionless parameter between $0-1$. We then binned the samples with 200 bins for each parameter and calculated the average total error for each 2d bin, again neglecting bins with less than 30 points. We then made use of principal component analysis (PCA) of the binned values and the total error to reduce the dimension. Lastly, we plotted the resulting PCA components for the binned values as a function of the total error. Figure 11 shows this calculation for pairs consisting of  $R_{var}$ and all other variables. We see that for all pairs there is a clear correlation with the error. Such correlations can potentially identify the parameter phase space region, in which the predictions are less reliable.

Another possible correlation is with the $w$ parameter from \cite[][see \ref{subsec:pre_process}]{McQuillan2014}. It is an important sanity check since we used samples with $w < 0.04$ as noise generators and we want to verify that the model indeed struggles to predict periodicity in those samples. This would reject the assumption of overfitting the model for those samples. The correlation is shown in Figure 12. It can be seen that up to $w\sim 0.7$, there is a clear monotonic correlation between the total error and $w$; high error correlates with low $w$ as expected. For $w>0.7$ there is interesting behavior with an opposite trend which is not fully understood. We revisit this inverse relationship in Section \ref{subsec:false_positive}. Notably, the error for these particular samples, which constitute 6.6\% of the data points, is comparatively small so we can safely reject the possibility of overfitting. 
The correlation result suggests that the total error indeed represents a physical constraint that affects the predictions. As such, the total error can serve as a reliable tool to constrain the results.

\begin{figure*}
    \centering
        \includegraphics[width=0.8\textwidth]{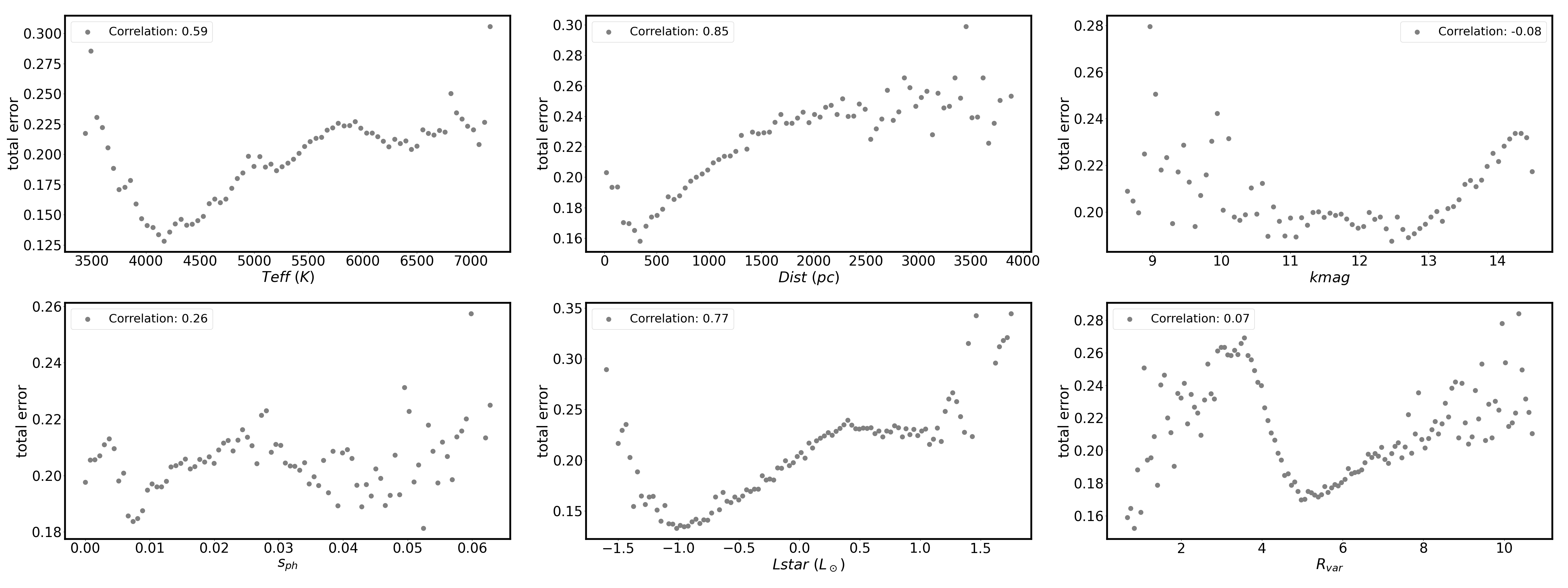}    
    \caption{Correlation between stellar parameters and the total error on each panel, the Pearson correlation is shown on the legend}
\end{figure*}

\begin{figure*}
    \includegraphics[width=\textwidth]{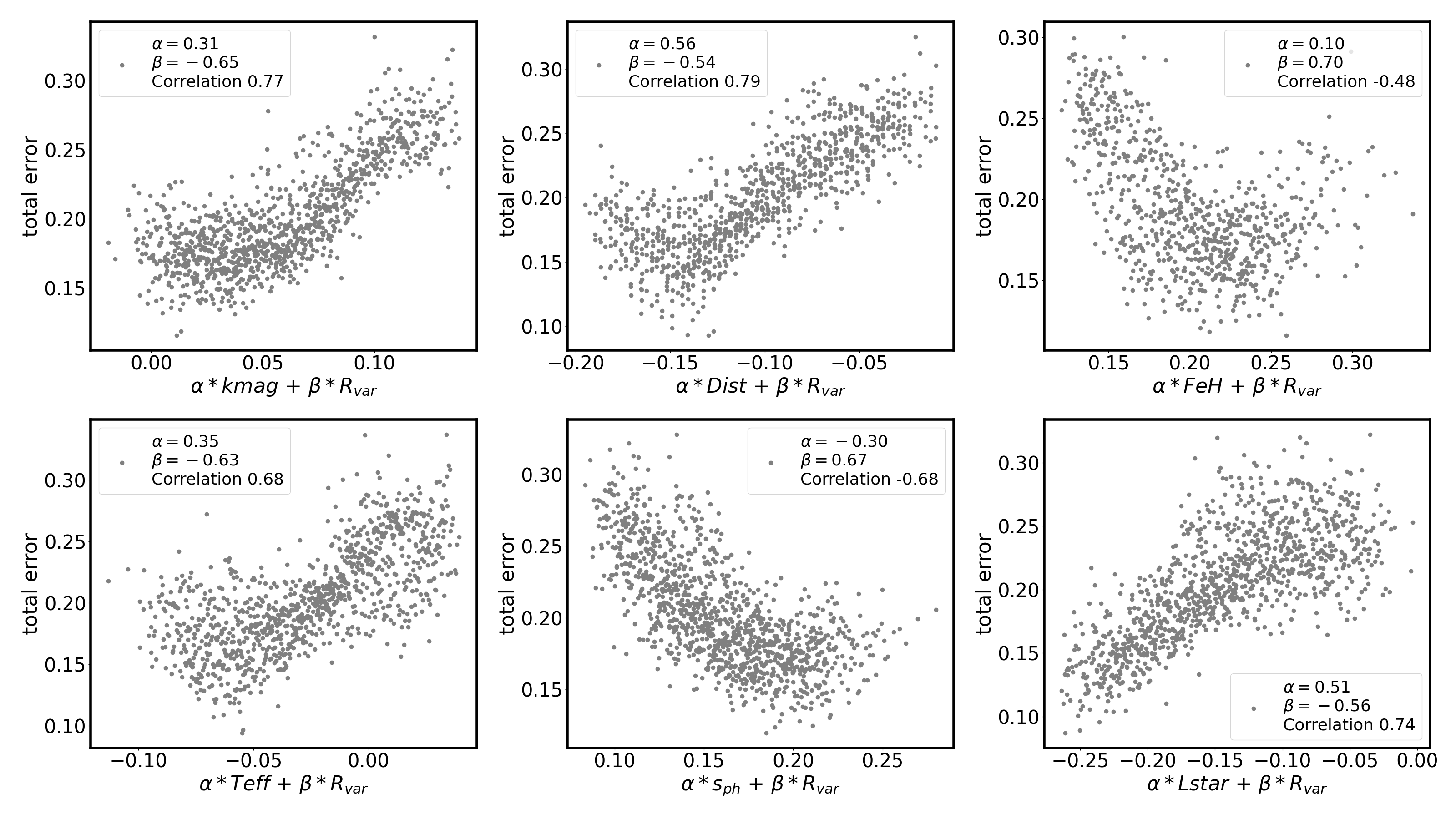}
    \caption{2d correlation between stellar parameters and the total error. On each panel the PCA parameters and Pearson Correlation are shown in the legend}
\end{figure*}

\begin{figure*}
\centering
    \includegraphics[width=0.6\textwidth]{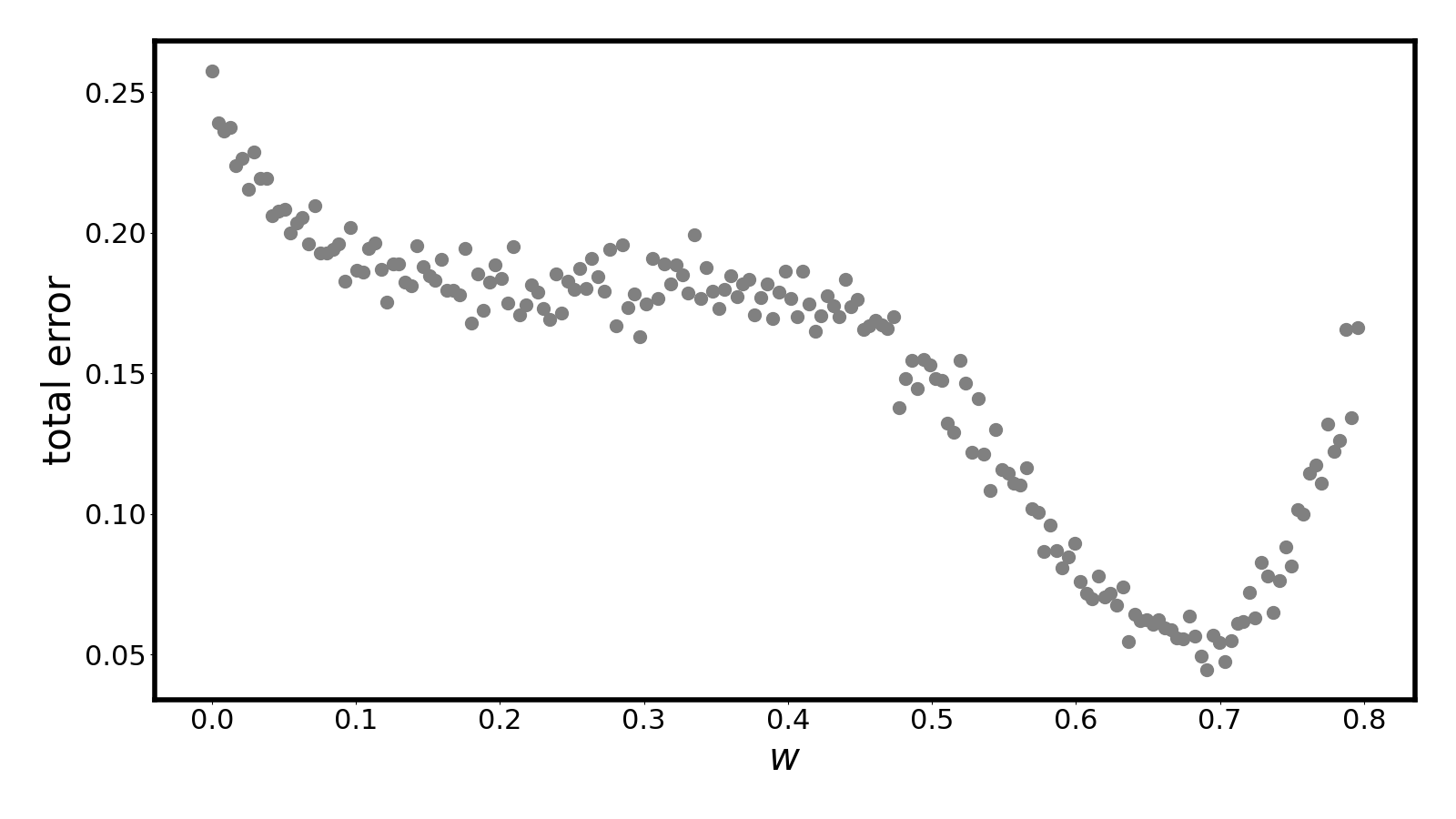}
    \caption{correlation between $w$ parameter from \cite{McQuillan2014} and the total error}
\centering
\end{figure*}

\subsubsection{Pollution}\label{subsec:poluttion}
In \ref{subsec:error_analysis} we discussed the effect of stellar parameters on the results but there are other factors that might bias our predictions. It is possible that the observed light curve does not represent a single object but multiple objects. This can be a result of crowding; since \textit{Kepler} pixels cover $4''$ x $4''$ patch of sky it is plausible that multiple objects would enter a single aperture. Indeed, correction for crowding is one of the main tasks of the pre-search data conditioning pipeline. The pipeline is, of course, not optimal, and indeed, using adaptive optics of a sample of \textit{Kepler} planet host candidate, \cite{Ziegler_2017} found a probability of $12.6\% \pm 0.9\%$ for nearby star within $0.15''-4''$. Another possible source of bias is binaries. A system of multiple stars can bias the model towards an incorrect period (the orbital period of the binary for example). Many works investigated the problem of identifying false positive signals in the case of planetary systems (\cite{Torres2011}, \cite{Desert2015}, \cite{Bryson2013}, \cite{Mullally2016}) which led to multiple vetting procedures (\cite{Coughlin2016}, \cite{colon2015}) but this is not the case for general \textit{Kepler} stars. However, \cite{Santos2021} did identify possible contaminants for a sample of general \textit{Kepler} stars.

We would like to test the sensitivity of our model to the different types of possible contamination. As a first step, we consider the contaminants specified in \cite{Santos2021}. Specifically, we look at samples that have potential pollution in the PDC-MAP signal as defined by \cite{Santos2021}. They found 1366 such samples (marked as No Rotation flag 6). When crossed-matched with our samples, we are left with 1046 samples. Figure 13 shows the distributions of periods and total errors for the sample of possible polluted samples and the entire samples. It can be seen that the distributions appear very similar. Kolmogorov-Smirnov (KS) test between the period distributions with a null hypothesis that the two samples were drawn from the same distribution, produced a p-value of $4.05*10^{-32}$ which implies that there is a high probability that the samples were not drawn from the same distribution. However, taking into account the statistical nature of such analysis and the fact that the average error of the possible pollution is similar to that of the entire sample, we conclude that it would be very hard to identify such contaminants based on the error alone. We therefore made a conservative choice and removed those samples from the predictions.

Next we consider binaries. We again rely on the analysis of \cite{Santos2021} which divided samples that were marked as CP/CB (classical pulsator/close-in binary) into four categories:
\begin{enumerate}
    \item \textbf{type 1}: Samples that showed high amplitude variations. 
    \item \textbf{type 2}: Objects whose light curves resemble that of a contact binary signal.
    \item \textbf{type 3}: $\delta$ Scuti and/or $\gamma$ Doradus candidates or alternatively stars whose light curves are polluted by a nearby star of this type.
    \item \textbf{type 4}:  Objects whose light curves resemble that of a heartbeat star or a close binary with tidally excited oscillation signals.
\end{enumerate}

In their paper, \cite{Santos2021} predicted periods only for type 1 CP/CD. In Figure 14 we compare the period and error distributions of all the CP/CD types. Interestingly, we find very different period distributions; while type 1 and type 3 show relatively short periods, type 2 shows relatively high periods, and type 4 shows periods similar to the entire sample. To investigate this more deeply, we analyze eclipsing binaries. Eclipsing Binaries (EB) are binaries whose orbital plane is edge-on with respect to the observer. Thus, when they rotate through the center of mass, an 'eclipse' is seen. This makes them relatively easy to identify using light curves. To date, there are 2920 known EBs in the Villanova\footnote{\url{http://keplerebs.villanova.edu/}} catalog (\cite{kirk2016}, \cite{Abdul-Masih2016}). It is known that due to tidal forces, those systems show a synchronization of the orbital period and the primary star period for short orbital periods \citep{hut1981}. \cite{Lurie_2017} found that most of the binaries with period $< 10$ Days are synchronized. On the other hand, \cite{Simonian2020} found that most of the rapid rotators are, in fact, binaries, so we expect most of the rapid rotators to be tidally synchronized binaries. Since we know the orbital period of EB systems, we can test the period predictions in this regime.

Figure 15 shows the orbital period vs the ratio between the orbital period and the predicted period for both our model and an ACF. The colors in the Figure correspond to classes that were assigned to each sample by \cite{Lurie_2017} using visual inspection of 2278 EBs. The difference between our model and the ACF is striking.  We see a clear separation between samples with starspots modulations (sp) and other types of stars in the predictions made by our model. For $3< P_{orb} < 10$, sp samples are very close to the synchronization line (the gray line at $P_{orb}/P_{rot}=1$) as expected. For $P_{orb} < 3$ Days and $P_{orb} > 10$ Days we do not see synchronization. While the latter is expected, the former can be assigned to the limitations of the model with very fast rotators, exactly as we saw in Figure 9. We also see that ellipsoidal variations are, in general, overestimated compared to the orbital period, making them appear below the synchronization line, and samples with no apparent periodicity (other than eclipses) are underestimated compared to the orbital period. The ACF created a totally different prediction scheme, falsely recognizing almost every ellipsoidal variation with its orbital periods. We can also see that the synchronization line continues after the $10$ Days threshold suggested by \cite{Lurie_2017}. This reveals a fundamental difference between classical and data-driven methods; our model was trained on data simulated from spots modulations and it implicitly differentiates stars with light curves that are different from the samples it was trained on. This phenomenon is not apparent in the ACF results since it is only able to identify periodicity, without any context. This also suggests that our model is able to distinguish between different stellar types. To test such a distinction, we show in the upper panel of Figure 16 the same plot of $P_{orb}$ vs $P_{orb}/P_{rot}$ colored by the model's confidence (upper left panel) and the total error as defined in \ref{subsec:error_analysis} (upper right panel). We see that while the total error does not separate the data well, the confidence of the model does separate it with very good correspondence to the classes shown in Figure 15. This implies that while different stellar types have the same consistency levels, and therefore a similar total error, the model learned to identify them since their variability structure is inherently different from that of the training data of the model. To find a threshold that can be used to filter the predictions, we constructed the following procedure. We defined a consistent prediction as one with an absolute deviation from the orbital period which is less than $40\%$, or less than a day. The specific choice of $40\%$ and 1 day was made to create a good separation between synchronized and unsynchronized samples. We then binned the dataset to integer values of orbital periods and, for each bin, we calculated the fraction of consistent samples according to the above criteria. We then took the bin with the largest jump in consistent samples to be the cutoff. As a sanity check, we did the same procedure for the comparison with \cite{McQuillan2014} predictions (using the periods from \cite{McQuillan2014}). The graphs of consistent fraction per bin for both scenarios can be seen in Appendix \ref{Appendix_D}. We found that in both scenarios, 3 days seem to be an optimal cutoff, so we adopted this value. The lower panel in Figure 16 shows the following groups for $P_{orb} > 3$ days:
\begin{itemize}
    \item synchronized samples (cyan). These are consistent samples with $3 < P_{orb} < 10$.
    \item Unsynchronized samples (red). These are non-consistent samples with $3 < P_{orb} < 10$.
    \item samples with $ P_{orb} > 10$ (gray).
\end{itemize}
The lower right panel shows the confidence histogram for each group and exhibits a very good separation as expected. We therefore decided to take the average confidence value of the unsynchronized group, $0.86$, as a filtering threshold.

\begin{figure*}
    \begin{centering}
    \begin{minipage}[b]{0.5\textwidth}
        \includegraphics[width=\textwidth]{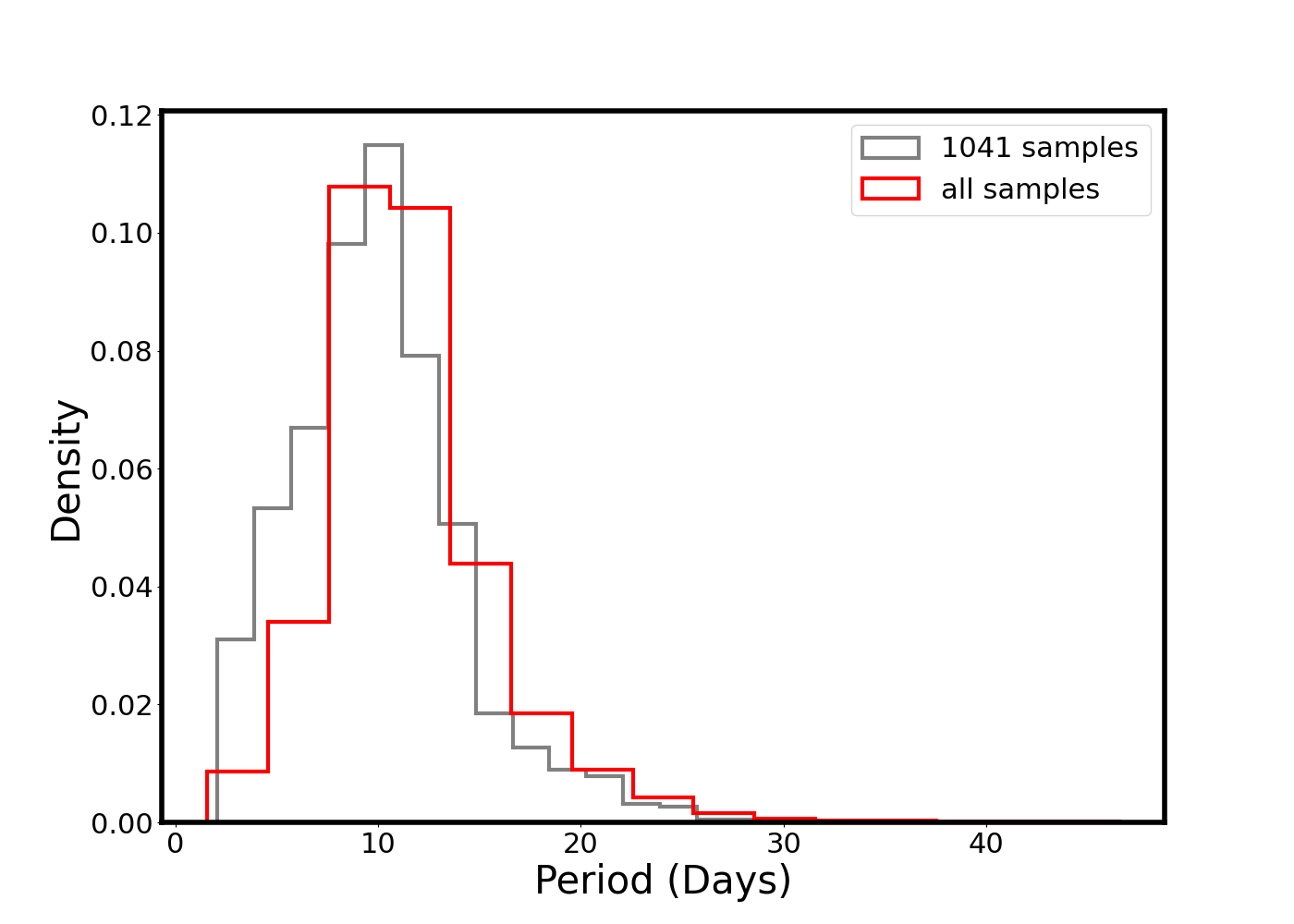}
    \end{minipage}
    \begin{minipage}[b]{0.5\textwidth}
       \includegraphics[width=\textwidth]{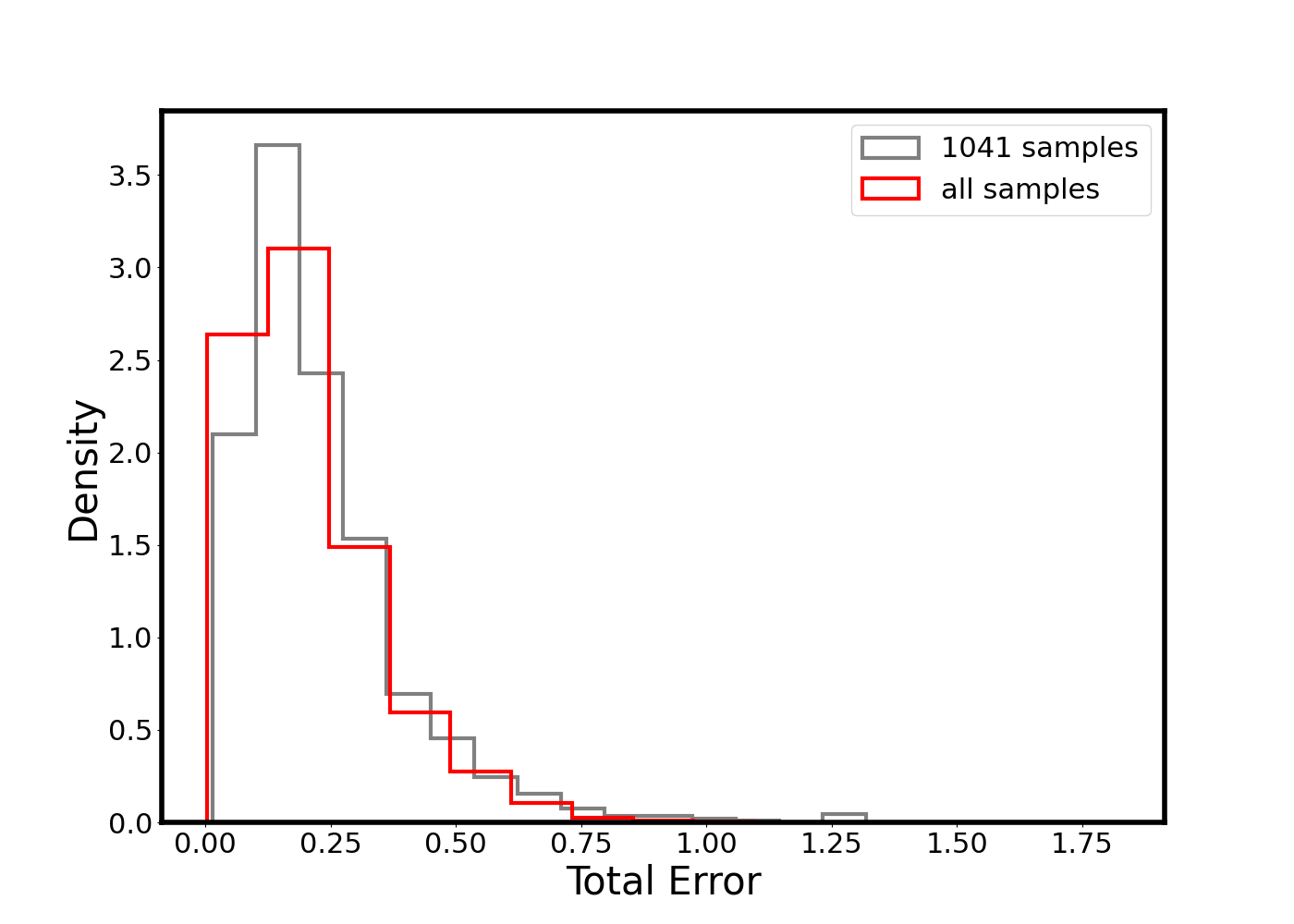}
    \end{minipage}
    \caption{distributions of the entire sample (red) vs possible contaminants (gray): No Rotation flag 6 in \cite{Santos2021}. the upper panel shows the period distributions and the lower panel shows the total error distributions.}
    \end{centering}      
\end{figure*}

\begin{figure*}
    \begin{centering}
    \begin{minipage}[b]{0.22\textwidth}
        \includegraphics[width=\textwidth]{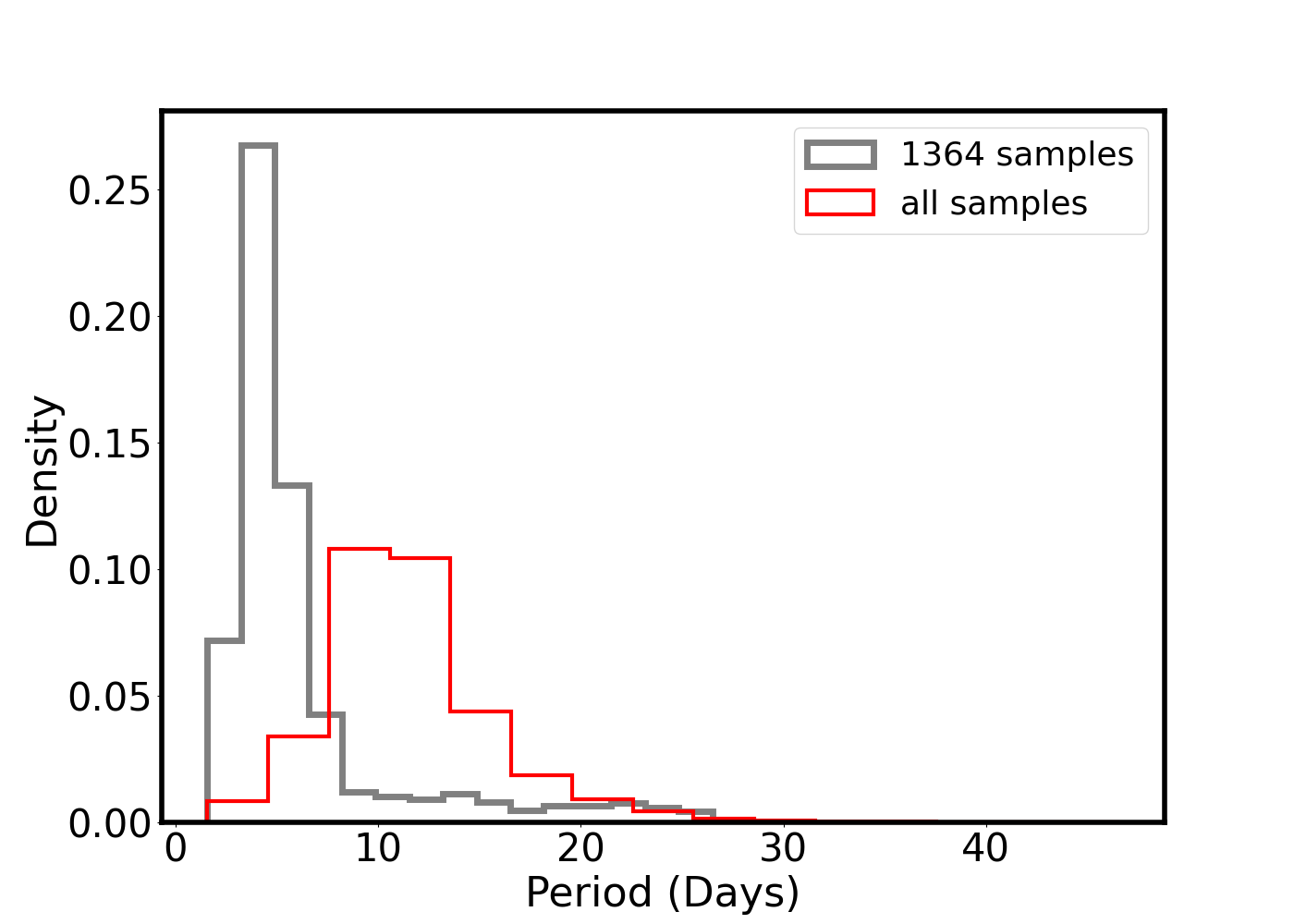}
    \end{minipage}
     \begin{minipage}[b]{0.22\textwidth}
        \includegraphics[width=\textwidth]{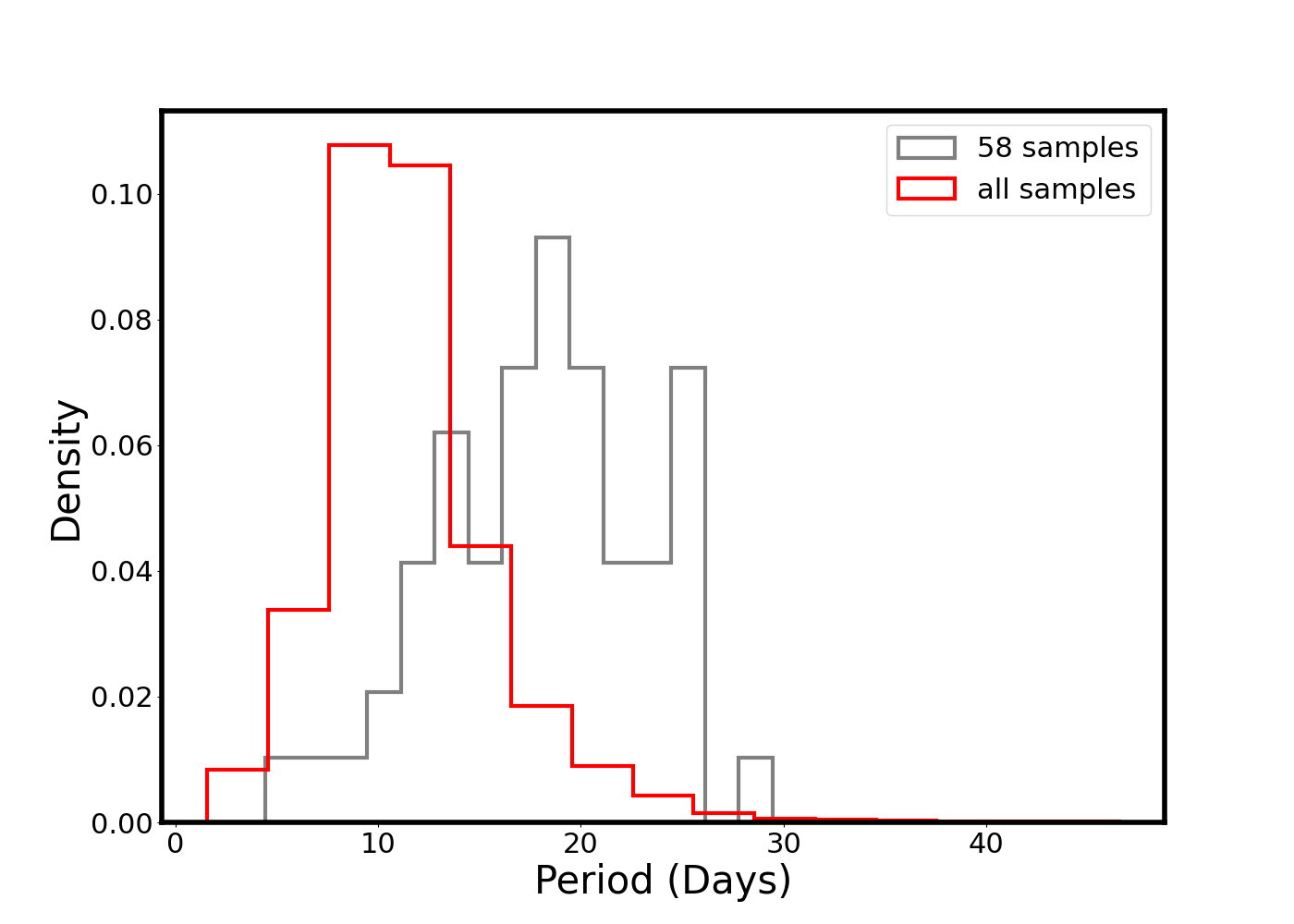}
    \end{minipage}
    \begin{minipage}[b]{0.22\textwidth}
        \includegraphics[width=\textwidth]{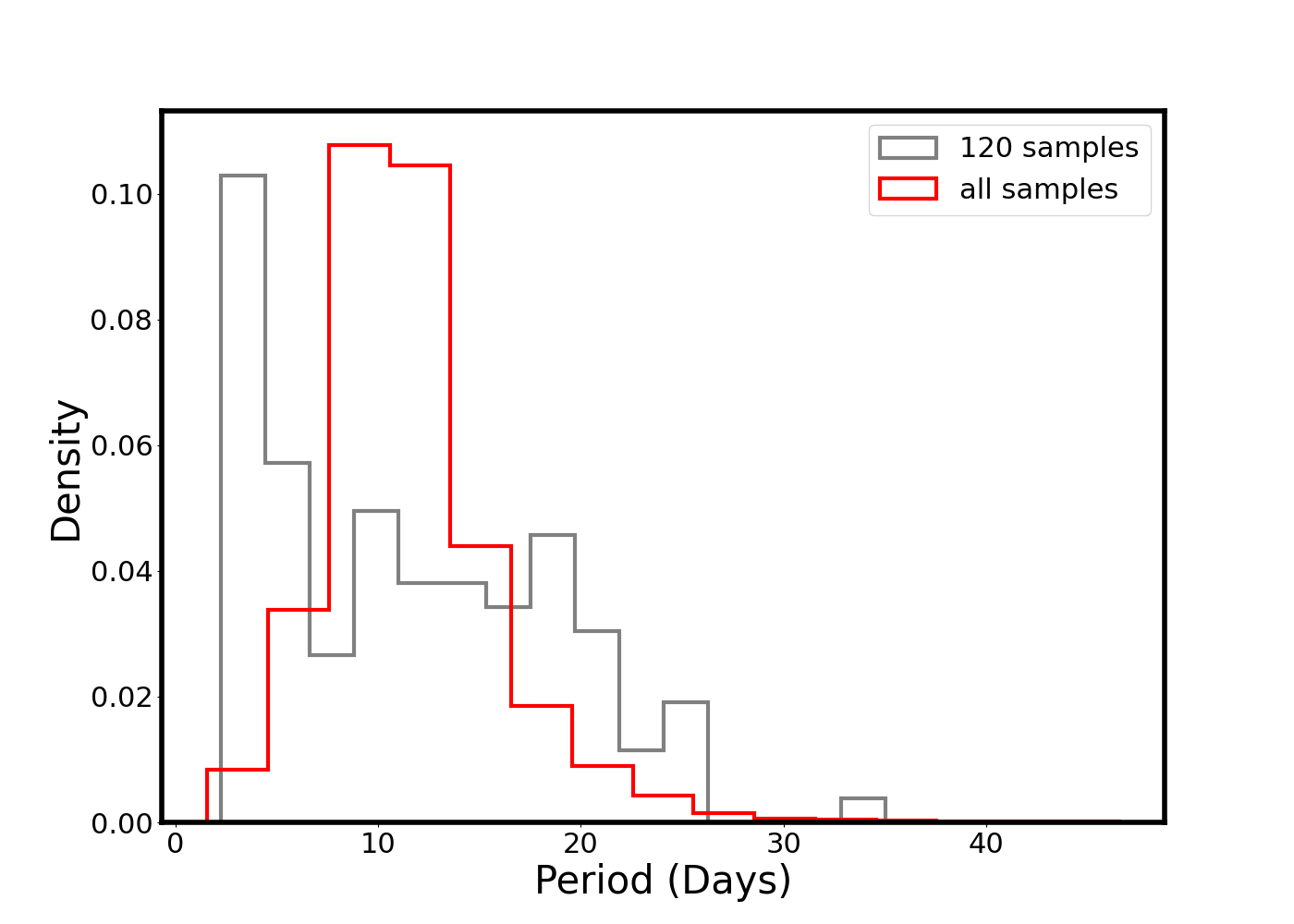}
    \end{minipage}
    \begin{minipage}[b]{0.22\textwidth}
        \includegraphics[width=\textwidth]{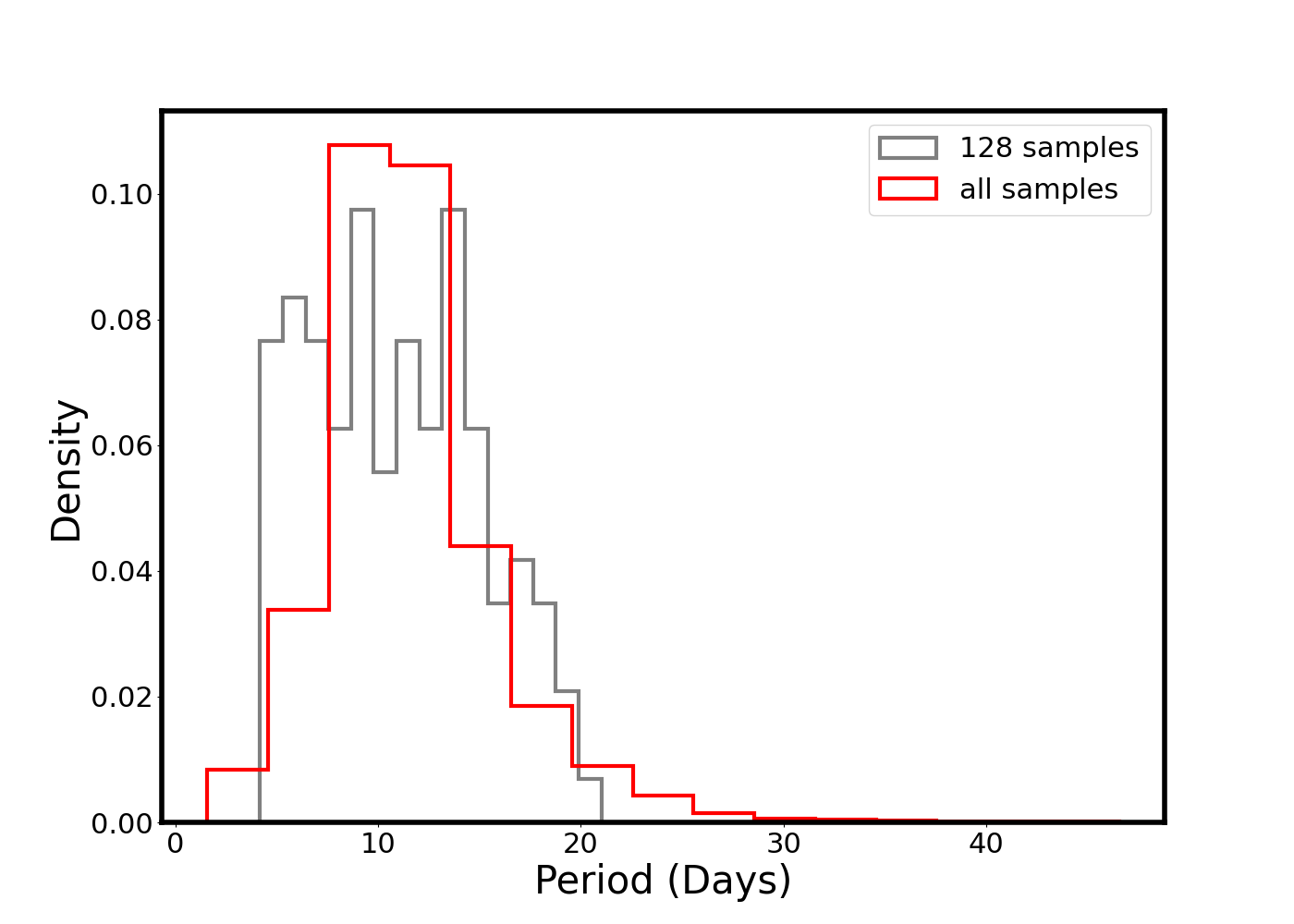}
    \end{minipage}
    \begin{minipage}[b]{0.22\textwidth}
        \includegraphics[width=\textwidth]{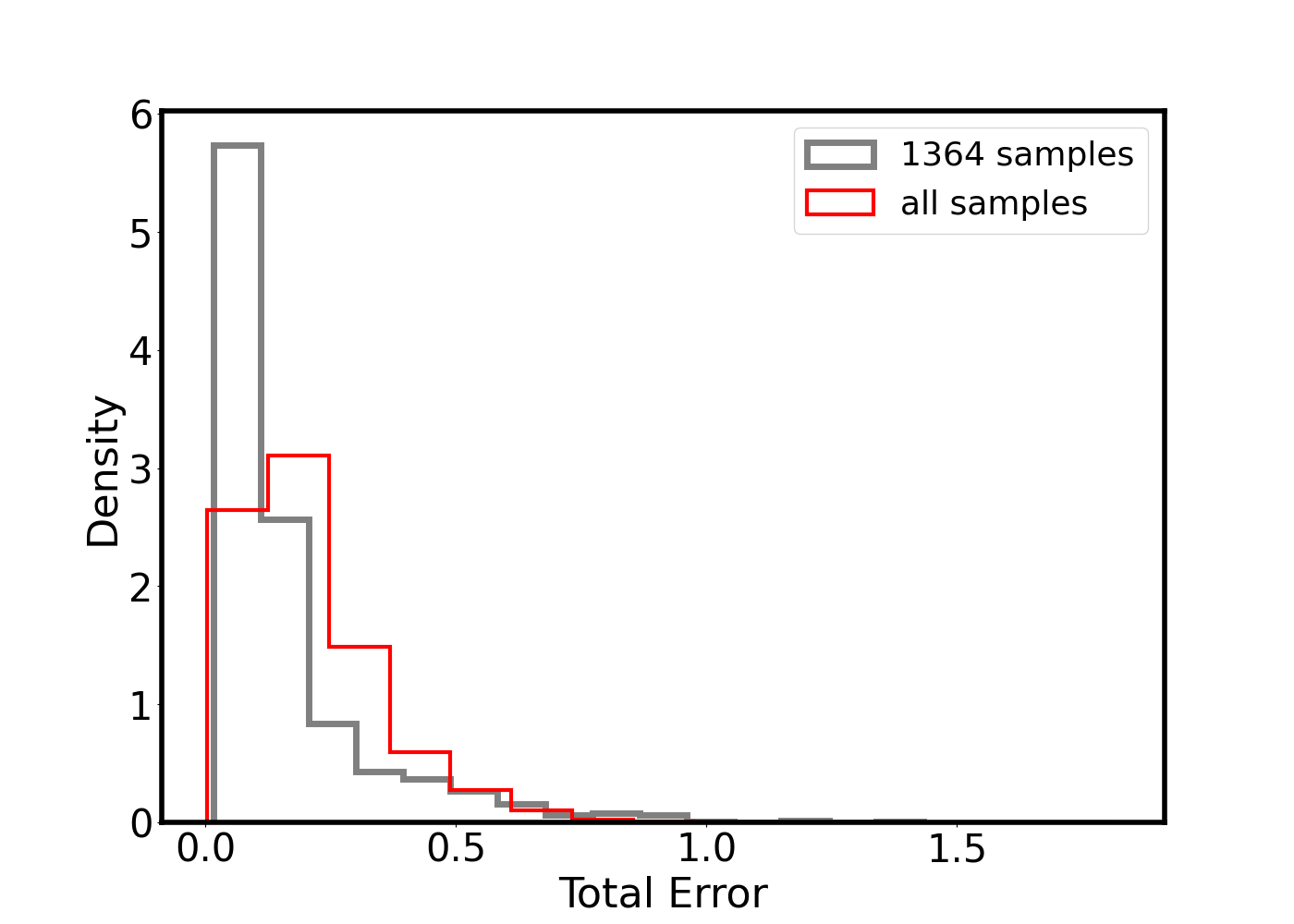}
    \end{minipage}
     \begin{minipage}[b]{0.22\textwidth}
        \includegraphics[width=\textwidth]{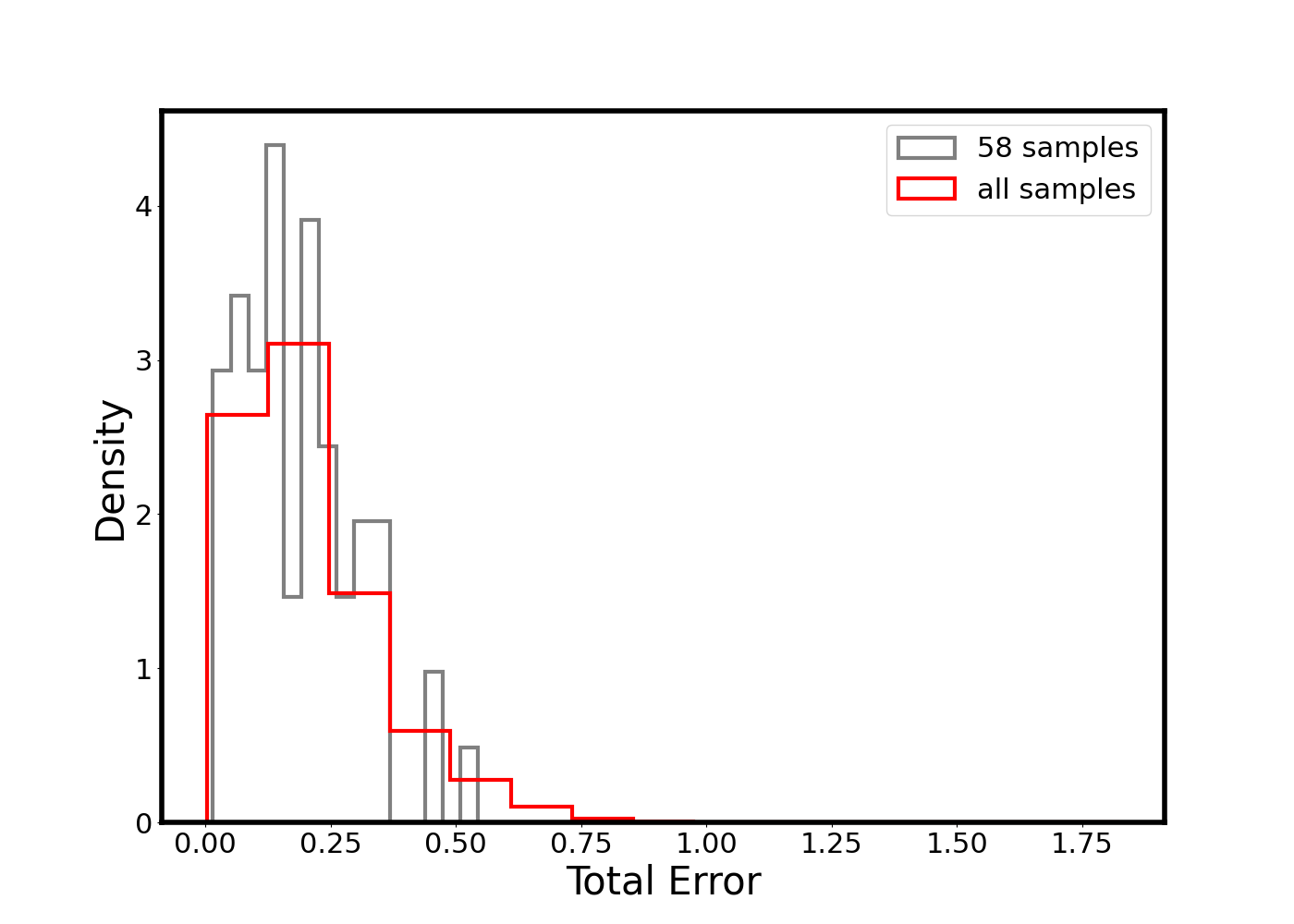}
    \end{minipage}
    \begin{minipage}[b]{0.22\textwidth}
        \includegraphics[width=\textwidth]{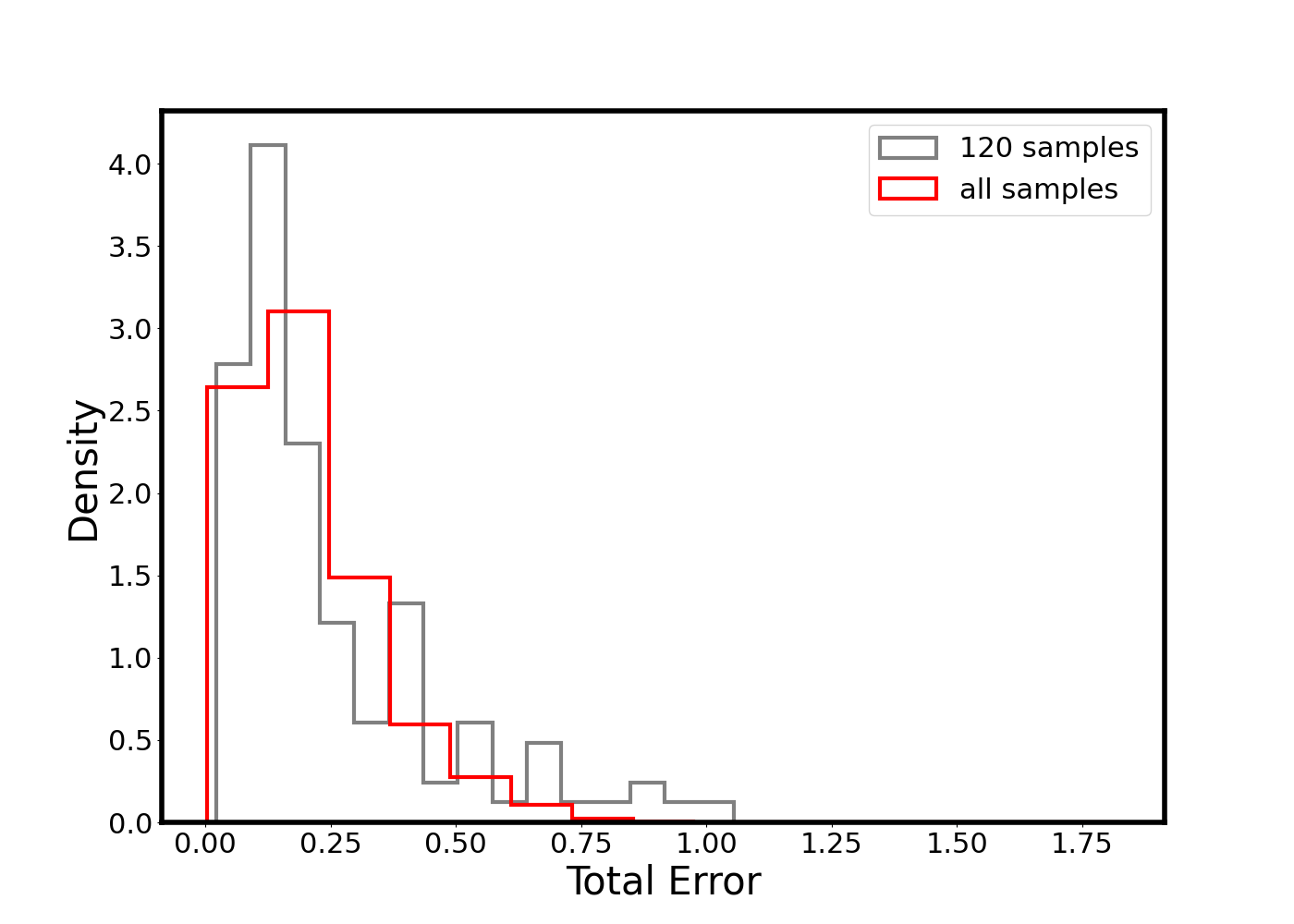}
    \end{minipage}
    \begin{minipage}[b]{0.22\textwidth}
        \includegraphics[width=\textwidth]{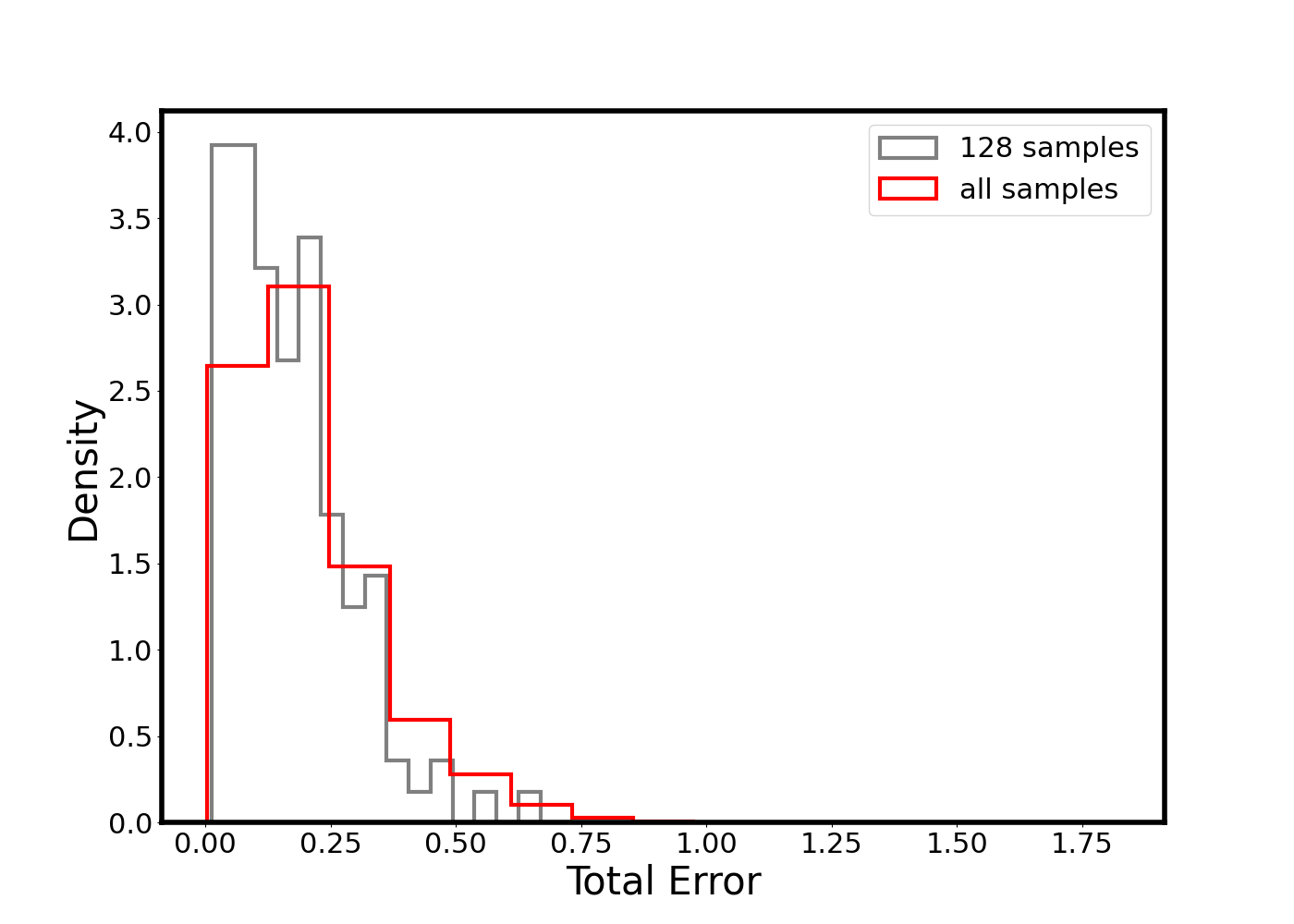}
    \end{minipage}
    \caption{distributions of the entire sample (red) vs possible contaminants (gray): 4 types of CP/CD contaminant. see text for details. the upper panels show the period distributions and the lower panels show the total error distributions.}
    \end{centering}      
\end{figure*}

\begin{figure*}
    \begin{centering}
    \begin{minipage}[b]{0.8\textwidth}
        \includegraphics[width=\textwidth]{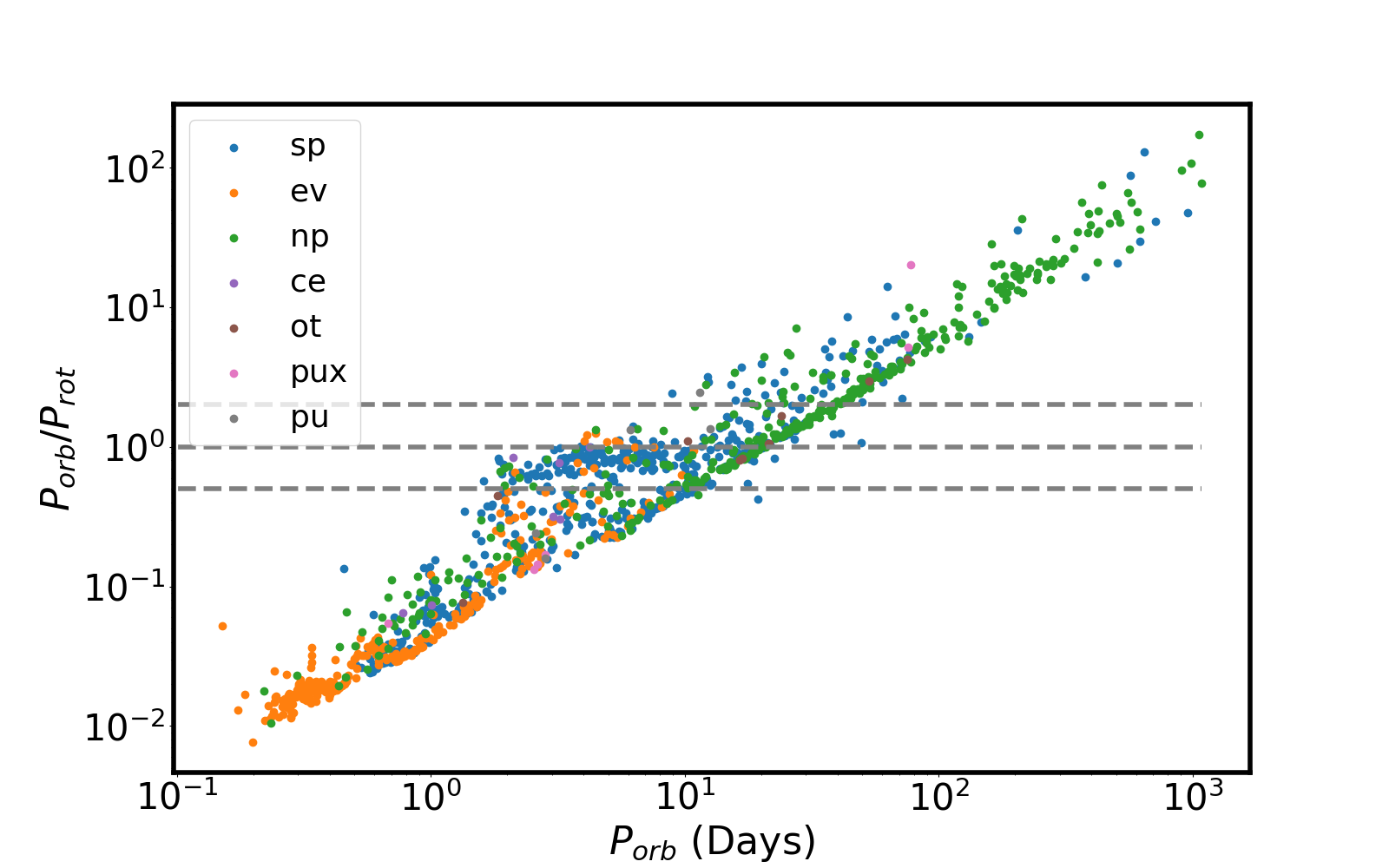}
    \end{minipage}
    \begin{minipage}[b]{0.8\textwidth}
      \includegraphics[width=\textwidth]{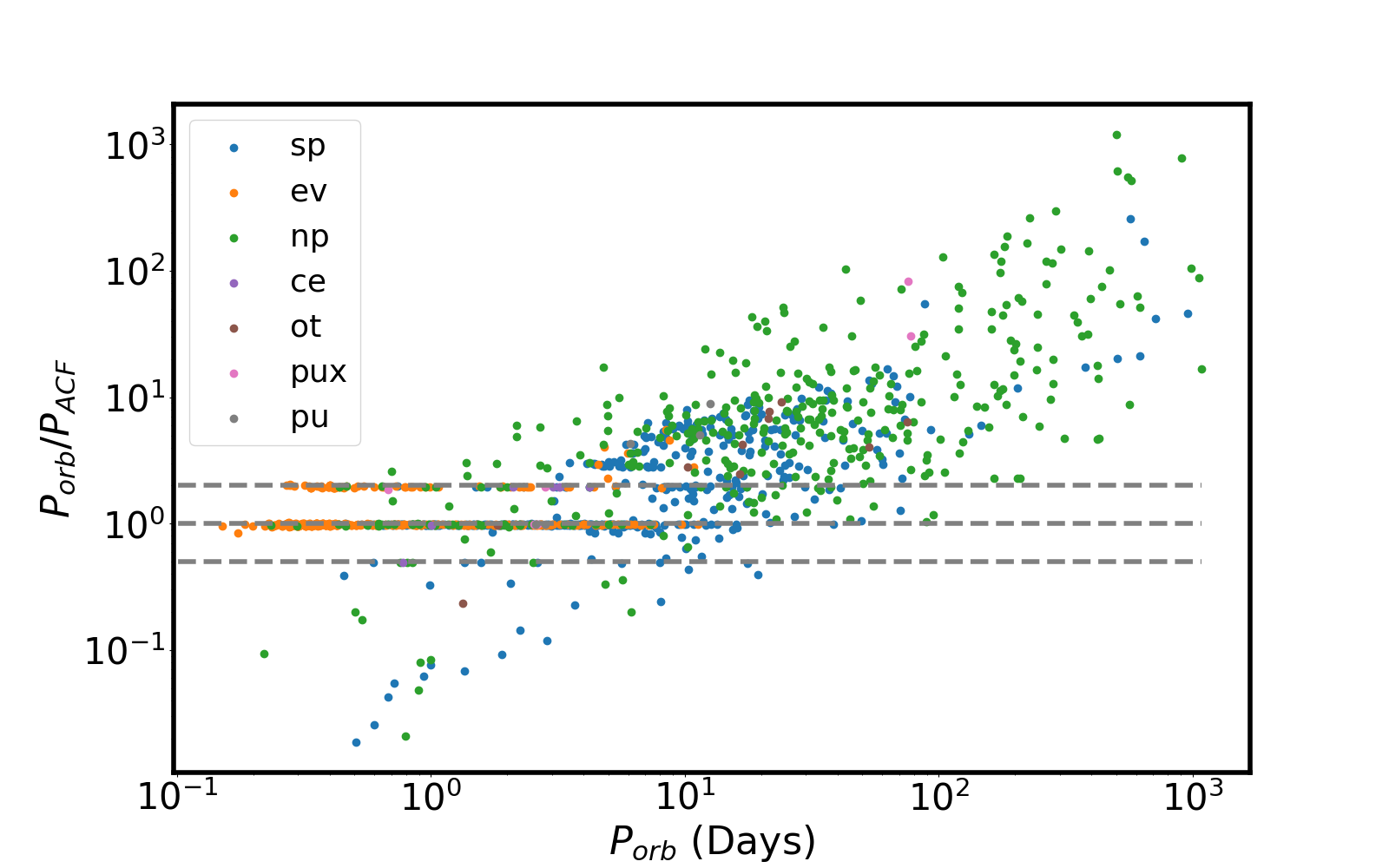}    
    \end{minipage}
    \end{centering}
\caption{Eclipsing binaries orbital period vs the ratio between the orbital period and stellar period. Horizontal lines corresponds to $P_{orb}/P_{rot} = 0.5, 1, 2$. The upper panel shows the stellar period predicted by LightPred and the lower panel shows the stellar period predicted by the ACF method. The colors represent stellar classes defined by \cite{Lurie_2017} - \textbf{sp} - starspots modulation;  \textbf{ev} - Ellipsoidal variations; \textbf{np} - no periodic out-of-eclipse variability; \textbf{ce} - Targets where starspot modulations appear to have been mistaken for ellipsoidal variation; \textbf{ot} - another periodic out-of-eclipse variability; \textbf{pux} - possible pulsator; \textbf{pu} - likely pulsator.}
\end{figure*}

\subsubsection{Filtering Predictions}
\label{subsec:false_positive}
Next, we want to investigate the different inconsistencies between our model and the predictions of \cite{McQuillan2014} as shown in Figure 9. For the fast rotators, as mentioned on \ref{subsec:poluttion}, we found the same inconsistency with EBs, and in both cases we found 3 days as a good cutoff for fast rotators. Three examples of fast rotators, as indicated by \citep{McQuillan2014}, can be seen in \ref{Appendix_D}. Therefore, we decided to remove all the known EBs with orbital period $<3$ days as well as samples where the ACF  predicted periods $<3$ days. Since fast rotators are dominated by binaries, and since we saw in Figure 15 that in this regime ACF identifies the binary orbital period rather than the stellar period, the two conditions can be understood as pointing to the same phenomena. We found a total of 3106 such samples.

Another group of inconsistent predictions is the samples around the 'half-period' line in Figure 9. Figure 17, shows on the left panel the same comparison as seen on the upper left panel in Figure 9 with different coloring; colors in Figure 17 represent the $w$ parameter that was given by \cite{McQuillan2014} as a score for periodicity. We see that the inconsistent samples have significantly lower $w$ values, meaning they are the least periodic samples in this subset. This is not surprising given that we saw the same pattern for the model confidence in Figure 9 and the fact that the total error and $w$ are correlated, as seen in Figure 12. Another interesting observation is related to high $w$ values. We now see that those values represent samples with fast predictions; In this regime, we expect tidally synchronized binaries to dominate. We also saw that ACF always finds the orbital period in those cases, while our model separates different stellar classes. This might explain the negative correlation we saw in Figure 12 for very high $w$ values.

To better examine this inconsistency, we visually examined 200 samples randomly sampled from this regime. During this examination, we found an interesting phenomenon that confused our predictions:
\cite{McQuillan2013} reported that spots at different hemispheres produce a lower correlation at half the period. This aligns with the findings of \cite{Berdyugina2003} that spots tend to form in active longitudes that are separated by $180\degree$.  To take this into account, they checked if a higher second peak exists in the ACF. If such a peak exists, they took it as the period indicator instead of the regular identification process. If our model takes the first peak of the ACF, those cases might result in a double period difference between our model and \cite{McQuillan2014}. In our visual inspection, we did find such examples where our model falsely detects half of the true period. One such example can be seen in Figure 18, lower right panel. Another interesting phenomenon are light curves that show dramatic changes in their periodicity; an example can be seen in Figure 18, lower left panel. It can be seen that the light curve in those cases has two distinct regions, one with high amplitude and one with lower. ACF would detect periods related to the high amplitude region only, regardless of its length, while our model usually gives a sort of 'average' prediction. It is not clear which result is better in general. In addition, we found a significant amount of cases where we believe our model gives a more accurate prediction. In those cases usually \cite{McQuillan2014} took the second peak of the ACF while the first one seemed to be more accurate. Two such examples can be seen in Figure 18, upper panels. Overall, we found that our model seemed more accurate in 57\% of the cases that we examined. We conclude that the only cases where we are sure our model is wrong, are cases with higher second ACF peak. To take this into account, we updated our predictions in the following way: in predictions where ACF shows a higher second peak and the difference between our predictions and \cite{McQuillan2014} predictions is a factor of 2 ($\pm 20\%$), we adapt the predictions of \cite{McQuillan2014}. We found 2231 such cases. The right panel of Figure 17 shows the same comparison after updating the predictions.

Lastly, we use our model confidence score to filter predictions. We saw in \ref{subsec:poluttion} that the confidence distributions of synchronized EBs and non-synchronized EBs are very different and suggested a value of 0.86 to potentially separate between spots-induced samples from other types of stars. Taking only predictions with confidence higher than 0.86 removed a total of 21173 samples. Figure 19, left panel, is identical to the upper panel of Figure 15 but after removing samples with $P_{orb} > 3$ and samples with confidence $< 0.86$. The right panel is the same but shows the orbital period vs the rotation period. In this case, we expect the synchronization line to be on the line with slope one. We see that the synchronization line is much clearer with most of the points below it being removed. We also see that the synchronization line disappears naturally at $P_{orb}>10$. In addition, we see that all the pulsators were removed and out of 350 Ellipsoids that were in the original sample, only five are left, of which four appear to be synchronized. For spots-modulated samples (sp), out of 285 samples with $P_{orb}>3$ in the original sample, 215 ($75\%$) were left after the filtering. For samples with no apparent periodicity (np), out of 295 samples with $P_{orb}>3$ in the original sample, 134 ($45\%$) were left after the filtering. This supports the assumption that the filtering process successfully removes variables that differ from spot-induced stars, mainly pulsators and ellipsoidals. \cite{Lurie_2017} used the ratio of the orbital period and the rotation period to define synchronization, finding that $72\%$ of the samples with orbital periods between 2 and 10 days have $0.92 < P_{orb}/P_{rot} < 1.2$. We see that in our results, most of the samples lie on a slightly lower line with $72\%$ of the samples with orbital periods between 3 and 10 days having $0.69 < P_{orb}/P_{rot} < 1.2$.

\begin{figure*}
    \centering
    \begin{tabular}{cc}
        \includegraphics[width=0.45\textwidth]{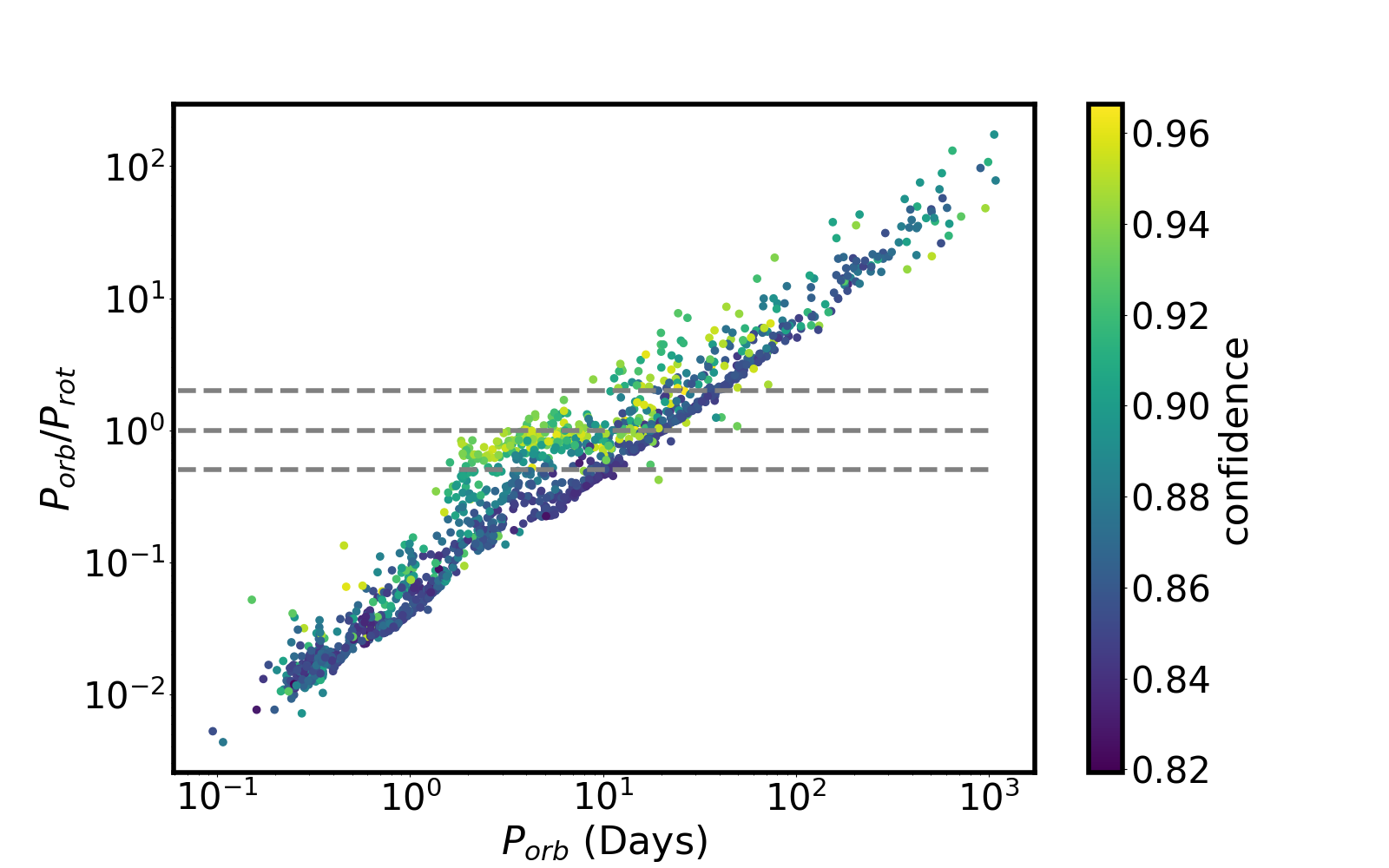} &
        \includegraphics[width=0.45\textwidth]{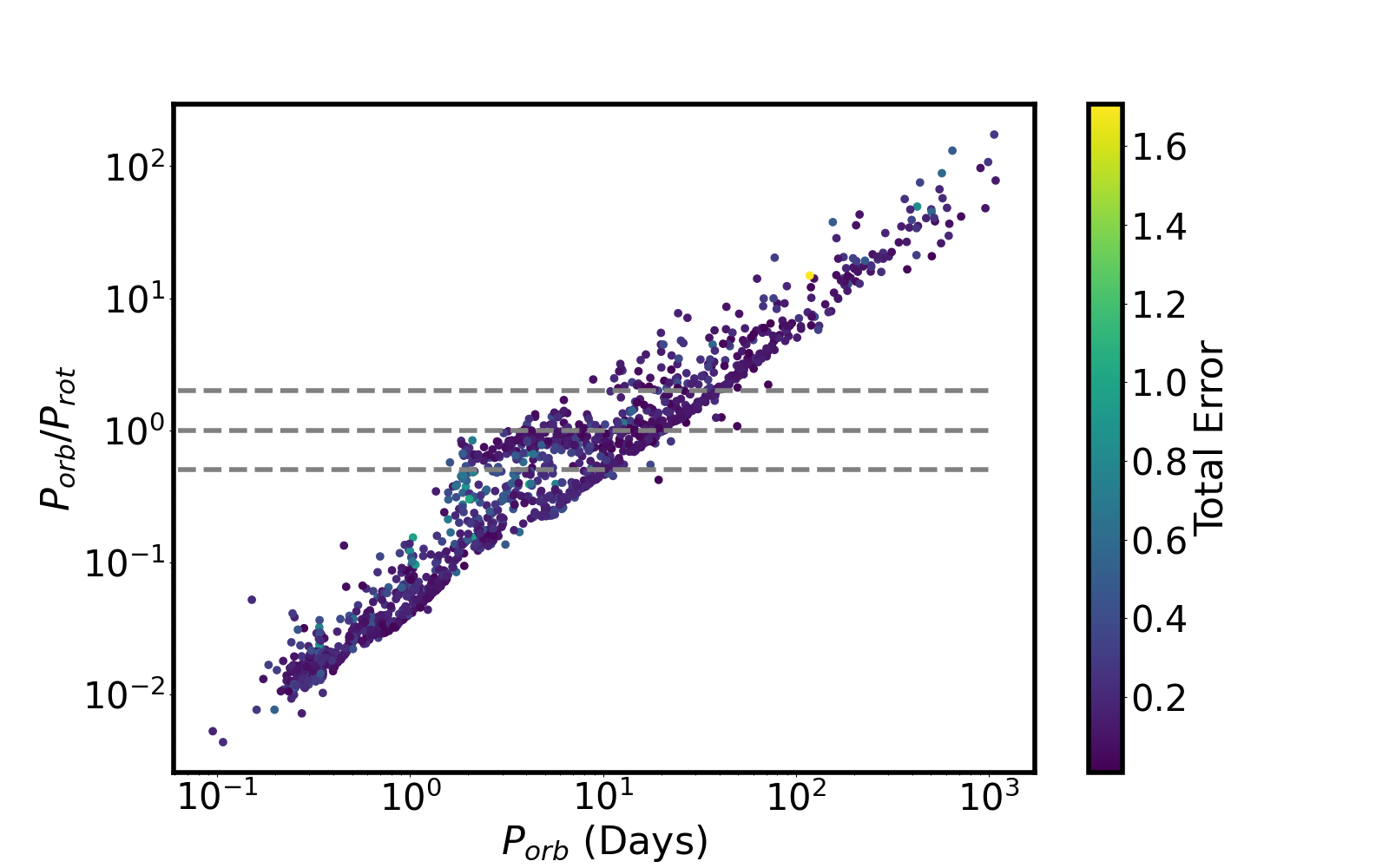} \\
        \includegraphics[width=0.45\textwidth]{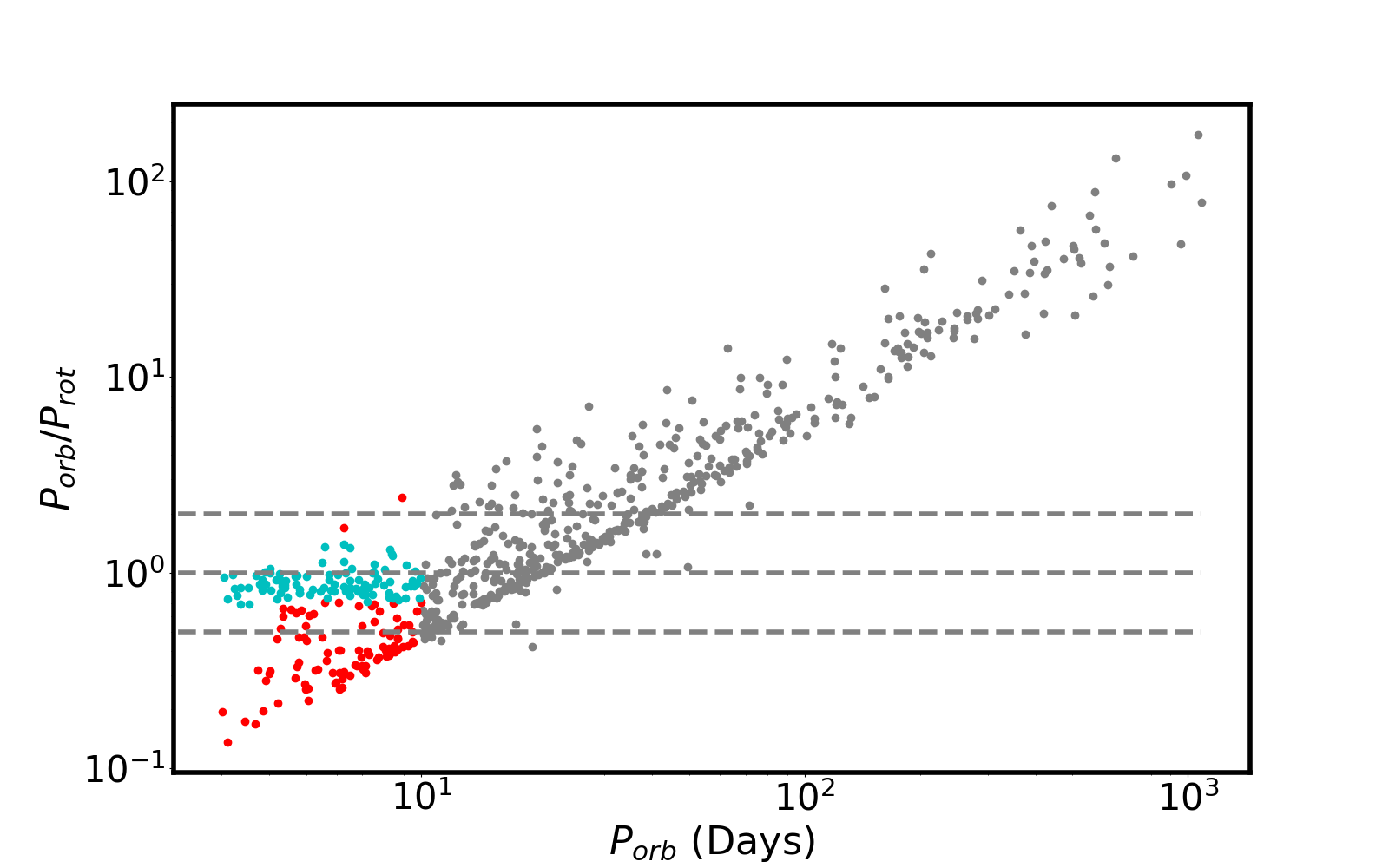} &
        \includegraphics[width=0.45\textwidth]{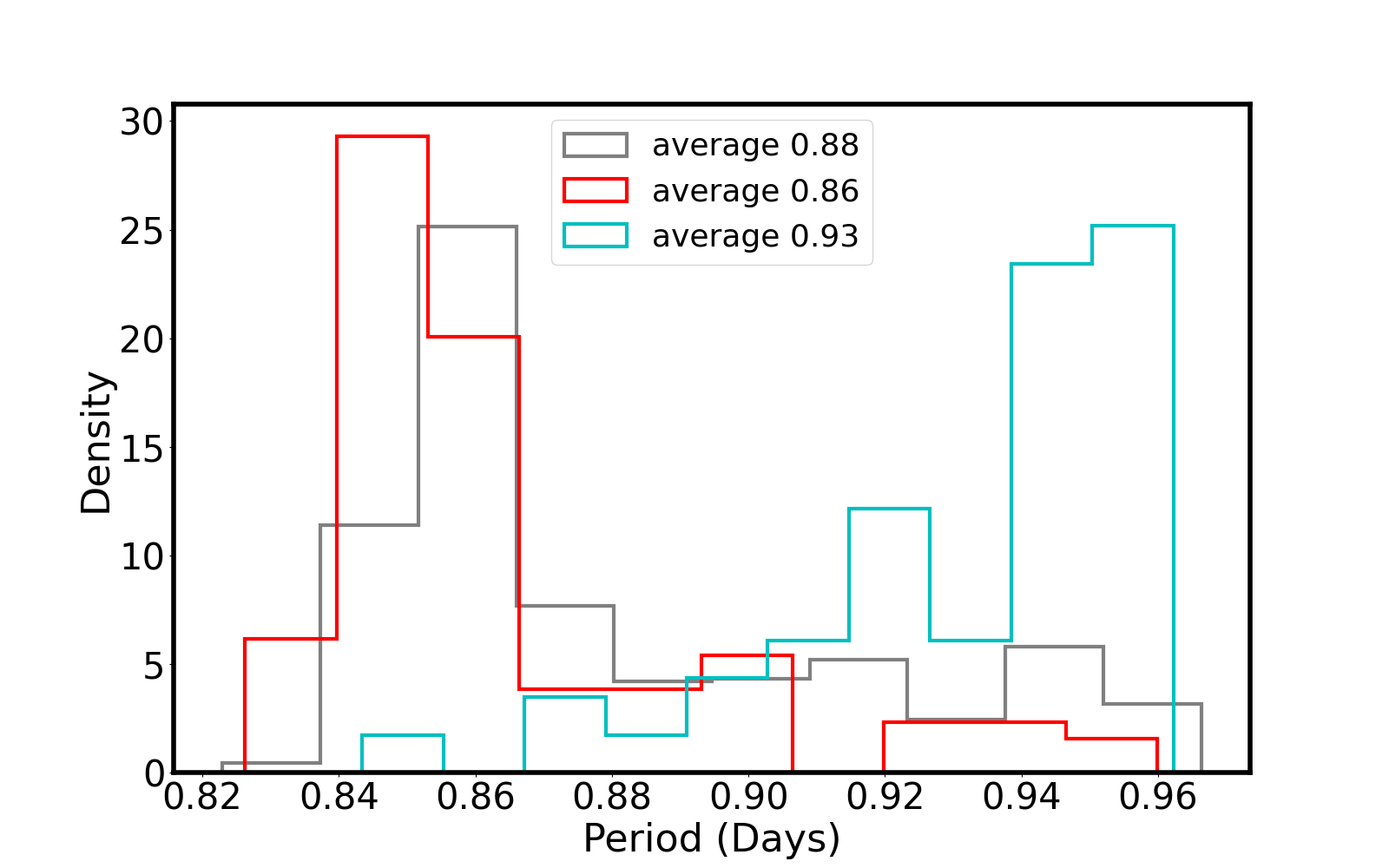}
    \end{tabular}
    \caption{Eclipsing binaries orbital period vs the ratio between orbital period and rotational period. Horizontal lines corresponds to $P_{orb}/P_{rot} = 0.5, 1, 2$. In the upper left panel, colors correspond to the model's confidence. In the upper right panel, colors correspond to the total error. The lower left panel shows the three classes defined in \ref{subsec:poluttion}. The lower right panel shows the histograms of confidence for the classes on the left panel.}
\end{figure*}

\begin{figure*}
    \centering
    \begin{tabular}{cc}
        \includegraphics[width=0.45\textwidth]{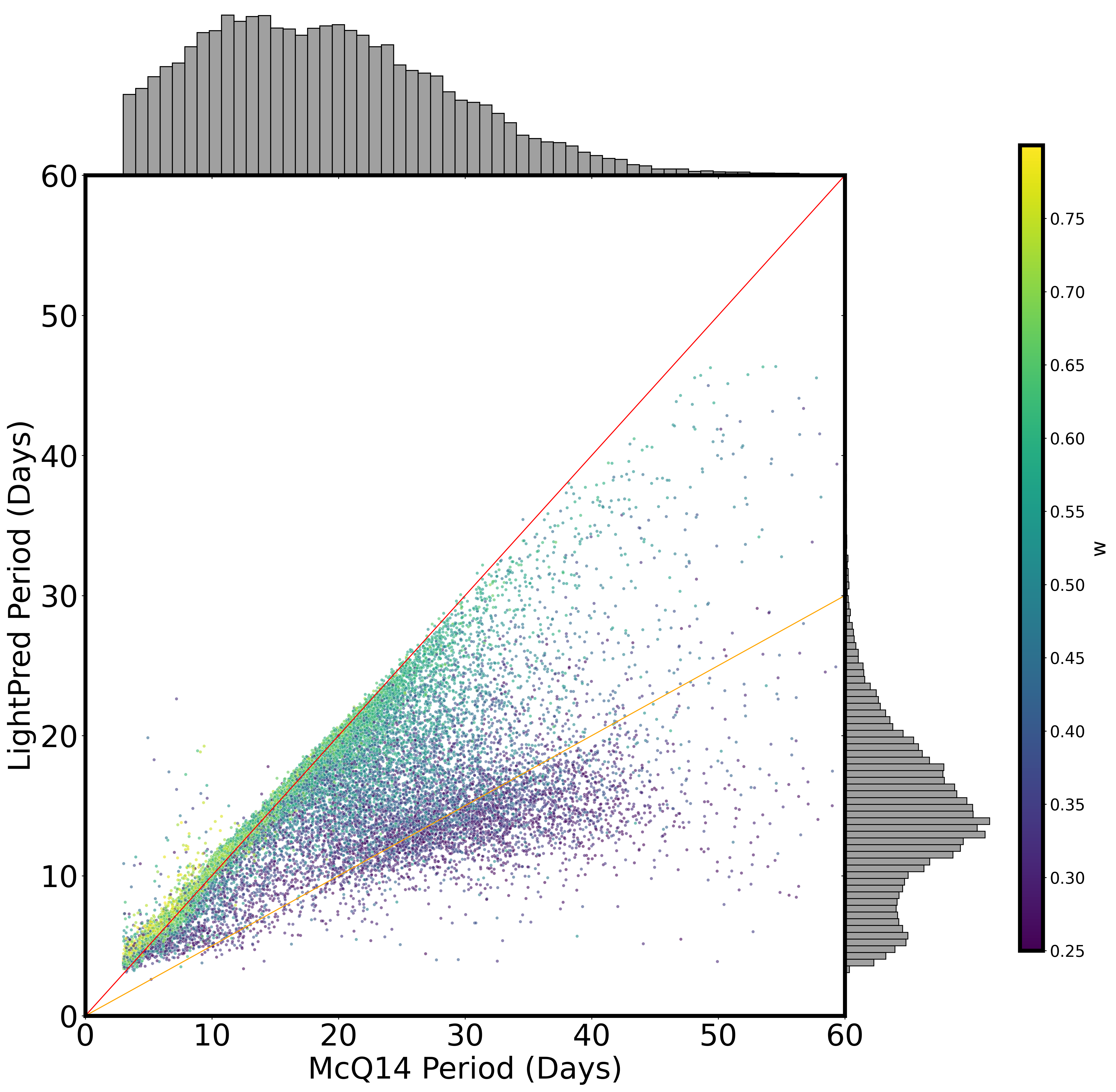} &
        \includegraphics[width=0.45\textwidth]{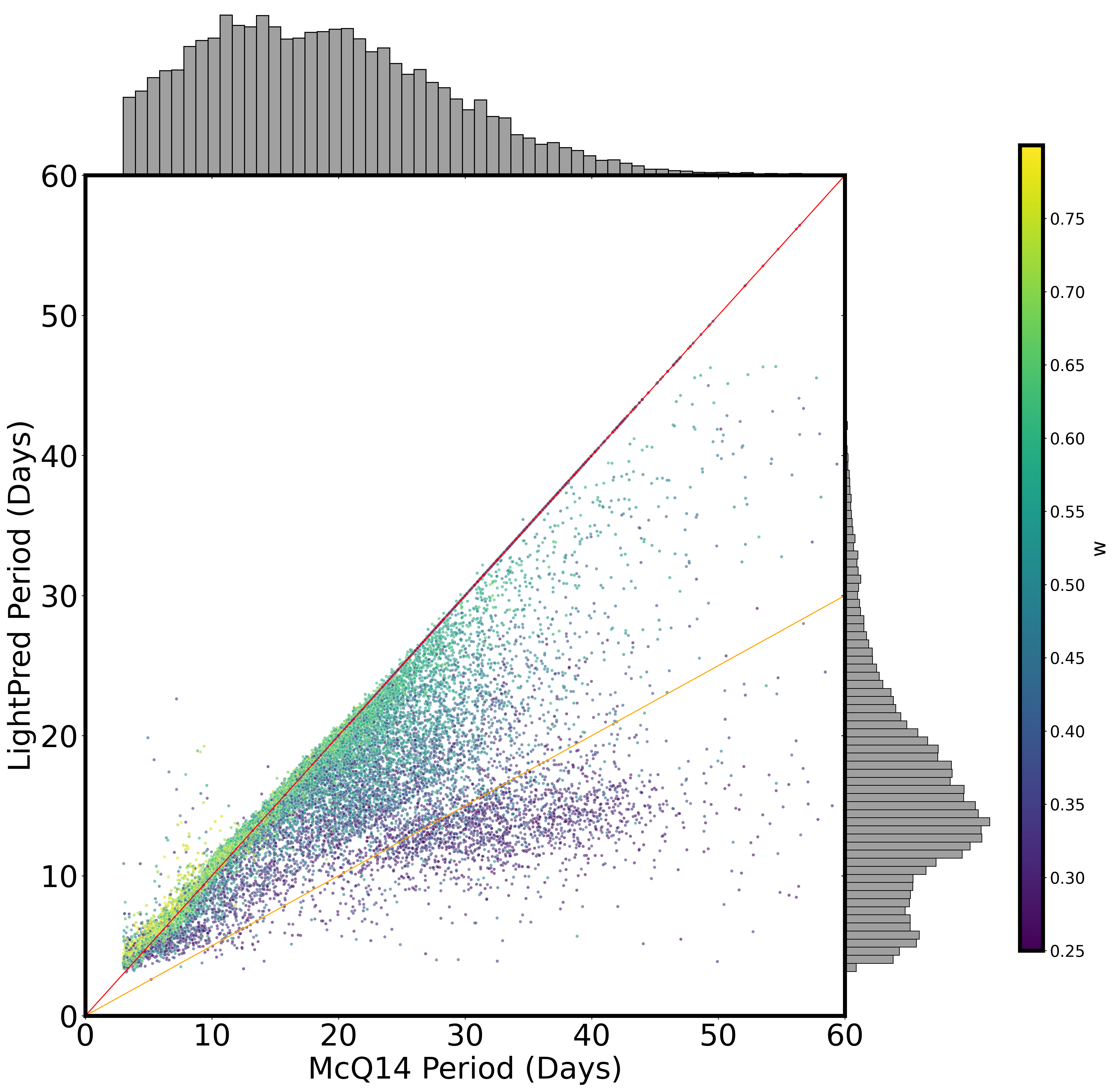} \\
    \end{tabular}
    \caption{Comparison of our predictions and \cite{McQuillan2014} predictions as seen in figure 9. Colors represent the $w$ parameter as given in \cite{McQuillan2014}. The left panel shows the results for $P_{McQ14} > 3$ and the right panel shows the results after the full selection process as described in \ref{subsec:catalog}.}
\end{figure*}

\begin{figure*}
    \begin{centering}
    \begin{minipage}[b]{0.45\textwidth}
        \includegraphics[width=\textwidth]{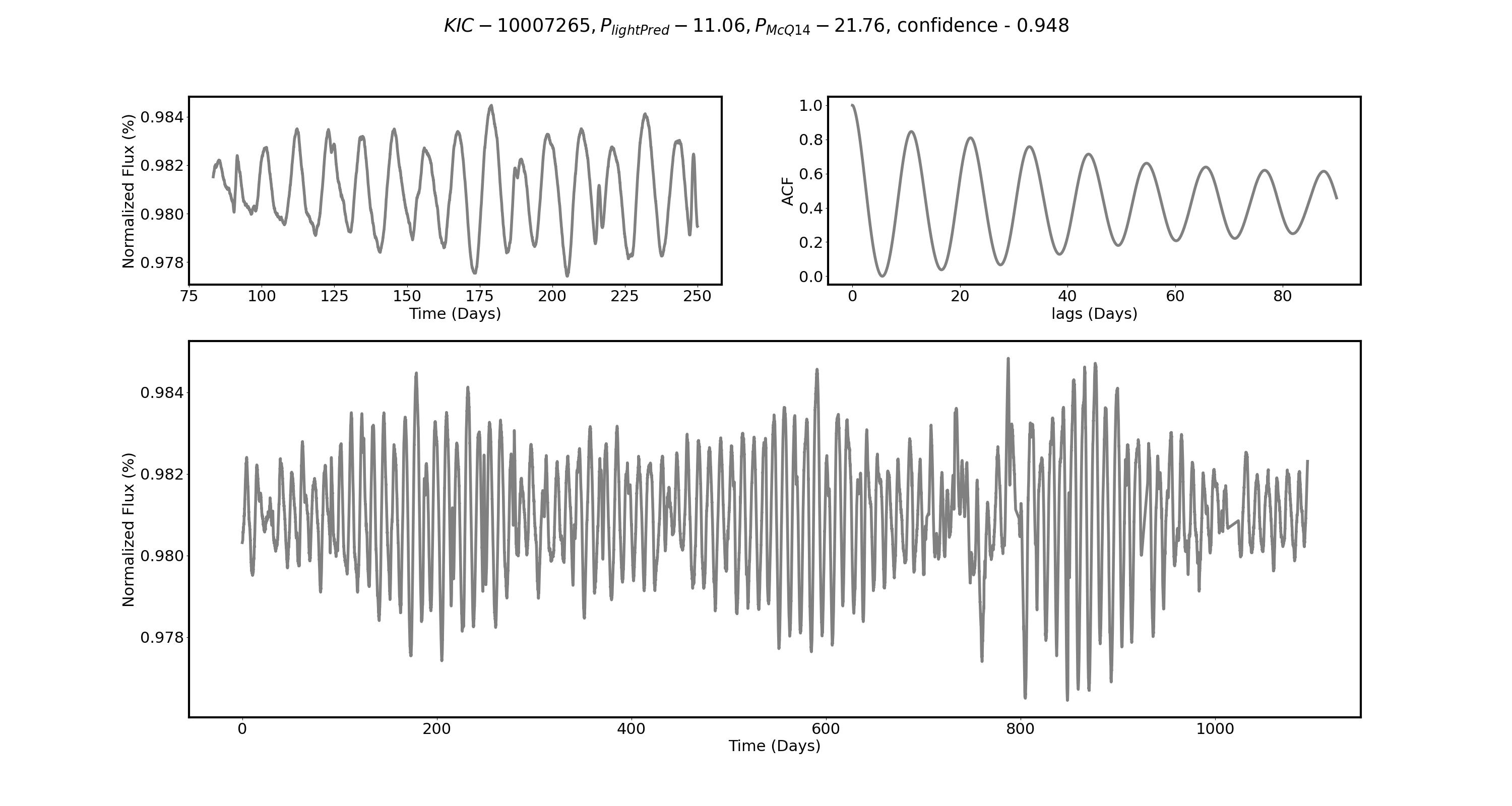}
    \end{minipage}
    \begin{minipage}[b]{0.45\textwidth}
       \includegraphics[width=\textwidth]{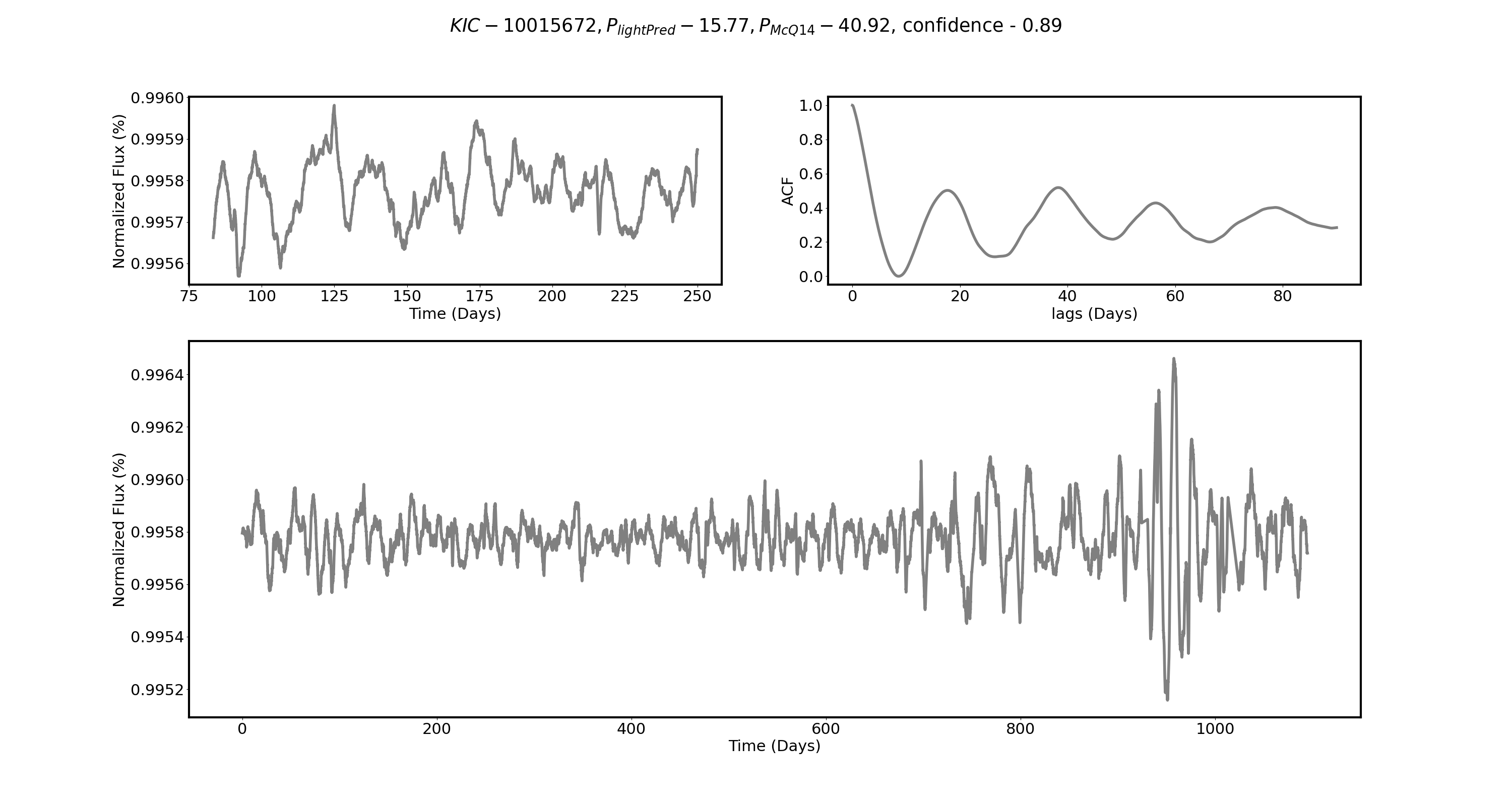}
    \end{minipage}
    \begin{minipage}[b]{0.45\textwidth}
       \includegraphics[width=\textwidth]{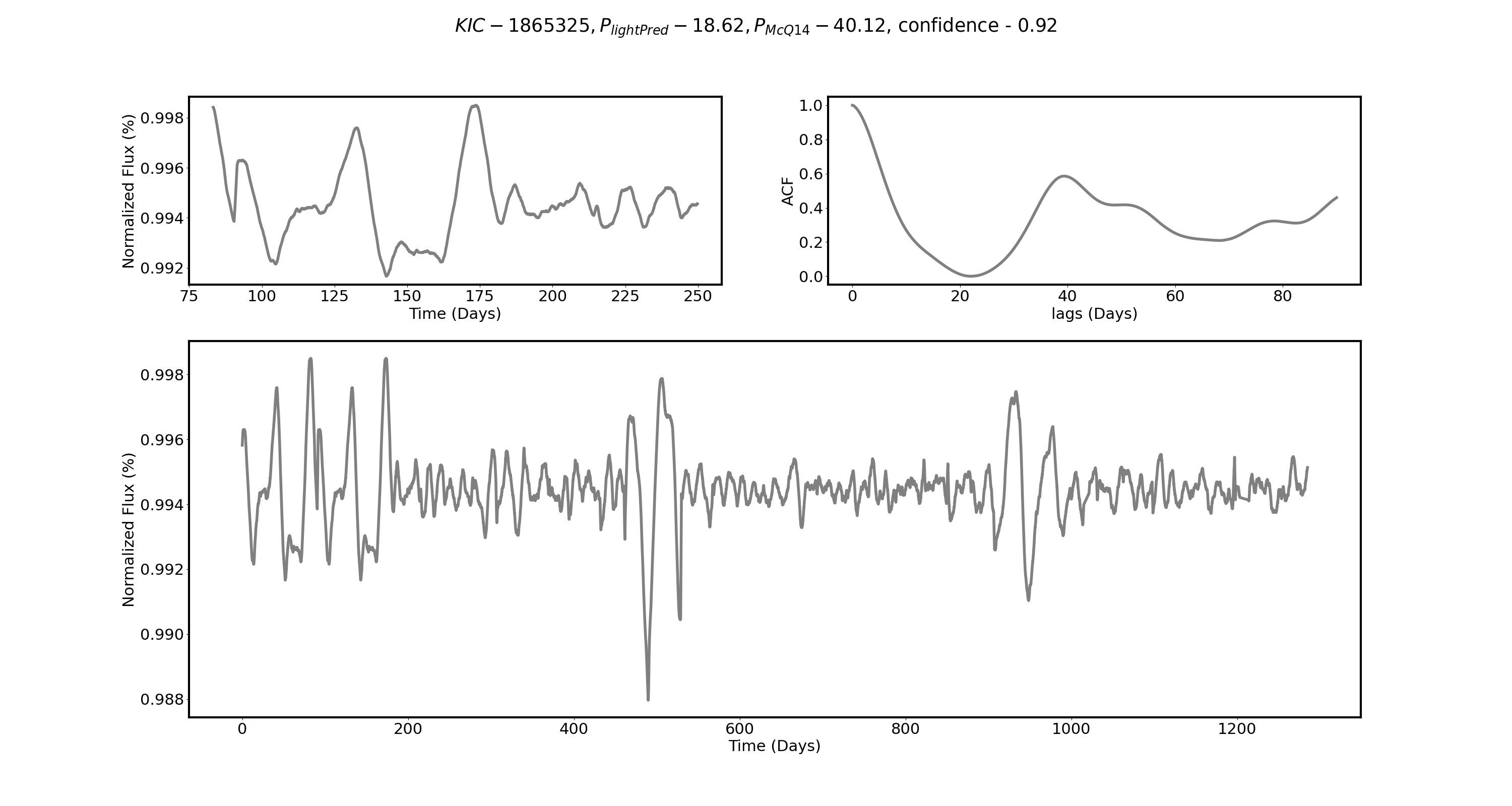}
    \end{minipage}
    \begin{minipage}[b]{0.45\textwidth}
       \includegraphics[width=\textwidth]{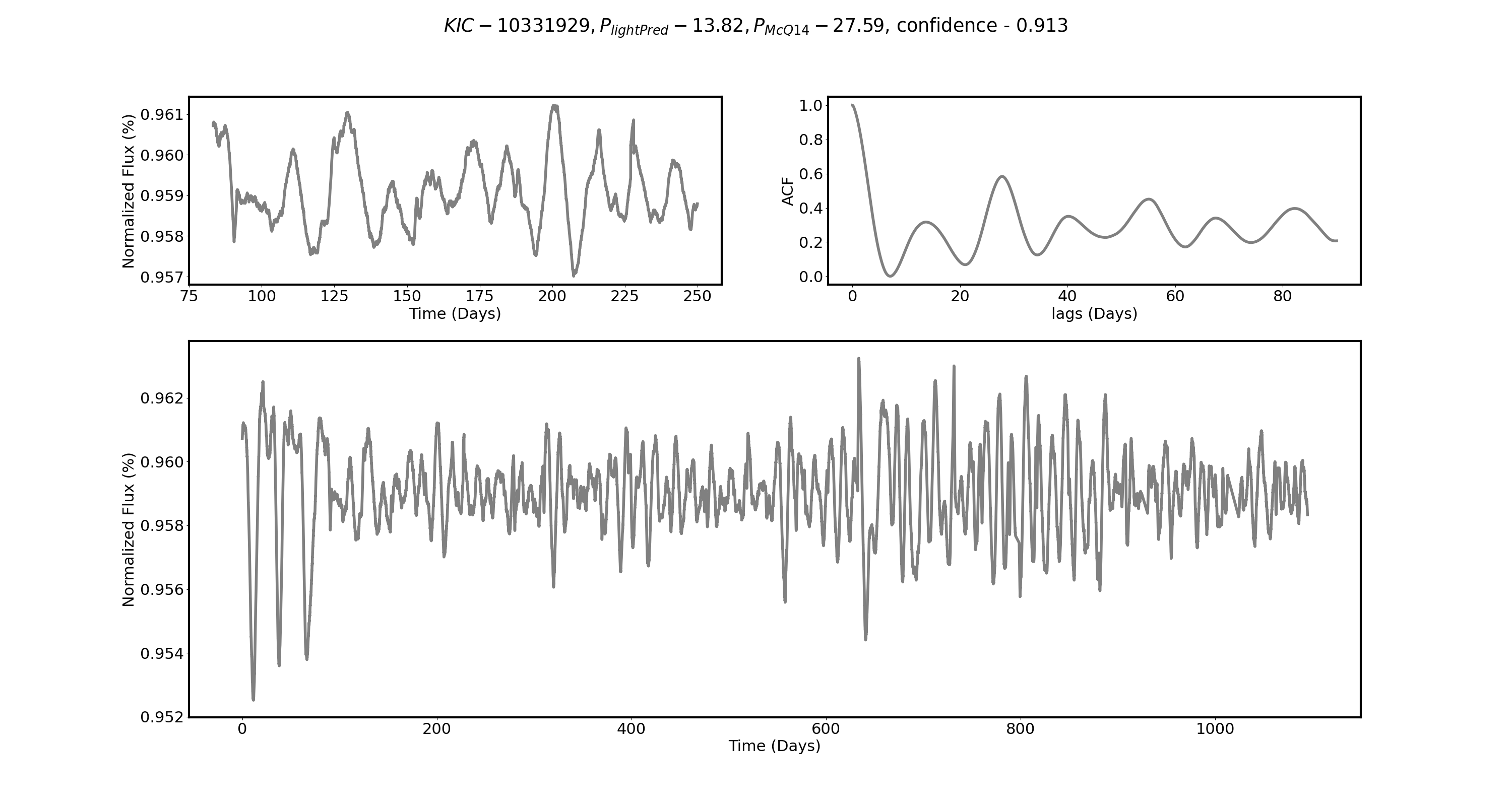}
    \end{minipage}
    \caption{Four light curve examples with inconsistency between our model and \cite{McQuillan2014}. For each sample, the lower panel shows the full light curve and the upper left panel shows a zoom-in slice of 166 days. The upper right panel shows the ACF of the light curve. The light curve is averaged over a window of one day. The titles show the \textit{Kepler} KIC, lightPred period, McQ14 period from \cite{McQuillan2014}, and the confidence of the model. }
    \end{centering}      
\end{figure*}

\begin{figure*}
    \begin{centering}
        \begin{minipage}[b]{0.45\textwidth}
            \includegraphics[width=\textwidth]{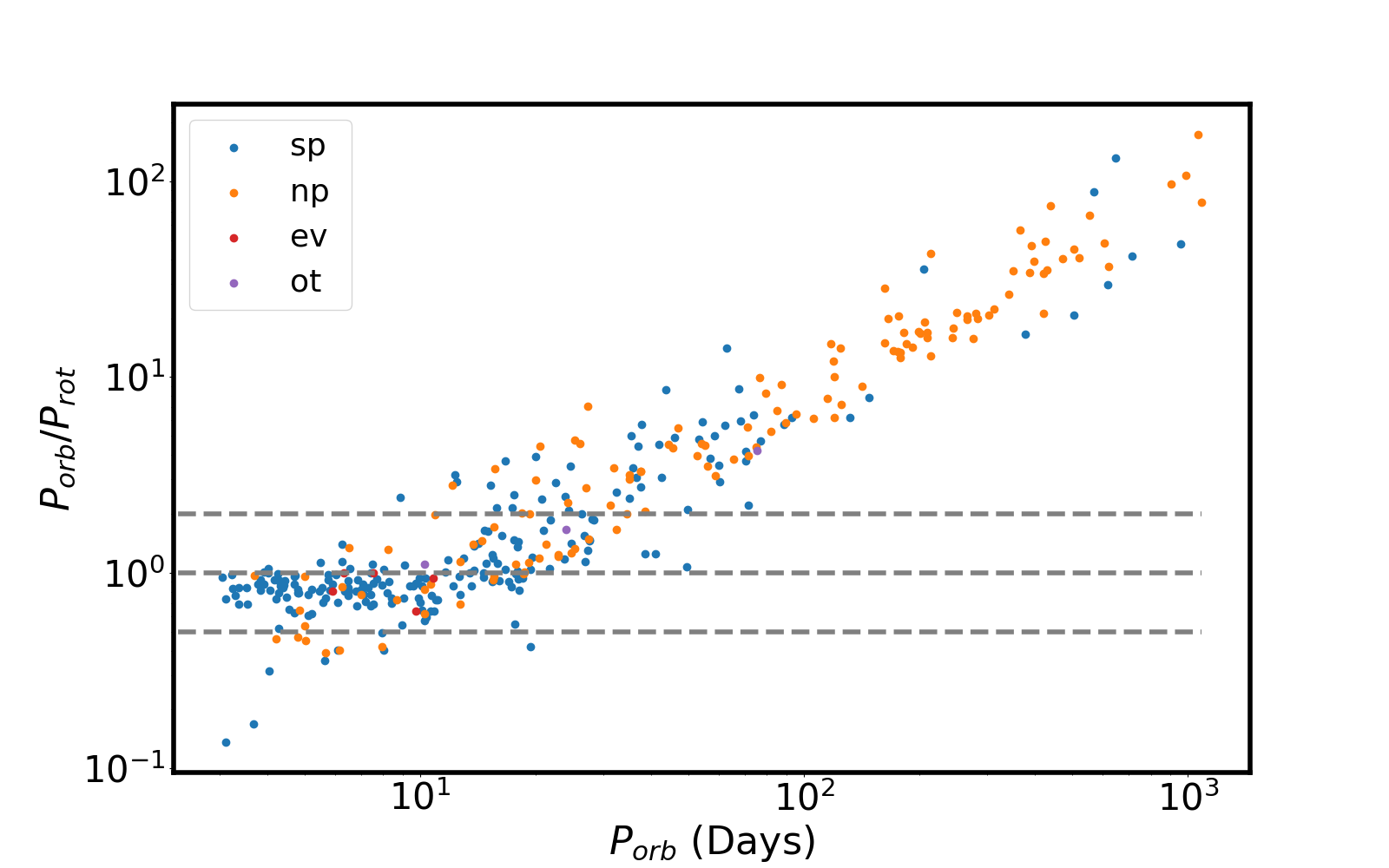}   
        \end{minipage}
        \begin{minipage}[b]{0.45\textwidth}
            \includegraphics[width=\textwidth]{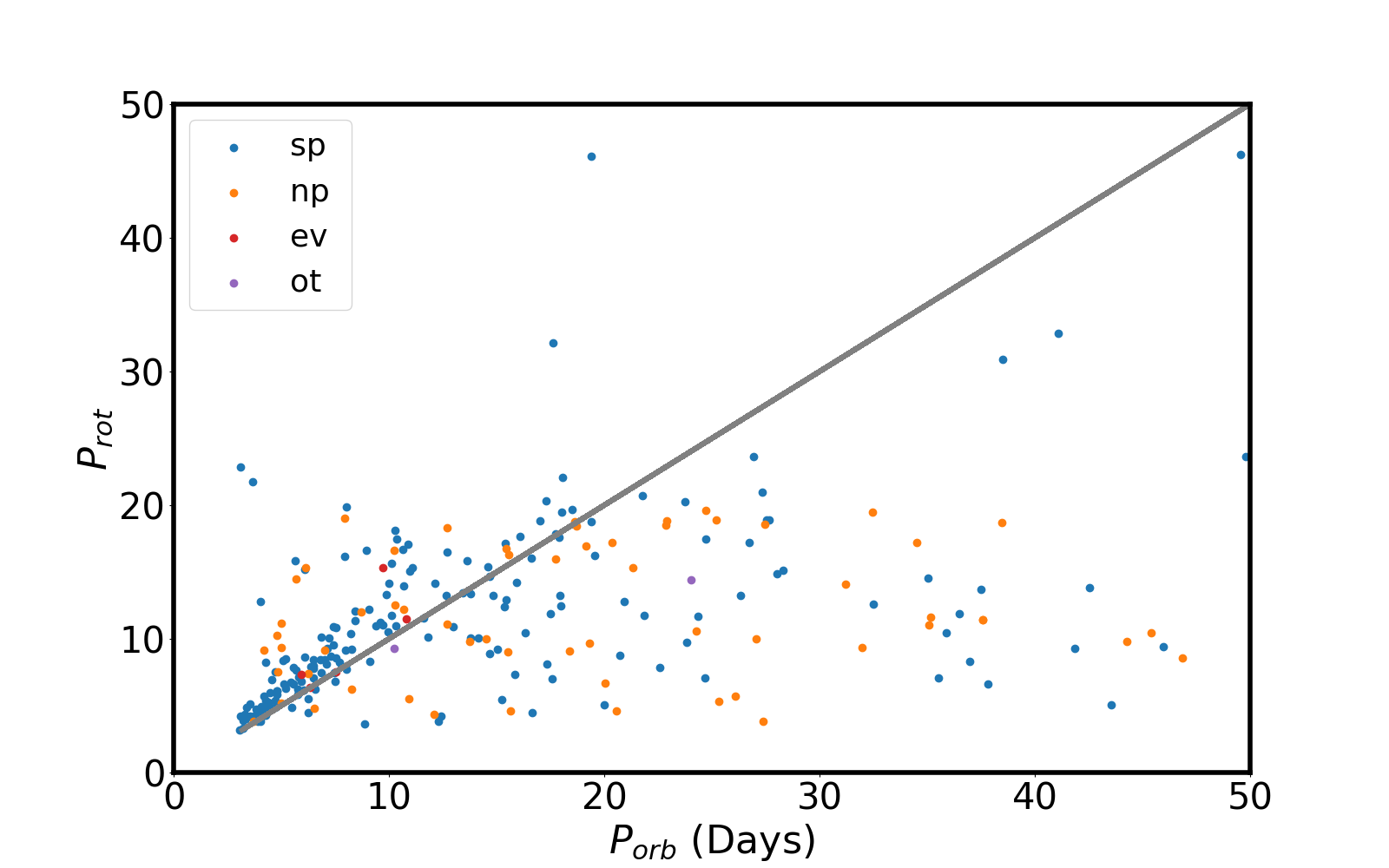}   
        \end{minipage}
    \end{centering}
\caption{Left panel: Eclipsing Binaries orbital period vs the ratio between orbital and rotation period. Right panel: Eclipsing Binaries orbital period vs rotation period. Both panels show the results after filtering low confidence and fast rotators as explained in \ref{subsec:false_positive}. Colors represents classes defined by \cite{Lurie_2017} and explained in Figure 15.
Horizontal lines corresponds to $P_{orb}/P_{rot} = 0.5, 1, 2$.}
\end{figure*}

\begin{table*}
    \begin{center}
    \begin{tabular}{||cccccc||}
    \hline
        \textbf{KID} & \textbf{Predicted Period} & \textbf{Observational Error (\%)} & \textbf{Model Confidence} & \textbf{Total Error} & \textbf{Flag} \\ \hline
      757450 & 19.67 & 0.049 & 0.942 & 0.052 & 1 \\ \hline
        891901 & 20.059 & 0.188 & 0.921 & 0.204 & \\ \hline
        891916 & 5.846 & 0.072 & 0.954 & 0.075 & \\ \hline
        892195 & 11.329 & 0.187 & 0.864 & 0.216 & 3 \\ \hline
        892203 & 6.693 & 0.29 & 0.88 & 0.329 & 3 \\ \hline
        892675 & 16.399 & 0.103 & 0.898 & 0.114 & 3 \\ \hline
        892678 & 16.474 & 0.308 & 0.889 & 0.347 & 3 \\ \hline
        892713 & 5.601 & 0.192 & 0.906 & 0.212 & \\ \hline
        892718 & 11.621 & 0.131 & 0.893 & 0.147 & 3 \\ \hline
        892772 & 10.528 & 0.229 & 0.871 & 0.263 & 3 \\ \hline
        892832 & 12.207 & 0.194 & 0.867 & 0.224 & 3 \\ \hline
        892834 & 13.578 & 0.023 & 0.959 & 0.024 & \\ \hline
        892882 & 20.415 & 0.052 & 0.95 & 0.055 & \\ \hline
        892911 & 12.094 & 0.305 & 0.866 & 0.352 & 3 \\ \hline
        892986 & 9.397 & 0.21 & 0.867 & 0.242 & 3 \\ \hline
        893004 & 11.916 & 0.228 & 0.867 & 0.262 & 3 \\ \hline
        893033 & 16.317 & 0.335 & 0.904 & 0.371 & \\ \hline
        893165 & 8.808 & 0.098 & 0.864 & 0.113 & 3\\ \hline
        893209 & 5.642 & 0.174 & 0.934 & 0.186 & \\ \hline
    \end{tabular}
    \caption{An example of the resulting predicted periods on \textit{Kepler} samples. The Flag column consists of the following values: 1 - confirmed or candidate planet host star; 2 -  Eclipsing Binary; 3 - samples where ACF algorithm failed to predict period on one or more segments.  The full table is available in machine-readable format.  \citep{kamai2024}}
\end{center}
\end{table*}

\subsection{Kepler Catalog}
\label{subsec:catalog}
We now briefly summarize all the selection processes we made towards obtaining the final sample. After the selections discussed in \ref{subsec:period_pred} we had predictions for 108096 samples. Next, we did the following:
\begin{itemize}
    \item We removed pulsators identified by \cite{Santos2021} - 1046 samples.
    \item We replaced period values with predictions of \cite{McQuillan2014} in case of higher second peak and double period difference - 2231 samples.
    \item We removed samples with period $< 3$ Days as predicted by ACF or \cite{McQuillan2014}, and EBs with orbital period $< 3$ Days - 3106 samples.
    \item We removed samples with confidence scores lower than 0.86 - 21173 samples.
\end{itemize}

This procedure left us with 82771 reliable predictions. An example of the final catalog 
is presented in Table 8. The full catalog is available in machine-readable format. 

\section{Discussion and summary}\label{sec:conclusions}
In this study, we introduced LightPred, a novel deep-learning model designed to extract stellar rotation periods from light curves, demonstrating the potential of deep learning in revolutionizing the analysis of stellar light curve data. By leveraging a dual-branch architecture that incorporates LSTM and Transformer components, LightPred excels at capturing both temporal dependencies and global patterns within light curves.

Our training process, combining simulated and real Kepler data, has allowed LightPred to outperform traditional methods like the Autocorrelation Function in terms of accuracy and robustness, particularly for noisy data. This advancement has led to the generation of the largest and most accurate catalog of stellar rotation periods for main-sequence stars to date, comprising over 80000 stars.

Furthermore, our analysis of eclipsing binaries confirms tidal synchronization in systems with orbital periods shorter than 10 days, further validating our approach. Another important result that emerged from the analysis is the ability to cluster the predictions to different stellar types based on the confidence level of the model. Using this property, we were able to filter out false positive predictions.

A comparison between our results and ACF results from \cite{McQuillan2014}, reveals a subgroup of slow rotating stars that shows non-consistent predictions among the methods. Those samples have challenging periodicity patterns and can be further analyzed.

Our findings illustrate the potential of a data-driven approach for the analysis of stellar lightcurves and pave the way for future research, including the development of improved spot models for simulation data, particularly to account for the behavior of fast-rotating young stars. We also anticipate further exploration of stellar inclination predictions, a more complex task that is crucial for a comprehensive understanding of stellar systems.

\section{acknowledgments}
We would like to thank the anonymous referee for helpful comments that improved the manuscript. We would like to acknowledge the support from the Minerva Center for life under extreme planetary conditions.

\bibliographystyle{aasjournal}
\bibliography{main}

\appendix

\section{Pre-processing examples}
\label{Appendix_A}
Examples of simulated light curves and the different preprocessing stages are shown in the figures. 
\begin{figure}[H]
    \centering
    \includegraphics[scale=0.3]{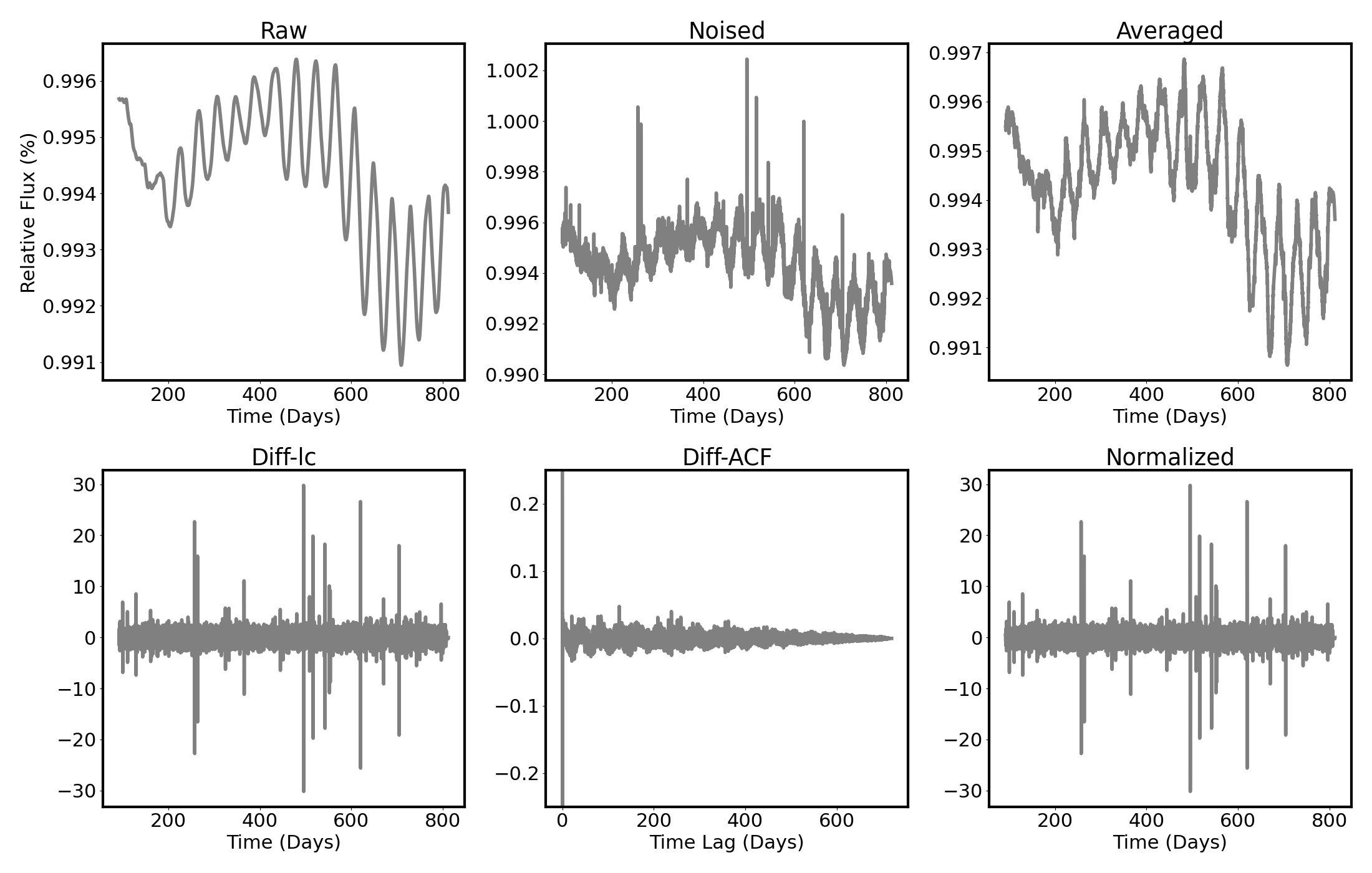}
\end{figure}
\begin{figure}[H]
    \centering
    \includegraphics[scale=0.3]{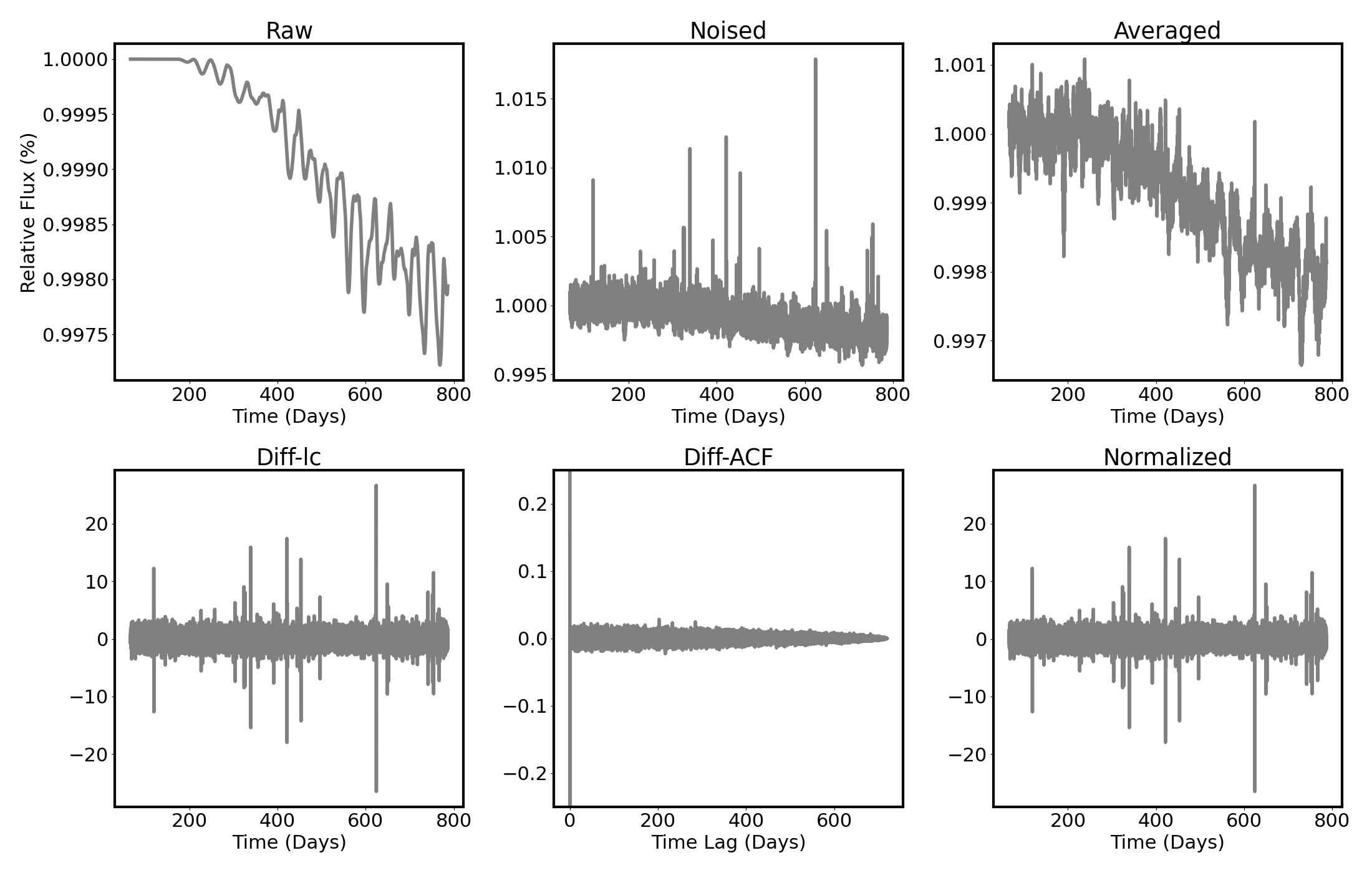}
\end{figure}
\begin{figure}[H]
    \centering
    \includegraphics[scale=0.3]{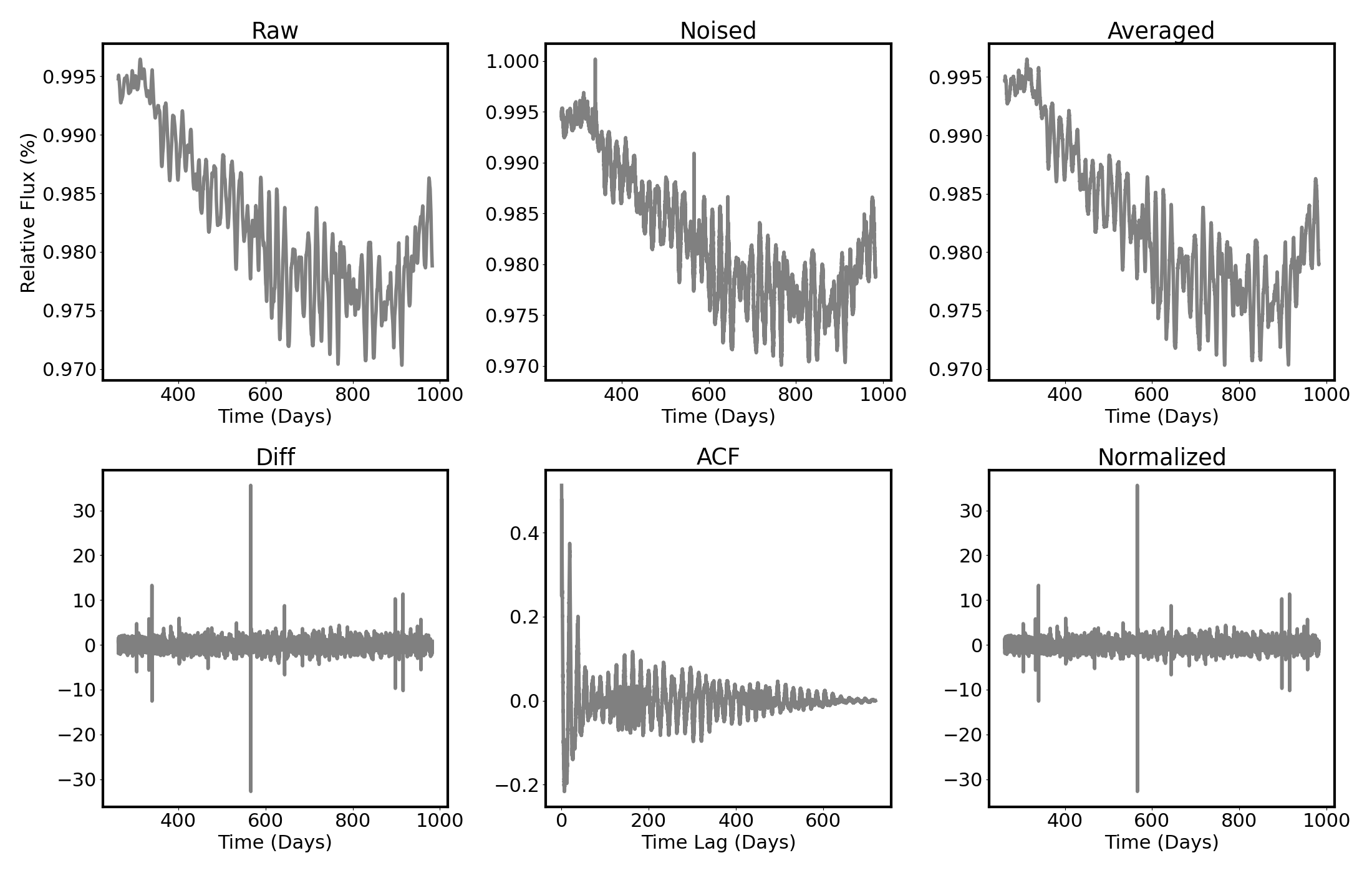}
    \noindent{Examples of simulated light curve and the different preprocessing stages.} 
\end{figure}

\section{LightPred Model Architecture}
\label{Appendix_B}
LightPred utilizes a dual-branch architecture to analyze light curve data comprehensively. The detailed architecture and its schematics are described and shown below, followed by a table of the model parameters.

\begin{itemize}
\item \textbf{Temporal Branch:}
\begin{itemize}
\item \textbf{Input:} The Diff-ACF (differenced autocorrelation function) of the light curve. This pre-processed input highlights short-term variations and periodic patterns related to stellar rotation.
\item \textbf{Convolutional Block:}
A 1D convolutional layer extracts local features and patterns from the Diff-ACF. Max pooling layer downsample the features to reduce dimensionality and computational cost while preserving essential information.
\item \textbf{Bi-directional LSTM Block:}
Long Short-Term Memory (LSTM) layers, arranged in a bi-directional fashion, process the convolutional features to capture both forward and backward temporal dependencies. LSTMs excel at modeling sequential data and remembering information over long durations, making them ideal for identifying the evolving nature of starspot patterns and periodicities in light curves.
\item \textbf{Self-Attention Block:}
A non-learnable self-attention mechanism refines the LSTM's final hidden states by weighting the importance of different time steps based on their relevance to the rotation period. This helps focus the model on the most informative segments of the Diff-ACF.
\end{itemize}

\item \textbf{Global Branch:}
\begin{itemize}
\item \textbf{Input:} The Diff-Lc (differenced light curve). This input provides a broader view of the overall flux variations and long-term trends in the light curve.
\item \textbf{AstroConformer Block:}
This block, a modified version of the Conformer architecture, combines multi-head self-attention with convolutional layers.
\begin{itemize}
\item Multi-head self-attention: Captures long-range dependencies and global context by allowing the model to weigh the importance of different time steps in relation to each other. This is crucial for identifying periodicities that may not be as apparent in the local features extracted by the temporal branch.
\item  Convolutional layers: Extract local features from the light curve, complementing the global context captured by self-attention.
\end{itemize}
\end{itemize}
\item \textbf{Prediction Head:}
\begin{itemize}
\item \textbf{Concatenation:} Features from both branches are concatenated, combining temporal and global representations.
\item \textbf{Fully Connected Layers:} Two dense layers with GELU activation functions and dropout regularization process the concatenated features. This allows the model to learn complex relationships between the features and the target outputs.
\item \textbf{Output:}
The model produces four values:
\begin{itemize}
\item Stellar rotation period
\item Stellar inclination
\item Confidence score for period prediction 
\item Confidence score for inclination prediction
\end{itemize}
\end{itemize}
\end{itemize}

\begin{figure}[H]
    \begin{center}
    \includegraphics[scale=0.45]{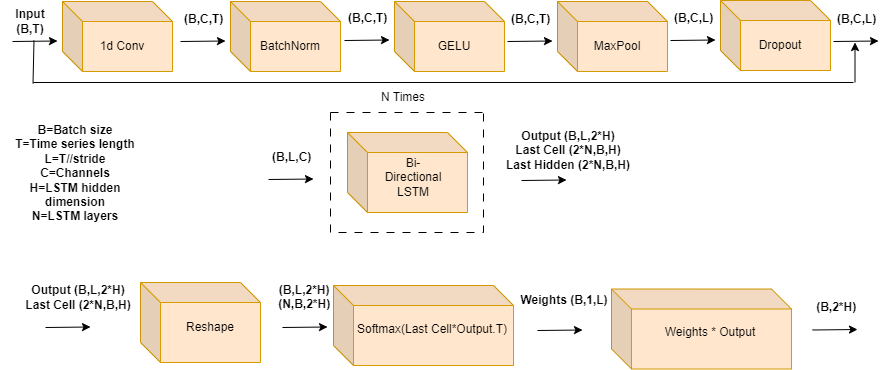}
    \noindent{\\Temporal block.}
    \end{center}
\end{figure}

\begin{figure}[H]
    \begin{center}
    \includegraphics[scale=0.45]{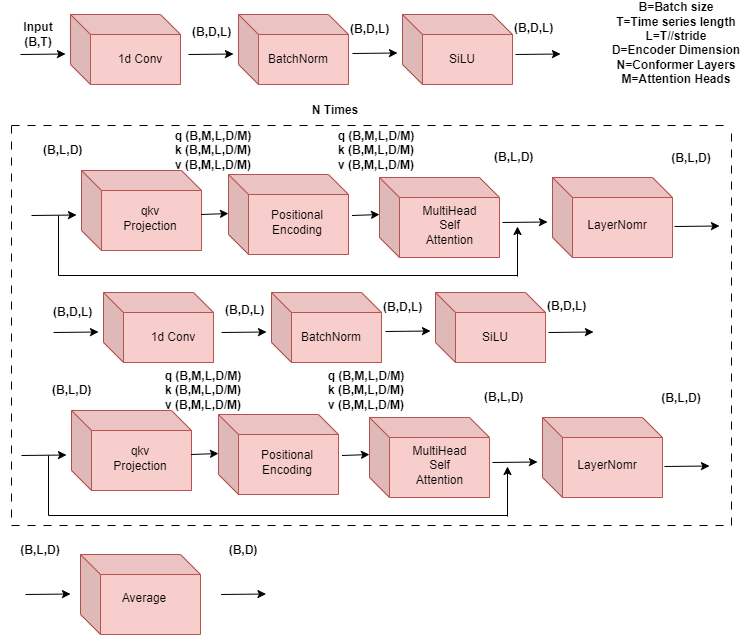}
    \noindent{\\Conformer Block.}
    \end{center}    
\end{figure}

\begin{figure}[H]
    \begin{center}        
    \includegraphics[scale=0.45]{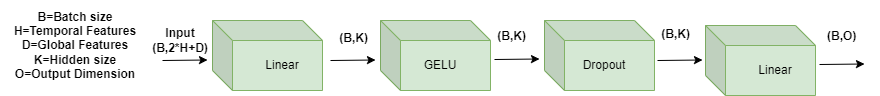}
    \noindent{\\Prediction head.}
    \end{center}    
\end{figure}

\begin{table}[!h]
\begin{center}
\begin{tabular}{||c c||} 
 \hline
 parameter & value  \\ [0.5ex] 
 \hline\hline
 kernel size  & 4 \\ 
 \hline
 stride-temporal & 4 \\
 \hline
 stride-conformer & 20 \\
 \hline
 channels-temporal  & 256 \\
 \hline
 Dropout-temporal & 0.35  \\
 \hline
 Dropout-conformer & 0.3  \\
 \hline
 LSTM layers & 5 \\
 \hline
 LSTM hidden size & 64 \\ 
 \hline
 conformer layers & 5 \\
 \hline
 attention heads & 8 \\
 \hline
 conformer encoder dim & 128 \\
 \hline
 fully connected hidden size & 128 \\ [1ex] 
 \hline
\end{tabular}
\noindent{\\parameters of the LightPred model}
\end{center}
\end{table}

\section{Example of the comparison between self-supervised and simulation-based training}
\label{Appendix_C}

\begin{figure}[H]
    \begin{center}
  \begin{minipage}[b]{0.45\textwidth}
    \includegraphics[width=\textwidth]{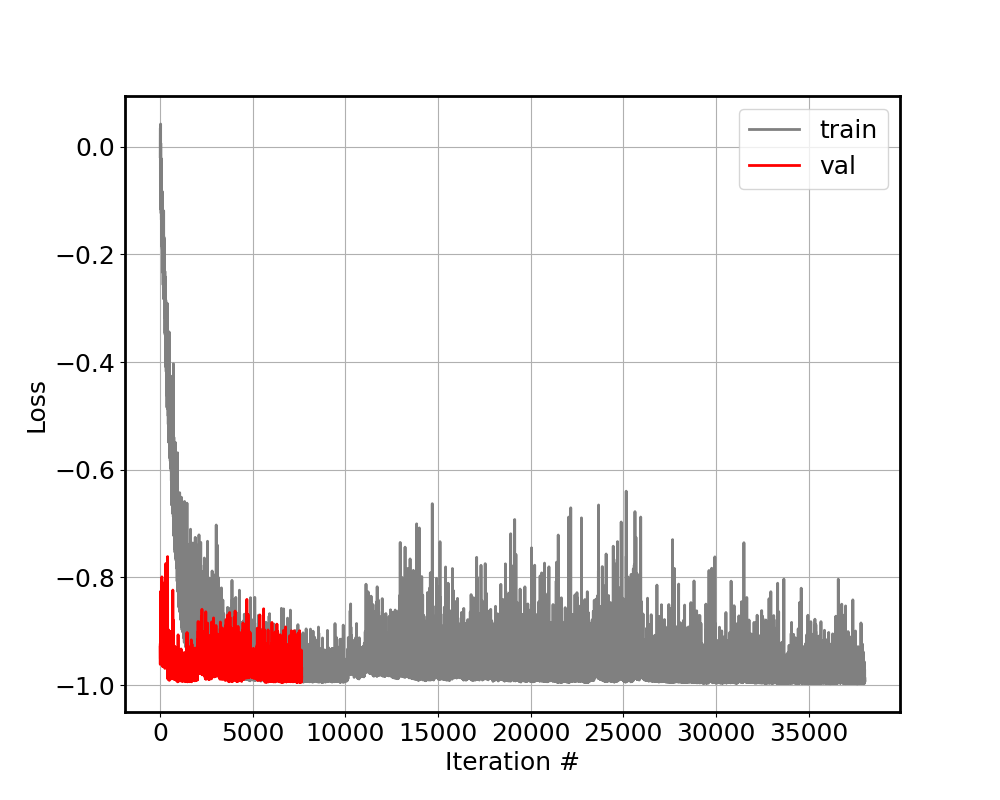}
  \end{minipage}
  \begin{minipage}[b]{0.45\textwidth}
    \includegraphics[width=\textwidth]{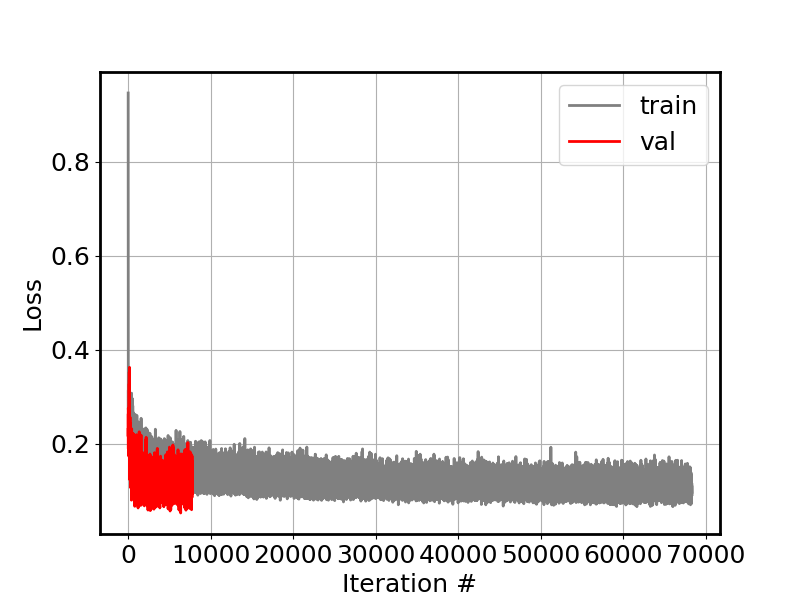}
  \end{minipage}
    \noindent{\\Self-supervised training (upper panel) and simulation-based training (lower panel). Grey represents the training dataset; Red represents the validation dataset.  }
    \end{center}
\end{figure}

\section{Fast Rotators}
\label{Appendix_D}

\begin{figure}[H]
    \begin{center}
  \begin{minipage}[b]{0.45\textwidth}
    \includegraphics[width=\textwidth]{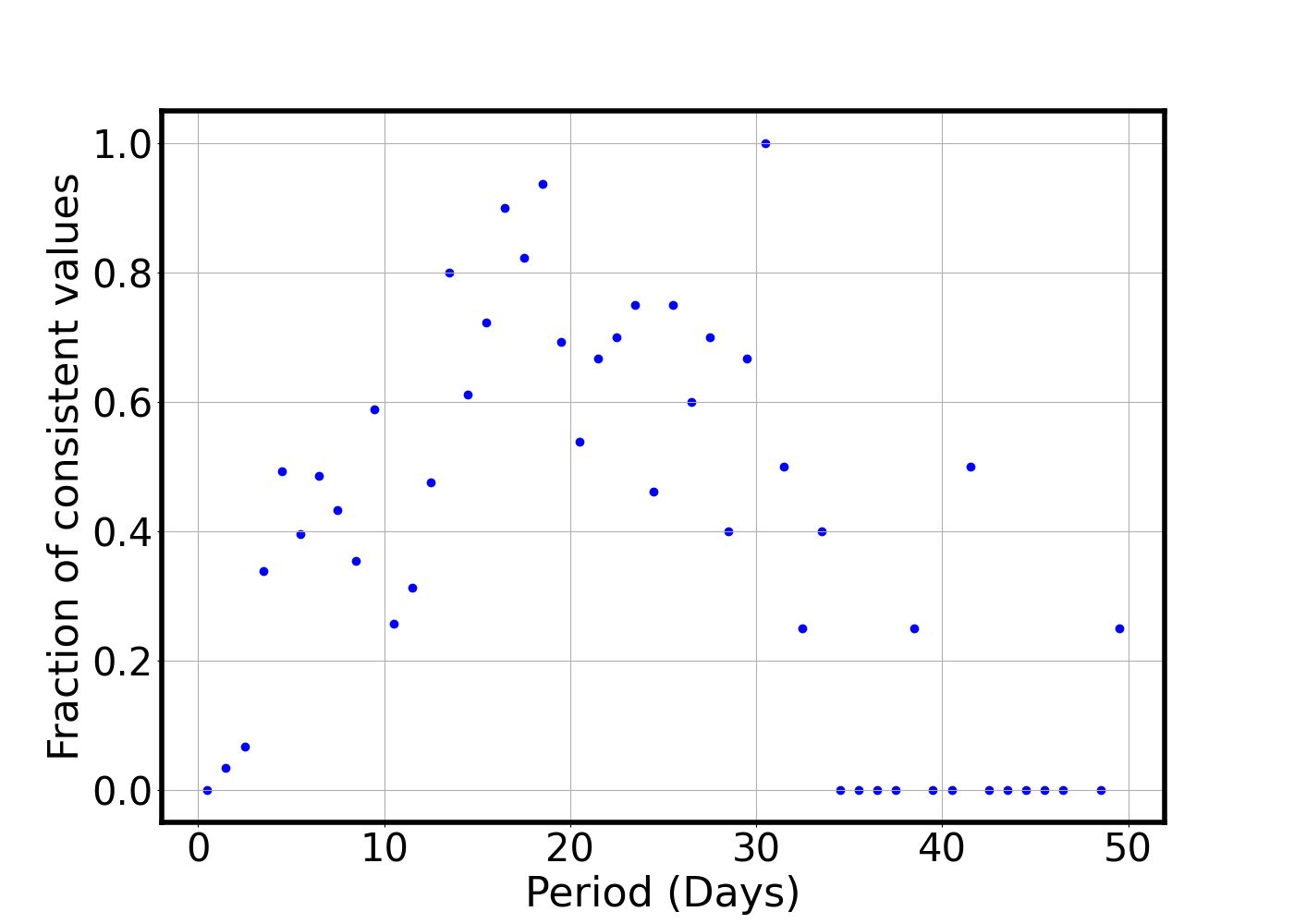}
  \end{minipage}
  \begin{minipage}[b]{0.45\textwidth}
    \includegraphics[width=\textwidth]{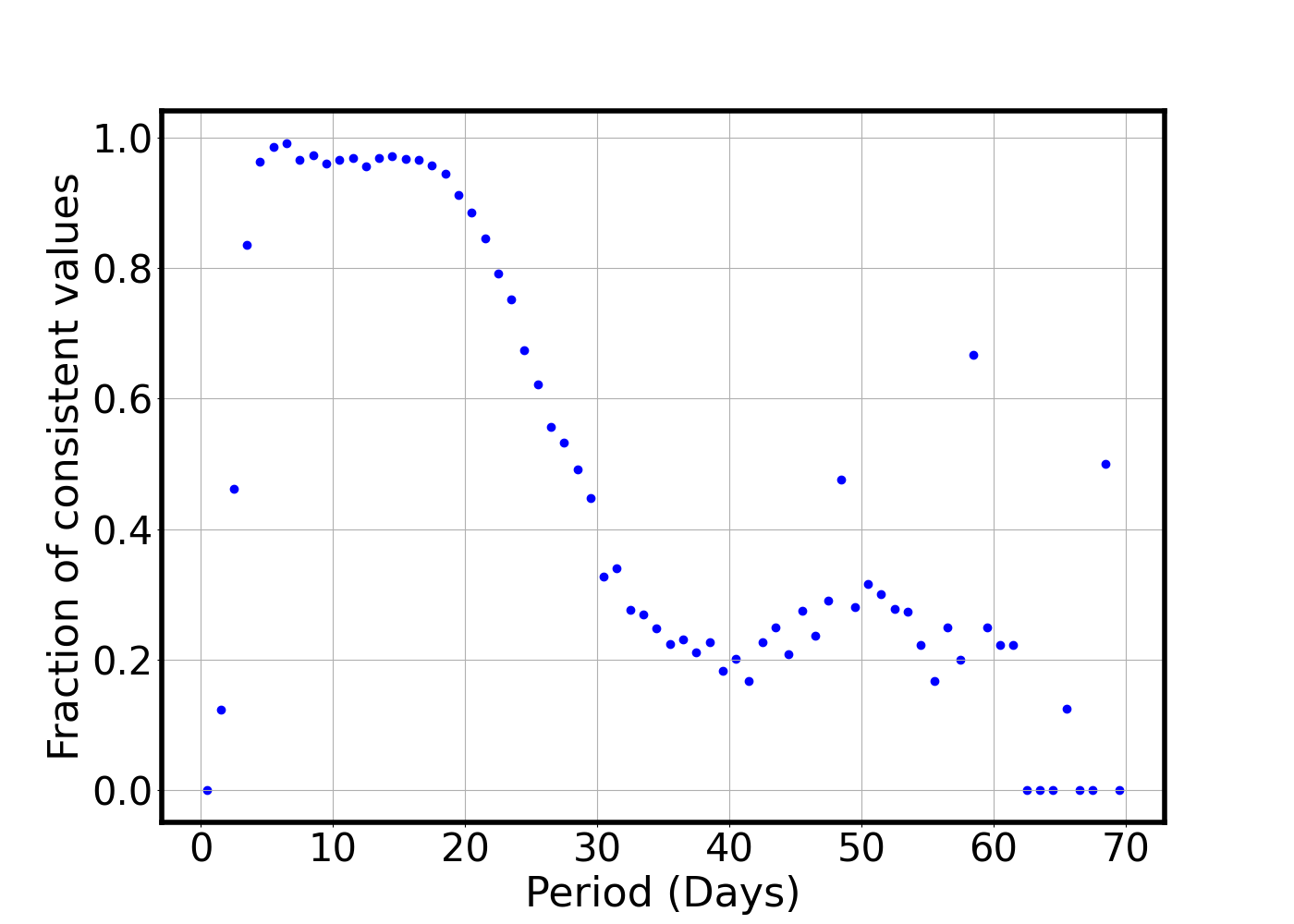}
  \end{minipage}
   \noindent{\\consistency curves of Eclipsing Binaries dataset (left) and \cite{McQuillan2014} Dataset (right).}
    \end{center}
\end{figure}

\begin{figure}[H]
    \begin{centering}
    \begin{minipage}[b]{0.3\textwidth}
        \includegraphics[width=\textwidth]{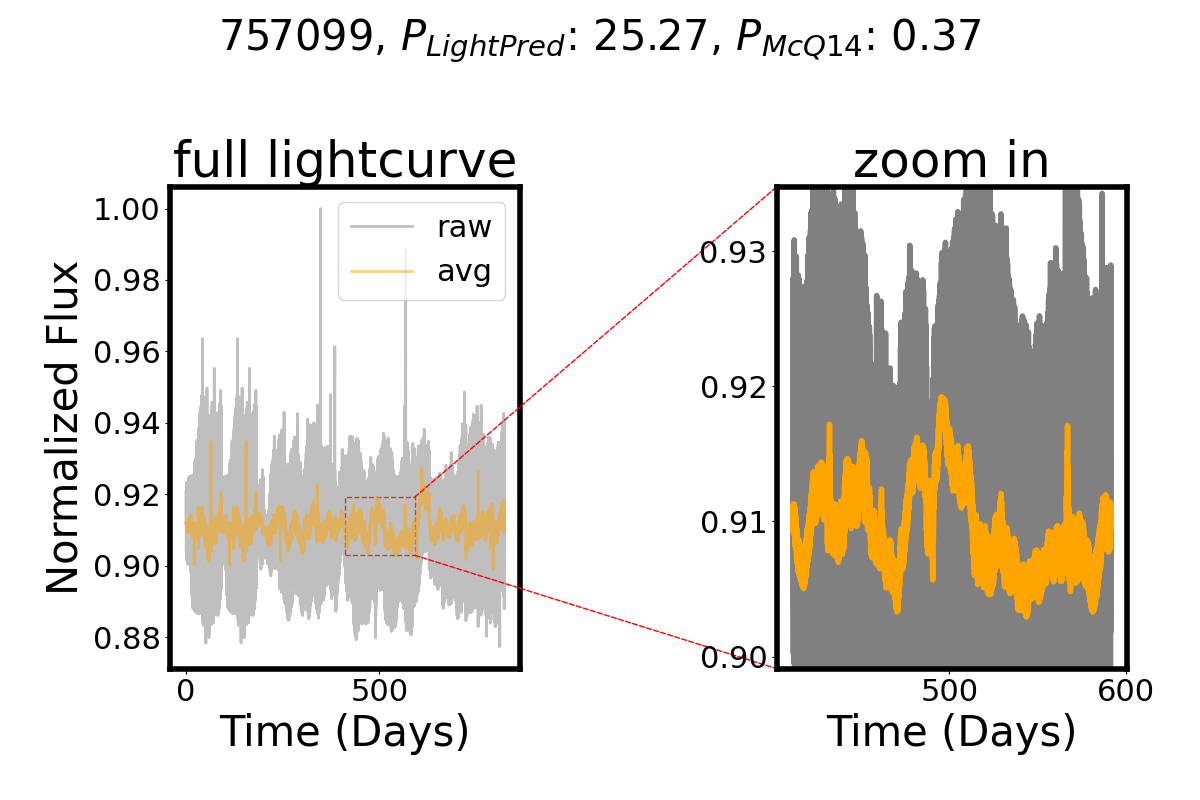}
    \end{minipage}
    \begin{minipage}[b]{0.3\textwidth}
       \includegraphics[width=\textwidth]{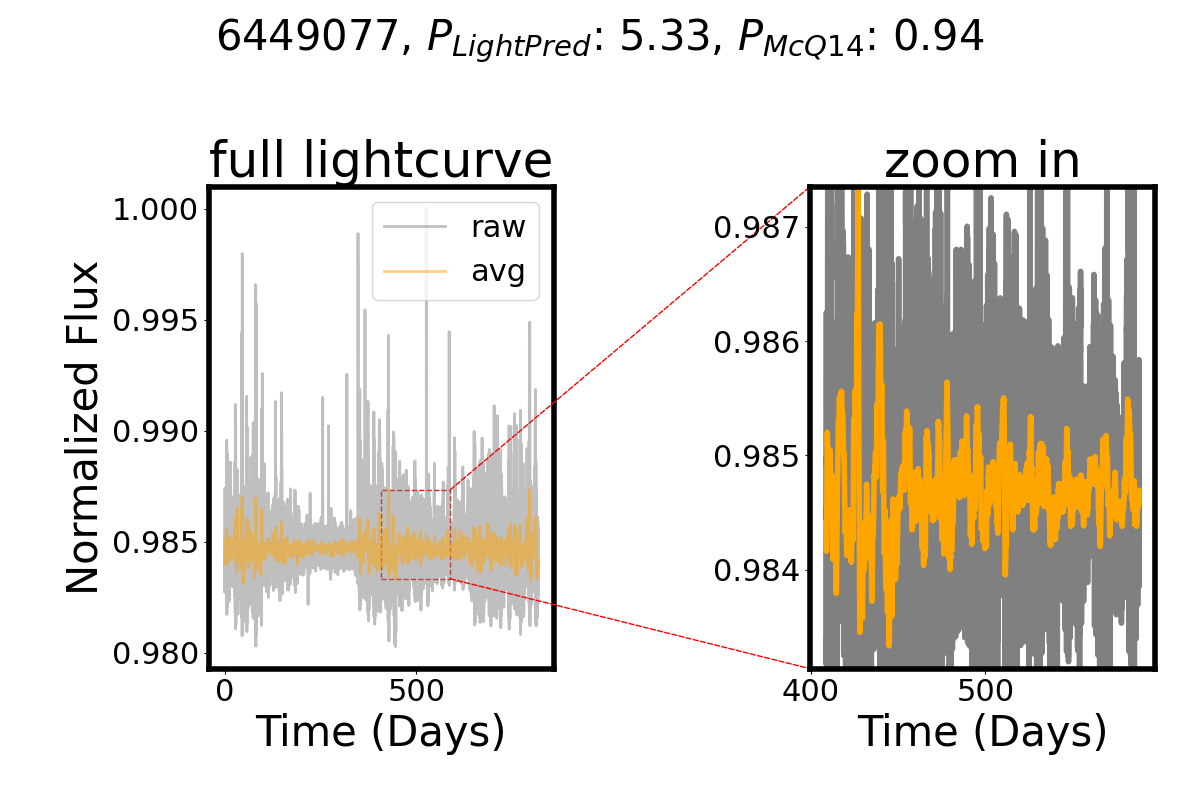}
    \end{minipage}
    \begin{minipage}[b]{0.3\textwidth}
       \includegraphics[width=\textwidth]{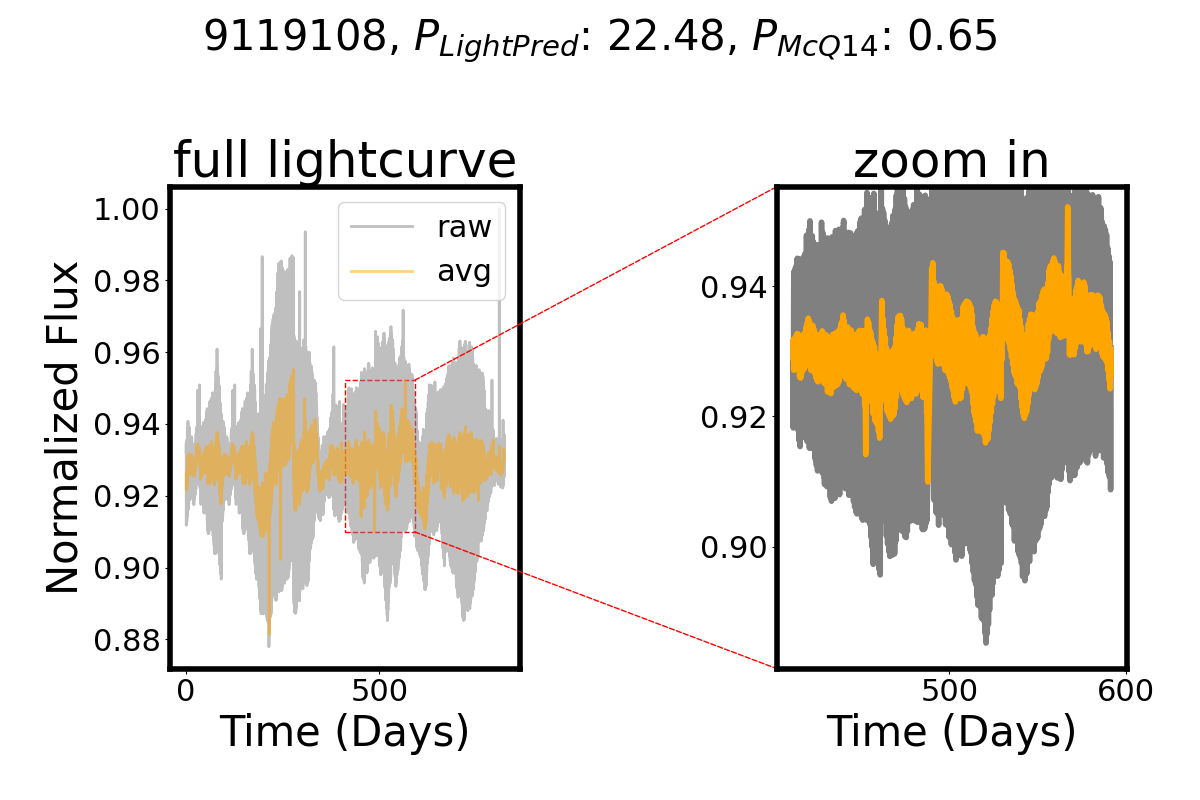}
    \end{minipage}
    \captionsetup{labelformat=empty}
    \noindent{\\Three potential wrong predictions of our model. 
    For each sample, the left panel shows the full light curve and the right panel shows a zoom-in slice of 180 days. The gray color represents the raw light curve.  The orange color represents an averaged light curve over a window of 1 day. The titles show the \textit{Kepler} KIC, lightPred period, and McQ14 period from\cite{McQuillan2014}. }
    \end{centering}      
\end{figure}

\end{document}